\documentclass[prd,aps,nofootinbib,floatfix,11pt]{revtex4}
\usepackage{amsmath,graphicx,epsfig,amssymb,dsfont,mathtools,}
\usepackage[usenames]{color}
\usepackage{ulem} 
\usepackage{bigstrut}
\usepackage{slashed}
\usepackage{multirow}
\usepackage{subfigure}

\allowdisplaybreaks

\begin{document}\title{SU(3) analysis of fully-light tetraquarks in heavy meson weak decays}
\author{Yu-Ji Shi~$^{1}$~\footnote{Email:shiyuji92@126.com},
   Ye Xing~$^{2}$~\footnote{Email:xingye\_guang@cumt.edu.cn} and 
  Zhen-Xing Zhao~$^{3}$~\footnote{Email:zhaozx19@imu.edu.cn}}

\affiliation{$^{1}$ Helmholtz-Institut f\"ur Strahlen- und Kernphysik and Bethe Center \\ for Theoretical Physics,
  Universit\"at Bonn, 53115 Bonn, Germany\\
  $^2$ School of Physics, China University of Mining and Technology, Xuzhou 221000, China\\
  $^{3}$ School of Physical Science and Technology, Inner Mongolia University, Hohhot 010021, China}

\begin{abstract}
We perform a SU(3) analysis for both semi-leptonic and non-leptonic heavy meson weak decays into a pseudoscalar meson and a fully-light tetraquark in 10 or 27 representation. A reduction of the SU(3) representation tensor for the fully-light tetraquarks is produced and all the flavor components for each representation tensor are listed. The decay channels we analysis include $B/D \to U/T~P~l\nu$, $B/D \to U/T~P $ and $B_c \to U/T~P/D$, with $U/T$ represents a fully-light tetraquark in 10 or 27 representation and $P$ is a pseudoscalar meson. Finally, among these results we list all the golden decay channels which are expected to have more possibilities to be observed in experiments. 

\end{abstract}
\maketitle

\section{Introduction}

The existence of multi-quark states, particularly the four-quark states is firstly predicted by M. GellMann according to the traditional quark model in the early 1960s.  It was not until 2003 that Belle Collaboration observed the first hidden charm four-quark candidate $X(3872)$~\cite{Choi:2003ue} in $B^{\pm} \to K^{\pm} X(X\to \pi^+\pi^- J/\psi)$ decays. Subsequently, on one side, more  four-quark states with hidden heavy quark flavor are observed, for instance $Z_c(3900)$~\cite{Ablikim:2013mio,Liu:2013dau} and $Z_b(10610)^{\pm}$~\cite{Oswald:2013tna} found by BESIII and Belle Collaborations. On the other side, however, the fully-light states $qq\bar q\bar q$$(q=u,d,s)$ are still unconfirmed by the experiments. The BES Collaboration has observed new signals $X(1835)$, $X(2120)$ and $X(2370)$ in the $J/\psi \to \gamma X$ channels~\cite{Ablikim:2005um,Ablikim:2010au,Ablikim:2016hlu,Ablikim:2019zyw}, which can be the four-quark state candidates while other candidates like glueballs, charmoniums or baryoniums ~\cite{Fermi:1959sa,Richard:1999qh,Zhu:2007wz} are still possible. Recently, further examining and studying on the light-quark exotic states by  BESIII, Belle-II  and LHCb Collaborations are in progress~\cite{Ablikim:2013wzq,Ablikim:2013emm,Ablikim:2013xfr,Aaij:2014jqa,Ablikim:2020das,Ablikim:2020coo,Ablikim:2020pgw,Kou:2018nap,Austregesilo:2018mno,Ablikim:2019hff,Aaij:2020tzn,Aaij:2020hpf,Aaij:2020hon}.

On the theoretical side, our understanding on these exotic states is still far from accomplished. In general, the nature of structure and properties of exotic hadrons may be quite different from the ordinary mesons. They can have quantum numbers that cannot be explained by the usual method used for ordinary hadrons. Therefore, the identification of the internal structure of the exotic states is always a critical problem, which requires a careful analysis of experimental observations and theoretical predictions, see e.g.~\cite{Esposito:2016noz,Guo:2017jvc} for a review.
In terms of the four-quark state, the ones with its quarks and antiquarks clustering into diquark-anti-diquark pairs is called a tetraquark which is combined by the color force, while the one with meson-molecule structure is combined by the electroweak force. In this paper, we will concentrate on the production of light four-quark states in the two-body weak decays of heavy mesons $B$, $B_c$ and $D$. Due to the rich yield of heavy mesons in the Heavy Flavor Factories, a considerable production of light four-quark states in these decay channels is expected.

Recently, most of the theoretical studied on the exotic states are relied on effective theories and models, for example the studies of mass spectrum within the simple quark model~\cite{Cui:2005az}, Iachello mass formula \cite{Santopinto:2006my} and QCD sum rules~\cite{Cui:2019roq}. However, the light quark flavor SU(3) symmetry is also a useful tool to analyze the decays of hadrons, which has been successfully applied for the ordinary meson or baryon case~\cite{Savage:1989ub,Gronau:1995hm,He:1998rq,Chiang:2004nm,Li:2007bh,Wang:2009azc,Cheng:2011qh,Hsiao:2015iiu,Lu:2016ogy,He:2016xvd,Wang:2017vnc,Wang:2017mqp,Wang:2017azm,Shi:2017dto,Wang:2018utj,He:2018php,He:2018joe,Wang:2020gmn,Xing:2018bqt,Xing:2019wil}. One advantage of SU(3) analysis is that it is irrelevant with the details of  the hadron structure, particularly whether the four-quark state is a bounded by diquark  and anti-diquark or  is a meson-molecule, as well as its explicit quantum number $J^{PC}$.  For this reason we will simply call the four-quark state in this work tetraquark. In this paper, we will choose the SU(3) symmetry analysis to study  the production of fully-light tetraquarks in the both semi-leptonic and non-leptonic $B$, $B_c$ and $D$ decays.  According to the reduction of SU(3) representation, a fully-light tetraquark can belong to a $27$ representation, a 10 representation, a $\overline {10}$ representation, four 8 representations or two singlets.  In this work, we will focus on the fully-light tetraquark in 27 or 10 representation, and study their production in the decays $B/D \to U/T~P~l\nu$, $B/D \to U/T~P $ and $B_c \to U/T~P/D$, with $U/T$ represents the 10 or 27 states and $P$ is a pseudoscalar meson. According to the SU(3) symmetry the relations among these decay channels can be obtained, which can be examined by the future experiments. This analysis is helpful to identify the decay modes that will be mostly useful to discover the fully-light tetraquark states.

The rest of this paper is organized as follows. In Sec. II, we give the representation of fully-light tetraquarks under the SU(3) symmetry. In Sec III we give the SU(3) representation of the Standard Model (SM) Hamiltonians which are relevant to semi-leptonic and non-leptonic $B$, $B_c$ and $D$ decays. Sec. IV is the SU(3) analysis of $B/D \to U/T~P~l\nu$, $B/D \to U/T~P $ and $B_c \to U/T~P/D$ decays. In Sec. V we list all the golden decay channels which have greater chance to be observed in the experiments than other channels. Finally, Sec. VI is a short summary.

\section{SU(3) irreducible representation of fully-light tetraquarks}
\label{sec:SU3irreducTetraq}
In general, according to the SU(3) flavor symmetry, the fully-light tetraquarks $q_1 q_2 {\bar q}_3{\bar q}_4$  are described by the inner-product  representation $3\otimes3\otimes\bar{3}\otimes\bar{3}$, which can be further reduced into 9 irreducible representaions
\begin{equation}
3\otimes3\otimes\bar{3}\otimes\bar{3}=27\oplus10\oplus\overline{10}\oplus8\oplus8\oplus8\oplus8\oplus1\oplus1.
\end{equation}
Explicitly, in the language of tensor reduction, an fully-light tetraquark can be represented by a rank $(2,2)$ tensor $H^{ij}_{kl}$, and the reduction reads as
\begin{align}
H_{kl}^{ij} =&\ (T_{27}){}_{\{kl\}}^{\{ij\}}+\frac{1}{2}\epsilon_{klm}(T_{10})^{\{ijm\}}+\frac{1}{2}\epsilon^{ijn}(T_{\overline{10}})_{\{kln\}}\nonumber\\
 & +\frac{1}{5}A_{klm}^{ijn}(T_{8}^{(1)})_{n}^{m}-\frac{1}{6}B_{klm}^{ijn}(T_{8}^{(2)})_{n}^{m}-\frac{1}{6}C_{klm}^{ijn}(T_{8}^{(3)})_{n}^{m}+\frac{1}{2}\epsilon^{ijn}\epsilon_{klm}(T_{8}^{(4)})_{n}^{m}\nonumber\nonumber\\
 & +\frac{1}{12}(\delta_{k}^{i}\delta_{l}^{j}+\delta_{l}^{i}\delta_{k}^{j})T_{1}^{(1)}-\frac{1}{6}(\delta_{k}^{i}\delta_{l}^{j}-\delta_{l}^{i}\delta_{k}^{j})T_{1}^{(2)},\label{TensorRedu}
\end{align}
where
\begin{align*}
A_{klm}^{ijn} & =\delta_{k}^{i}\delta_{m}^{j}\delta_{l}^{n}+\delta_{k}^{j}\delta_{m}^{i}\delta_{l}^{n}+\delta_{l}^{i}\delta_{m}^{j}\delta_{k}^{n}+\delta_{l}^{j}\delta_{m}^{i}\delta_{k}^{n},\\
B_{klm}^{ijn} & =\delta_{k}^{i}\delta_{m}^{j}\delta_{l}^{n}+\delta_{k}^{j}\delta_{m}^{i}\delta_{l}^{n}-\delta_{l}^{i}\delta_{m}^{j}\delta_{k}^{n}-\delta_{l}^{j}\delta_{m}^{i}\delta_{k}^{n},\\
C_{klm}^{ijn} & =\delta_{k}^{i}\delta_{m}^{j}\delta_{l}^{n}-\delta_{k}^{j}\delta_{m}^{i}\delta_{l}^{n}+\delta_{l}^{i}\delta_{m}^{j}\delta_{k}^{n}-\delta_{l}^{j}\delta_{m}^{i}\delta_{k}^{n}.
\end{align*}
Here the coefficient of each term in Eq.~(\ref{TensorRedu}) can always be rescaled by redefining the corresponding irreducible representation tensors. Explicitly, these tensors read as
\begin{align}
(T_{10})^{\{ijm\}} & =H_{kl}^{\{ij}\epsilon^{m\}kl},~~~~(T_{\overline{10}})_{\{kln\}} =\epsilon_{ij\{n}H_{kl\}}^{ij},\nonumber\\
(T_{8}^{(1)})_{n}^{m} & =H_{\{in\}}^{\{im\}}-\frac{1}{3}\delta_{n}^{m}H_{\{ij\}}^{\{ij\}},~~~~
(T_{8}^{(2)})_{n}^{m}  =H_{nj}^{\{mj\}}-H_{jn}^{\{mj\}},\nonumber\\
(T_{8}^{(3)})_{n}^{m}  &=H_{\{nj\}}^{mj}-H_{\{nj\}}^{jm},~~~~
(T_{8}^{(4)})_{n}^{m}  =\frac{1}{2}\epsilon_{ijn}\epsilon^{klm}H_{[kl]}^{[ij]}-\frac{1}{3}\delta_{n}^{m}H_{[ij]}^{[ij]},\nonumber\\
T_{1}^{(1)} & =H_{\{ij\}}^{\{ij\}},~~~~T_{1}^{(2)} =H_{[ij]}^{[ij]},\nonumber\\
(T_{27}){}_{\{kl\}}^{\{ij\}} & =H_{\{kl\}}^{\{ij\}}-\frac{1}{5}A_{klm}^{ijn}(T_{8}^{(1)})_{n}^{m}-\frac{1}{12}(\delta_{k}^{i}\delta_{l}^{j}+\delta_{l}^{i}\delta_{k}^{j})T_{1}^{(1)}.
\end{align}

By writing down all the components of these tensors, one can find out the flavor structure of each tetraquark. For example, a tetraquark with flavor $ss{\bar u}{\bar d}$  belongs to the 10 or 27 representation. The corresponding components are
\begin{equation}
(T_{10})^{333} =ss\bar{u}\bar{d}-ss\bar{d}\bar{u}~,~~~
(T_{27})_{12}^{33}  =\frac{1}{2}\left(ss\bar{u}\bar{d}+ss\bar{d}\bar{u}\right).
\end{equation}
Note that in the $(T_{10})^{333}$, the $u, d$ quarks are anti-symmetric in the flavor space, which means that without angular momentum, they form a spin-0 structure. On the other hand,  in the $(T_{27})_{12}^{33}$  $u, d$ are symmetric and thus form a spin-1 structure. All the components of the tensors in $27, 10, \overline{10}, 8\  \text{and} \ 1$ representations are listed in Appendix~\ref{flavorWFs}. For simplicity, we can rename each component according to its electric charge $Q$ and isospin $I_z$, namely $U^Q_{I_z}$ for the 10 representation and $T^Q_{I_z}$ for the 27 representation. The components for the 10 representation are denoted as 
\begin{align}
(T_{10})^{111}=U_{3/2}^{++},~
(T_{10})^{112}=U_{1/2}^{+},~
(T_{10})^{113}=U_{1}^{+},~
(T_{10})^{122}=U_{-1/2}^{0},~
(T_{10})^{222}=U_{-3/2}^{-},\nonumber\\
(T_{10})^{123}=U_{0}^{0},~
(T_{10})^{133}=U_{1/2}^{0},~
(T_{10})^{223}=U_{-1}^{-},~
(T_{10})^{233}=U_{-1/2}^{-},~
(T_{10})^{333}=U_{0}^{-}.
\end{align}
\begin{table}
\begin{center}
\caption{Independent components of 27 representation tensor}
\label{tab:T27compo}
\begin{tabular}{|ccc|c|c|}
\hline 
\hline 
$I_{z}$ & $Y$ & Q & $T_{27}$ components & Notation $T_{I_z}^Q$\tabularnewline
\hline 
2 & 0 & ++ & $(T_{27})_{22}^{11}$ & $T_{2}^{++}$\tabularnewline
1 & 0 & + & $(T_{27})_{22}^{12}$, $(T_{27})_{12}^{11}$ & $T_{1}^{+}$, $T_{1}^{\prime+}$\tabularnewline
0 & 0 & 0 & $(T_{27})_{11}^{11}$, $(T_{27})_{12}^{12}$, $(T_{27})_{22}^{22}$ & $T_{0}^{0}$, $T_{0}^{\prime0}$, $T_{0}^{\prime\prime0}$\tabularnewline
-1 & 0 & $-$ & $(T_{27})_{11}^{12}$, $(T_{27})_{12}^{22}$ & $T_{-1}^{-}$, $T_{-1}^{\prime-}$\tabularnewline
-2 & 0 & $- -$ & $(T_{27})_{11}^{22}$ & $T_{-2}^{--}$\tabularnewline
3/2 & 1 & ++ & $(T_{27})_{23}^{11}$ & $T_{3/2}^{++}$\tabularnewline
1/2 & 1 & + & $(T_{27})_{13}^{11}$, $(T_{27})_{23}^{12}$ & $T_{1/2}^{+}$, $T_{1/2}^{\prime+}$\tabularnewline
-1/2 & 1 & 0 & $(T_{27})_{13}^{12}$, $(T_{27})_{23}^{22}$ & $T_{-1/2}^{0}$, $T_{-1/2}^{\prime0}$\tabularnewline
-3/2 & 1 & $-$ & $(T_{27})_{13}^{22}$ & $T_{-3/2}^{-}$\tabularnewline
1 & 2 & ++ & $(T_{27})_{33}^{11}$ & $T_{1}^{++}$\tabularnewline
0 & 2 & + & $(T_{27})_{33}^{12}$ & $T_{0}^{+}$\tabularnewline
-1 & 2 & 0 & $(T_{27})_{33}^{22}$ & $T_{-1}^{0}$\tabularnewline
3/2 & -1 & + & $(T_{27})_{22}^{13}$ & $T_{3/2}^{+}$\tabularnewline
1/2 & -1 & 0 & $(T_{27})_{12}^{13}$, $(T_{27})_{22}^{23}$ & $T_{1/2}^{0}$, $T_{1/2}^{\prime0}$\tabularnewline
-1/2 & -1 & $-$ & $(T_{27})_{11}^{13}$, $(T_{27})_{12}^{23}$ & $T_{-1/2}^{-}$, $T_{-1/2}^{\prime-}$\tabularnewline
-3/2 & -1 & $- -$ & $(T_{27})_{11}^{23}$ & $T_{-3/2}^{--}$\tabularnewline
1 & -2 & 0 & $(T_{27})_{22}^{33}$ & $T_{1}^{0}$\tabularnewline
0 & -2 & $-$ & $(T_{27})_{12}^{33}$ & $T_{0}^{-}$\tabularnewline
-1 & -2 & $- -$ & $(T_{27})_{11}^{33}$ & $T_{-1}^{--}$\tabularnewline
\hline 
\end{tabular}
\end{center}
\end{table}
\begin{figure}
\includegraphics[width=0.8\columnwidth]{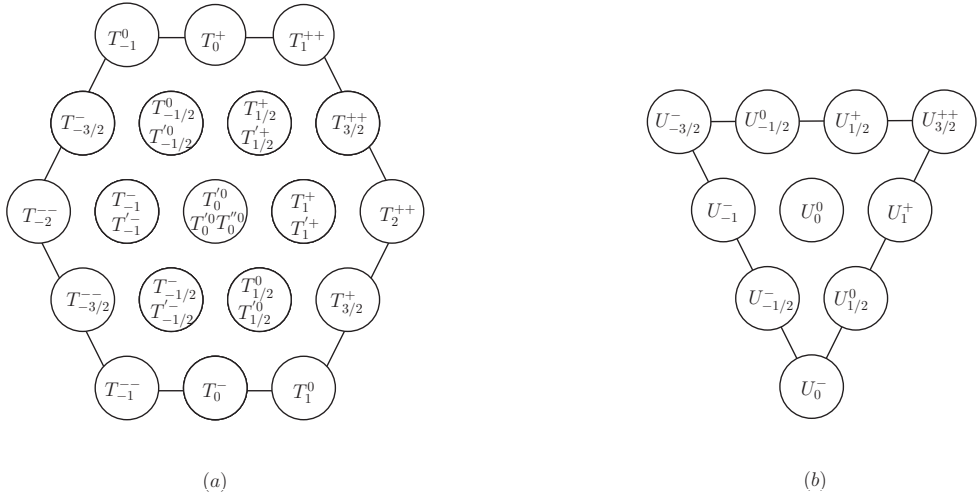} 
\caption{Weight diagrams of the fully-light tetraquarks in 27 (a) and 10 (b) representations.}
\label{fig:weitghtDiag} 
\end{figure}
For the 27 representation, the notation of all the independent components are listed in Table~\ref{tab:T27compo}, where the $(T_{27})^{i3}_{k3}$ components are absent because they are related with other components as
\begin{equation}
(T_{27})^{i3}_{k3}=-(T_{27})^{i1}_{k1}-(T_{27})^{i2}_{k2}
\end{equation}
due to the traceless condition. Particularly we have $(T_{10})^{333}=U_0^-$ and $(T_{27})_{12}^{33}=T_0^-$. The weight diagrams of the 27 and 10 states are shown in Fig.~\ref{fig:weitghtDiag}

\section{\bf Electroweak effective Hamiltonian in SU(3) representation}

\subsection{\bf Effective Hamiltonian for semi-leptonic $b$ or $c$ decays}
The electroweak effective Hamiltonian for semi-leptonic $b$ or $c$ decays in the SM is
\begin{align}
\mathcal{H}_{\mathrm{eff}}^b&=\frac{G_{F}}{\sqrt{2}}\left[V_{q^{\prime} b} \bar{q}^{\prime} \gamma^{\mu}\left(1-\gamma_{5}\right) b \bar{l} \gamma_{\mu}\left(1-\gamma_{5}\right) v\right]+h . c .~,\nonumber\\
\mathcal{H}_{\mathrm{eff}}^c&=\frac{G_{F}}{\sqrt{2}}\left[V_{c q}^{*} \bar{q} \gamma^{\mu}\left(1-\gamma_{5}\right) c \bar{\nu} \gamma_{\mu}\left(1-\gamma_{5}\right) l\right]+h . c .~,
\end{align}
where $q=d,s$ and $q^{\prime}=u$. In the SU(3) representation the $\mathcal{H}_{\mathrm{eff}}^b$ corresponds to a triplet operator $H_3$ with the components as $\left(H_{3}\right)^{1}=V_{ub} \text { and }\left(H_{3}\right)^{2,3}=0$. $\mathcal{H}_{\mathrm{eff}}^c$ also corresponds to a triplet operator $H_{3}^{\prime}$, with the components as $\left(H_{3}^{\prime}\right)^{1}=0,~\left(H_{3}^{\prime}\right)^{2}=V_{c d}^{*} \text { and }\left(H_{3}^{\prime}\right)^{3}=V_{c s}^{*}$.

\subsection{\bf Effective Hamiltonian for non-leptonic $b$ decays}
The electroweak effective Hamiltonian for non-leptonic $b$ decays in the SM is~\cite{Buchalla:1995vs,Ciuchini:1993vr,Deshpande:1994pc}:
 \begin{eqnarray}
 {\cal H}^b_{eff}  = \frac{G_{F}}{\sqrt{2}}
     \big\{ V_{ub} V_{uq}^{*}  [
     C_{1}  O_{1}
  +  C_{2}  O_{2}]
 - V_{tb} V_{tq}^{*}   {\sum\limits_{i=3}^{10}} C_{i}  O_{i} \big\}+ \mbox{h.c.},
 \label{eq:hamiltonian}
\end{eqnarray}
where  $O_{i}$ is the four-quark operator and $C_i$ is its Wilson coefficient.  The explicit forms of the $O_i$s read as 
\begin{eqnarray}
 O_1 = (\bar q^i u^j)_{V-A} (\bar u^j b^i)_{V-A}, && O_2 = (\bar q u)_{V-A} (\bar u b)_{V-A}, \nonumber\\
 O_3= (\bar q b)_{V-A} \sum_{q'} (\bar q'q')_{V-A}, && O_4= (\bar q^i b^j)_{V-A} \sum_{q'} (\bar q^{\prime j}q^{\prime i})_{V-A}, \nonumber\\
 O_5= (\bar q b)_{V-A} \sum_{q'} (\bar q'q')_{V+A}, && O_6= (\bar q^i b^j)_{V-A} \sum_{q'} (\bar q^{\prime j}q^{\prime i})_{V+A}, \nonumber\\
 O_7=\frac{3}{2} (\bar q b)_{V-A} \sum_{q'} e_{q'}(\bar q'q')_{V+A}, && O_8= \frac{3}{2}(\bar q^i b^j)_{V-A} \sum_{q'}e_{q'} (\bar q^{\prime j}q^{\prime i})_{V+A}, \nonumber\\
 O_9=\frac{3}{2} (\bar q b)_{V-A} \sum_{q'}e_{q'}  (\bar q'q')_{V-A}, &&  O_{10}= \frac{3}{2}(\bar q^i b^j)_{V-A} \sum_{q'}e_{q'}  (\bar q^{\prime j}q^{\prime i})_{V-A}, \label{operators}
\end{eqnarray}
$q=d,s$ and $q'=u,d,s$. 
The subscript $V\mp A$ denotes a left or right-handed $\gamma_\mu (1\mp\gamma_5)$ current.  

In the  SU(3) representation, the tree operators $O_{1,2}$ and the electroweak penguin  operators $O_{7-10}$ can
be decomposed in terms of three  SU(3) representation operators: a vector $H_{\bf \bar 3}^i$, a traceless
tensor antisymmetric in upper indices $(H_{\bf6})^{[ij]}_{k}$, and a
traceless tensor symmetric in  upper indices
$(H_{\bf{\overline{15}}})^{\{ij\}}_{k}$.  
In the case of $\Delta S=0 (b\to d)$ transitions, the non-zero components of these effective Hamiltonian are~\cite{Savage:1989ub,He:2000ys,Hsiao:2015iiu}:
\begin{eqnarray}
 (H_{\bf \bar3})^2=1,\;\;\; (H_{6})^{12}_1=-(H_{6})^{21}_1=(H_{6})^{23}_3=-(H_{6})^{32}_3=1,\nonumber\\
 2(H_{\overline{15}})^{12}_1= 2(H_{\overline{15}})^{21}_1=-3(H_{\overline{15}})^{22}_2 = 
 -6(H_{\overline{15}})^{23}_3=-6(H_{\overline{15}})^{32}_3=6.\label{eq:H3615_bb}
\end{eqnarray}
In the case of $\Delta S=-1(b\to s)$
transitions, the nonzero components of $H_{\bf{\bar 3}}$, $H_{\bf 6}$ and
$H_{\bf{\overline{15}}}$ can be  obtained from Eq.~\eqref{eq:H3615_bb}
by changement  $2\leftrightarrow 3$ corresponding to the $d \leftrightarrow s$ exchange. On the other hand,  the QCD penguin operators $O_{3-6}$  also belong to the ${\bf  \bar 3}$ representation, while the electromagnetic moment operator $O_{7\gamma}$ can be effectively incorporated into the $O_{7-10}$. The color magnetic moment  operator $O_{8g}$ is an SU(3) triplet and thus is not considered here~\cite{He:2000ys}.

\subsection{\bf Effective Hamiltonian for non-leptonic $c$ decays}

For the non-leptonic $c$ decays, the effective Hamiltonian with $\Delta C= 1$ is
\begin{eqnarray}
{\cal H}^c_{eff} &=& \frac{G_F}{\sqrt2}\big\{V_{cs} V_{ud}^* [C_1O_1^{sd}+C_2O_2^{sd}]+V_{cd} V_{ud}^* [C_1O_1^{dd}+C_2O_2^{dd}] \nonumber\\
&&+ V_{cs} V_{us}^* [C_1O_1^{ss}+C_2O_2^{ss}]+V_{cd} V_{us}^* [C_1O_1^{ds}+C_2O_2^{ds}]  \big\}, 
\end{eqnarray}
where the highly suppressed penguin contributions have been neglected, and 
\begin{eqnarray}
O_1^{sd} &=& [\bar s^i  \gamma_{\mu}(1-\gamma_5) c^j ][\bar u^i \gamma^{\mu}(1-\gamma_5) d^j], \;\; 
O_2^{sd} = [\bar s  \gamma_{\mu}(1-\gamma_5) c][\bar u \gamma^{\mu}(1-\gamma_5) d], 
\end{eqnarray}
while the operators containing other light flavors can be obtained by replacing the $d,s$ quark fields. 
Similar to the $b$ decays, the tree operators of $c$ decays  transform
under the flavor SU(3) symmetry as ${ \bar  3}\otimes {  3}\otimes {  \bar
3}={  \bar 3}\oplus {   \bar3}\oplus {  6}\oplus {   {\overline {15}}}$. 

For the Cabibbo allowed $c\to s  u \bar d$ transition, the amplitudes are proportional to $V_{cs}V^*_{ud}$ and the decay operators are
\begin{eqnarray}
(H_{  6})^{31}_2=-(H_{  6})^{13}_2=1,\;\;\;
 (H_{\overline {15}})^{31}_2= (H_{\overline {15}})^{13}_2=1.\label{eq:H3615_c_allowed}
\end{eqnarray}
For the  doubly Cabbibo suppressed $c\to d  u \bar s$ transition, the amplitudes are proportional to $V_{cd}V^*_{us}$ and the decay operators are
\begin{eqnarray}
(H_{  6})^{21}_3=-(H_{  6})^{12}_3=1,\;\;
 (H_{\overline {15}})^{21}_3= (H_{\overline {15}})^{12}_3=1. \label{eq:H3615_c_doubly_suprressed}
\end{eqnarray}
For the singly Cabbibo suppressed decays proportional to $V_{cs}V_{us}^*$, we have
\begin{eqnarray}
(H_{  6})^{31}_3 =-(H_{  6})^{13}_3 =1,\;\;\;
 (H_{\overline {15}})^{31}_3= (H_{\overline  {15}})^{13}_3 = 1, \label{eq:H3615_cc_singly_suppressed}
\end{eqnarray}
and For the singly Cabbibo suppressed decays proportional to $V_{cd}V_{ud}^*$, we have
\begin{eqnarray}
(H_{  6})^{12}_2 =-(H_{  6})^{21}_2 =1,\;\;\;
(H_{\overline  {15}})^{12}_2=(H_{\overline  {15}})^{21}_2= - 1. \label{eq:H3615_cc_singly_suppressed}
\end{eqnarray}
Note that  since $V_{cd}V_{ud}^* = - V_{cs}V_{us}^* -V_{cb}V_{ub}^* \approx - V_{cs}V_{us}^*$ (with $10^{-3}$ deviation), the contributions from the $\bar 3$ representation vanish, and  the nonzero components are only from 6 and $\overline{15}$ representations.

\subsection{Hadron Multiplets}

In this subsection we display the $SU(3)$ representation of the heavy mesons $B, D, B_c$ and pseudoscalar $P$.
The $B_c$ meson contains no light quark and it is a singlet. The heavy mesons containing one heavy quark are flavor SU(3) anti-triplets:
\begin{eqnarray}
(B_i) = (B^-(b \bar u), \overline B^0(b  \bar d), \overline B_s^0(b \bar s))\;, \;\;
(D_i)=  (D^0(c \bar u), D^+ (c \bar d), D^+_s (c \bar s))\;,
\end{eqnarray}

The light pseudoscalar $P$ mesons are mixture of octets and singlets, and its representation contains nine hadrons
\begin{eqnarray}
 P=\begin{pmatrix}
 \frac{\pi^0}{\sqrt{2}}+\frac{\eta_8}{\sqrt{6}}+\frac{\eta_1}{\sqrt3}  &\pi^+ & K^+\\
 \pi^-&\frac{\eta_8}{\sqrt{6}}-\frac{\pi^0}{\sqrt{2}}+\frac{\eta_1}{\sqrt3}&{K^0}\\
 K^-&\overline K^0 &\frac{\eta_1}{\sqrt3}-2\frac{\eta_8}{\sqrt{6}}
 \end{pmatrix}, 
\end{eqnarray}
where  $\eta_8$ and $\eta_1$ $\eta$ and $\eta'$  into $\eta$ and $\eta'$  with the mixing angle $\theta$
\begin{equation}
\eta= \cos\theta\eta_8+ \sin\theta \eta_1, ~~~~
\eta'=-\sin\theta\eta_8+ \cos\theta \eta_1. 
\end{equation}
Note that $\eta_8$ and $\eta_1$ are   not physical states, practically one can choose a basis $\eta_q$ and $\eta_s$ for the mixing, which read as \cite{He:2018joe} 
\begin{eqnarray}
P=\begin{pmatrix}
 \frac{\pi^0+\eta_q}{\sqrt{2}}  &\pi^+ & K^{+}\\
 \pi^-& \frac{-\pi^0+\eta_q}{\sqrt{2}}&{K^{0}}\\
 K^{-}&\overline K^{0} & \eta_s
 \end{pmatrix},  \label{eq:new_pseudo_scalar}
\end{eqnarray}
with 
\begin{eqnarray}
\eta_8=\frac{1}{\sqrt{3}}\eta_q- \sqrt{\frac{2}{3}}\eta_s,\;\;\; \eta_1= \sqrt{\frac{2}{3}}\eta_q+ \frac{1}{\sqrt{3}}\eta_s.
\end{eqnarray}

\section{fully-light tetraquarks in non-leptonic heavy meson decays}
\subsection{Semi-leptonic $B/D$ decays}
\begin{figure}
\includegraphics[width=0.2\columnwidth]{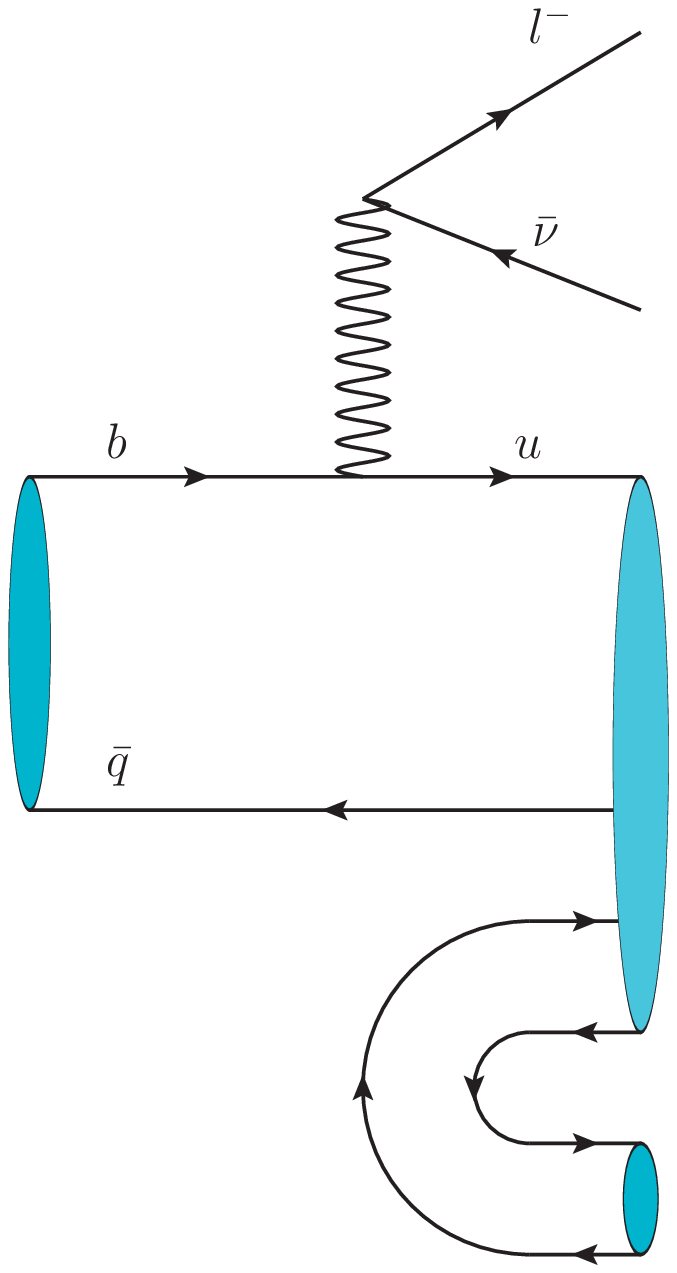} 
\caption{Feynman diagrams for the semi-leptonic $B$ decays, where the final states contain a fully-light tetraquark in 10 or 27 representation.}
\label{fig:SemiLept} 
\end{figure}

We firstly consider the two-body semi-leptonic decay  $B \to T_{10}/T_{27} P \ l{\bar \nu}$ and $D \to T_{10}/T_{27} P \ l{\bar \nu}$ where the final state containing an fully-light tetraquark in 10 or 27 representation, a pseudoscalar meson $P$ and leptons $l{\bar \nu}$ . The effective Hamiltonian for $B \to T_{10}/T_{27} P \ l{\bar \nu}$ is
\begin{align}
{\cal H}_{eff} = & a_1 B^{[ij]}  H_{3}^{k} (\overline T_{10})_{ikl}P_{j}^{l} l{\bar \nu} +a_{2}B_{i}H_{3}^{k}(\overline{T}_{27})_{kj}^{il}P_{l}^{j}l{\bar \nu},
\end{align}
where
\begin{equation}
B^{[ij]} = \epsilon^{ijk} B_k,~~~~(\overline T_{10})_{ijk}=(T_{10})^{\{ijk\}},~~~~(\overline T_{27})_{kl}^{ij}=(T_{27})_{ij}^{kl}. 
\end{equation}
The Feynman diagram corresponding to these two terms is shown in Fig.~\ref{fig:SemiLept}.
The decay amplitudes for $B \to T_{10} P \ l{\bar \nu}$ and $B \to T_{27} P \ l{\bar \nu}$ are listed in Table~\ref{tab:ampBtoT10Mlnu} and Table~\ref{tab:ampBtoT27Mlnu} respectively.
\begin{table}
\caption{Semileptonic $B$ decays into a light tetraquark $U_{I_{z}}^{Q}$ in
the 10 representation and a light meson.}
\label{tab:ampBtoT10Mlnu}%
\begin{tabular}{|c|c|c|c|}
\hline 
\hline
channel  & amplitude  & channel  & amplitude \tabularnewline
\hline 
$B^{-}\to U_{1/2}^{+}K^{-}l^{-}\bar{\nu}$  & $a_{1}V_{\text{ub}}$ & $\overline{B}^{0}\to U_{1}^{+}\eta_{q}l^{-}\bar{\nu}$  & $\frac{a_{1}V_{\text{ub}}}{\sqrt{2}}$\tabularnewline
$B^{-}\to U_{1}^{+}\pi^{-}l^{-}\bar{\nu}$  & $-a_{1}V_{\text{ub}}$ & $\overline{B}^{0}\to U_{1}^{+}\eta_{s}l^{-}\bar{\nu}$  & $-a_{1}V_{\text{ub}}$\tabularnewline
$B^{-}\to U_{-1/2}^{0}\overline{K}^{0}l^{-}\bar{\nu}$  & $a_{1}V_{\text{ub}}$ & $\overline{B}^{0}\to U_{0}^{0}\pi^{+}l^{-}\bar{\nu}$  & $a_{1}V_{\text{ub}}$\tabularnewline
$B^{-}\to U_{0}^{0}\pi^{0}l^{-}\bar{\nu}$  & $\frac{a_{1}V_{\text{ub}}}{\sqrt{2}}$ & $\overline{B}^{0}\to U_{1/2}^{0}K^{+}l^{-}\bar{\nu}$  & $a_{1}V_{\text{ub}}$\tabularnewline
$B^{-}\to U_{0}^{0}\eta_{q}l^{-}\bar{\nu}$  & $-\frac{a_{1}V_{\text{ub}}}{\sqrt{2}}$ & $\overline{B}_{s}^{0}\to U_{3/2}^{++}\pi^{-}l^{-}\bar{\nu}$  & $a_{1}V_{\text{ub}}$\tabularnewline
$B^{-}\to U_{0}^{0}\eta_{s}l^{-}\bar{\nu}$  & $a_{1}V_{\text{ub}}$ & $\overline{B}_{s}^{0}\to U_{1/2}^{+}\pi^{0}l^{-}\bar{\nu}$  & $-\sqrt{2}a_{1}V_{\text{ub}}$\tabularnewline
$B^{-}\to U_{1/2}^{0}K^{0}l^{-}\bar{\nu}$  & $-a_{1}V_{\text{ub}}$ & $\overline{B}_{s}^{0}\to U_{1}^{+}K^{0}l^{-}\bar{\nu}$  & $a_{1}V_{\text{ub}}$\tabularnewline
$\overline{B}^{0}\to U_{3/2}^{++}K^{-}l^{-}\bar{\nu}$  & $-a_{1}V_{\text{ub}}$ & $\overline{B}_{s}^{0}\to U_{-1/2}^{0}\pi^{+}l^{-}\bar{\nu}$  & $-a_{1}V_{\text{ub}}$\tabularnewline
$\overline{B}^{0}\to U_{1/2}^{+}\overline{K}^{0}l^{-}\bar{\nu}$  & $-a_{1}V_{\text{ub}}$ & $\overline{B}_{s}^{0}\to U_{0}^{0}K^{+}l^{-}\bar{\nu}$  & $-a_{1}V_{\text{ub}}$\tabularnewline
$\overline{B}^{0}\to U_{1}^{+}\pi^{0}l^{-}\bar{\nu}$  & $\frac{a_{1}V_{\text{ub}}}{\sqrt{2}}$ &  & \tabularnewline
\hline 
\end{tabular}
\end{table}
\begin{table}
\caption{Semileptonic $B$ decays into a light tetraquark $T_{I_{z}}^{Q}$ in
the 27 representation and a light meson.}
\label{tab:ampBtoT27Mlnu}%
\begin{tabular}{|c|c|c|c|c|c|}
\hline 
\hline 
channel  & amplitude  & channel  & amplitude  & channel  & amplitude \tabularnewline
\hline 
$B^{-}\to T_{0}^{0}\pi^{0}l^{-}\bar{\nu}$  & $\frac{a_{2}V_{\text{ub}}}{\sqrt{2}}$ & $\overline{B}^{0}\to T_{1}^{+}\pi^{0}l^{-}\bar{\nu}$  & $\frac{a_{2}V_{\text{ub}}}{\sqrt{2}}$ & $\overline{B}_{s}^{0}\to T_{0}^{\prime0}K^{+}l^{-}\bar{\nu}$  & $-a_{2}V_{\text{ub}}$\tabularnewline
$B^{-}\to T_{0}^{0}\eta_{q}l^{-}\bar{\nu}$  & $\frac{a_{2}V_{\text{ub}}}{\sqrt{2}}$ & $\overline{B}^{0}\to T_{1}^{+}\eta_{q}l^{-}\bar{\nu}$  & $\frac{a_{2}V_{\text{ub}}}{\sqrt{2}}$ & $\overline{B}_{s}^{0}\to T_{1/2}^{+}\pi^{0}l^{-}\bar{\nu}$  & $\frac{a_{2}V_{\text{ub}}}{\sqrt{2}}$\tabularnewline
$B^{-}\to T_{0}^{0}\eta_{s}l^{-}\bar{\nu}$  & $-a_{2}V_{\text{ub}}$ & $\overline{B}^{0}\to T_{0}^{\prime0}\pi^{+}l^{-}\bar{\nu}$  & $a_{2}V_{\text{ub}}$ & $\overline{B}_{s}^{0}\to T_{1/2}^{+}\eta_{q}l^{-}\bar{\nu}$  & $\frac{a_{2}V_{\text{ub}}}{\sqrt{2}}$\tabularnewline
$B^{-}\to T_{-1}^{-}\pi^{+}l^{-}\bar{\nu}$  & $a_{2}V_{\text{ub}}$ & $\overline{B}^{0}\to T_{1/2}^{0}K^{+}l^{-}\bar{\nu}$  & $a_{2}V_{\text{ub}}$ & $\overline{B}_{s}^{0}\to T_{1/2}^{+}\eta_{s}l^{-}\bar{\nu}$  & $-a_{2}V_{\text{ub}}$\tabularnewline
$B^{-}\to T_{-1/2}^{-}K^{+}l^{-}\bar{\nu}$  & $a_{2}V_{\text{ub}}$ & $\overline{B}^{0}\to T_{2}^{++}\pi^{-}l^{-}\bar{\nu}$  & $a_{2}V_{\text{ub}}$ & $\overline{B}_{s}^{0}\to T_{-1/2}^{0}\pi^{+}l^{-}\bar{\nu}$  & $a_{2}V_{\text{ub}}$\tabularnewline
$B^{-}\to T_{1}^{+}\pi^{-}l^{-}\bar{\nu}$  & $a_{2}V_{\text{ub}}$ & $\overline{B}^{0}\to T_{1}^{\prime+}\pi^{0}l^{-}\bar{\nu}$  & $-\frac{a_{2}V_{\text{ub}}}{\sqrt{2}}$ & $\overline{B}_{s}^{0}\to T_{1}^{\prime+}K^{0}l^{-}\bar{\nu}$  & $-a_{2}V_{\text{ub}}$\tabularnewline
$B^{-}\to T_{0}^{\prime0}\pi^{0}l^{-}\bar{\nu}$  & $-\frac{a_{2}V_{\text{ub}}}{\sqrt{2}}$ & $\overline{B}^{0}\to T_{1}^{\prime+}\eta_{q}l^{-}\bar{\nu}$  & $\frac{a_{2}V_{\text{ub}}}{\sqrt{2}}$ & $\overline{B}_{s}^{0}\to T_{3/2}^{++}\pi^{-}l^{-}\bar{\nu}$  & $a_{2}V_{\text{ub}}$\tabularnewline
$B^{-}\to T_{0}^{\prime0}\eta_{q}l^{-}\bar{\nu}$  & $\frac{a_{2}V_{\text{ub}}}{\sqrt{2}}$ & $\overline{B}^{0}\to T_{1}^{\prime+}\eta_{s}l^{-}\bar{\nu}$  & $-a_{2}V_{\text{ub}}$ & $\overline{B}_{s}^{0}\to T_{1/2}^{\prime+}\pi^{0}l^{-}\bar{\nu}$  & $-\frac{a_{2}V_{\text{ub}}}{\sqrt{2}}$\tabularnewline
$B^{-}\to T_{0}^{\prime0}\eta_{s}l^{-}\bar{\nu}$  & $-a_{2}V_{\text{ub}}$ & $\overline{B}^{0}\to T_{3/2}^{+}K^{0}l^{-}\bar{\nu}$  & $a_{2}V_{\text{ub}}$ & $\overline{B}_{s}^{0}\to T_{1/2}^{\prime+}\eta_{q}l^{-}\bar{\nu}$  & $\frac{a_{2}V_{\text{ub}}}{\sqrt{2}}$\tabularnewline
$B^{-}\to T_{1/2}^{0}K^{0}l^{-}\bar{\nu}$  & $a_{2}V_{\text{ub}}$ & $\overline{B}^{0}\to T_{3/2}^{++}K^{-}l^{-}\bar{\nu}$  & $a_{2}V_{\text{ub}}$ & $\overline{B}_{s}^{0}\to T_{1/2}^{\prime+}\eta_{s}l^{-}\bar{\nu}$  & $-a_{2}V_{\text{ub}}$\tabularnewline
$B^{-}\to T_{1/2}^{+}K^{-}l^{-}\bar{\nu}$  & $a_{2}V_{\text{ub}}$ & $\overline{B}^{0}\to T_{1/2}^{\prime+}\overline{K}^{0}l^{-}\bar{\nu}$  & $a_{2}V_{\text{ub}}$ & $\overline{B}_{s}^{0}\to T_{1}^{++}K^{-}l^{-}\bar{\nu}$  & $a_{2}V_{\text{ub}}$\tabularnewline
$B^{-}\to T_{-1/2}^{0}\overline{K}^{0}l^{-}\bar{\nu}$  & $a_{2}V_{\text{ub}}$ & $\overline{B}_{s}^{0}\to T_{0}^{0}K^{+}l^{-}\bar{\nu}$  & $-a_{2}V_{\text{ub}}$ & $\overline{B}_{s}^{0}\to T_{0}^{+}\overline{K}^{0}l^{-}\bar{\nu}$  & $a_{2}V_{\text{ub}}$\tabularnewline
\hline 
\end{tabular}
\end{table}

The effective Hamiltonian for $D \to T_{10}/T_{27} P \ l{\bar \nu}$ is
\begin{align}
{\cal H}_{eff} = & b_1 D^{[ij]}  \left(H_{3}^{\prime}\right)^{k} (\overline T_{10})_{ikl}P_{j}^{l} l{\bar \nu} +b_{2}D_{i} \left(H_{3}^{\prime}\right)^{k}(\overline{T}_{27})_{kj}^{il}P_{l}^{j}l{\bar \nu},
\end{align}
where $D^{[ij]} = \epsilon^{ijk} D_k$. The corresponding Feynman diagram has exactly the same topology as that of $B \to T_{10}/T_{27} P \ l{\bar \nu}$. The decay amplitudes are listed in Table~\ref{tab:ampDtoT10Mlnu} and Table~\ref{tab:ampDtoT27Mlnu} respectively.
\begin{table}
\caption{Semileptonic $D$ decays into a light tetraquark $U_{I_{z}}^{Q}$ in
the 10 representation and a light meson.}
\label{tab:ampDtoT10Mlnu}%
\begin{tabular}{|c|c|c|c|c|c|}
\hline 
\hline 
channel  & amplitude  & channel  & amplitude  & channel  & amplitude \tabularnewline
\hline 
$D^{0}\to U_{-1/2}^{0}K^{-}l^{+}\nu$  & $b_{1}\left(V_{\text{cd}}\right){}^{*}$ & $D^{+}\to U_{1/2}^{+}K^{-}l^{+}\nu$  & $-b_{1}\left(V_{\text{cd}}\right){}^{*}$ & $D^{+}\to U_{-1/2}^{-}K^{+}l^{+}\nu$  & $b_{1}\left(V_{\text{cd}}\right){}^{*}$\tabularnewline
$D^{0}\to U_{-3/2}^{-}\overline{K}^{0}l^{+}\nu$  & $b_{1}\left(V_{\text{cd}}\right){}^{*}$ & $D^{+}\to U_{-1/2}^{0}\overline{K}^{0}l^{+}\nu$  & $-b_{1}\left(V_{\text{cd}}\right){}^{*}$ & $D_{s}^{+}\to U_{1/2}^{+}\pi^{-}l^{+}\nu$  & $b_{1}\left(V_{\text{cd}}\right){}^{*}$\tabularnewline
$D^{0}\to U_{0}^{0}\pi^{-}l^{+}\nu$  & $-b_{1}\left(V_{\text{cd}}\right){}^{*}$ & $D^{+}\to U_{0}^{0}\pi^{0}l^{+}\nu$  & $\frac{b_{1}\left(V_{\text{cd}}\right){}^{*}}{\sqrt{2}}$ & $D_{s}^{+}\to U_{-1/2}^{0}\pi^{0}l^{+}\nu$  & $-\sqrt{2}b_{1}\left(V_{\text{cd}}\right){}^{*}$\tabularnewline
$D^{0}\to U_{-1}^{-}\pi^{0}l^{+}\nu$  & $\frac{b_{1}\left(V_{\text{cd}}\right){}^{*}}{\sqrt{2}}$ & $D^{+}\to U_{0}^{0}\eta_{q}l^{+}\nu$  & $\frac{b_{1}\left(V_{\text{cd}}\right){}^{*}}{\sqrt{2}}$ & $D_{s}^{+}\to U_{-3/2}^{-}\pi^{+}l^{+}\nu$  & $-b_{1}\left(V_{\text{cd}}\right){}^{*}$\tabularnewline
$D^{0}\to U_{-1}^{-}\eta_{q}l^{+}\nu$  & $-\frac{b_{1}\left(V_{\text{cd}}\right){}^{*}}{\sqrt{2}}$ & $D^{+}\to U_{0}^{0}\eta_{s}l^{+}\nu$  & $-b_{1}\left(V_{\text{cd}}\right){}^{*}$ & $D_{s}^{+}\to U_{0}^{0}K^{0}l^{+}\nu$  & $b_{1}\left(V_{\text{cd}}\right){}^{*}$\tabularnewline
$D^{0}\to U_{-1}^{-}\eta_{s}l^{+}\nu$  & $b_{1}\left(V_{\text{cd}}\right){}^{*}$ & $D^{+}\to U_{-1}^{-}\pi^{+}l^{+}\nu$  & $b_{1}\left(V_{\text{cd}}\right){}^{*}$ & $D_{s}^{+}\to U_{-1}^{-}K^{+}l^{+}\nu$  & $-b_{1}\left(V_{\text{cd}}\right){}^{*}$\tabularnewline
$D^{0}\to U_{-1/2}^{-}K^{0}l^{+}\nu$  & $-b_{1}\left(V_{\text{cd}}\right){}^{*}$ &  &  &  & \tabularnewline
\hline 
$D^{0}\to U_{0}^{0}K^{-}l^{+}\nu$  & $b_{1}\left(V_{\text{cs}}\right){}^{*}$ & $D^{+}\to U_{1}^{+}K^{-}l^{+}\nu$  & $-b_{1}\left(V_{\text{cs}}\right){}^{*}$ & $D^{+}\to U_{0}^{-}K^{+}l^{+}\nu$  & $b_{1}\left(V_{\text{cs}}\right){}^{*}$\tabularnewline
$D^{0}\to U_{1/2}^{0}\pi^{-}l^{+}\nu$  & $-b_{1}\left(V_{\text{cs}}\right){}^{*}$ & $D^{+}\to U_{0}^{0}\overline{K}^{0}l^{+}\nu$  & $-b_{1}\left(V_{\text{cs}}\right){}^{*}$ & $D_{s}^{+}\to U_{1}^{+}\pi^{-}l^{+}\nu$  & $b_{1}\left(V_{\text{cs}}\right){}^{*}$\tabularnewline
$D^{0}\to U_{-1}^{-}\overline{K}^{0}l^{+}\nu$  & $b_{1}\left(V_{\text{cs}}\right){}^{*}$ & $D^{+}\to U_{1/2}^{0}\pi^{0}l^{+}\nu$  & $\frac{b_{1}\left(V_{\text{cs}}\right){}^{*}}{\sqrt{2}}$ & $D_{s}^{+}\to U_{0}^{0}\pi^{0}l^{+}\nu$  & $-\sqrt{2}b_{1}\left(V_{\text{cs}}\right){}^{*}$\tabularnewline
$D^{0}\to U_{-1/2}^{-}\pi^{0}l^{+}\nu$  & $\frac{b_{1}\left(V_{\text{cs}}\right){}^{*}}{\sqrt{2}}$ & $D^{+}\to U_{1/2}^{0}\eta_{q}l^{+}\nu$  & $\frac{b_{1}\left(V_{\text{cs}}\right){}^{*}}{\sqrt{2}}$ & $D_{s}^{+}\to U_{1/2}^{0}K^{0}l^{+}\nu$  & $b_{1}\left(V_{\text{cs}}\right){}^{*}$\tabularnewline
$D^{0}\to U_{-1/2}^{-}\eta_{q}l^{+}\nu$  & $-\frac{b_{1}\left(V_{\text{cs}}\right){}^{*}}{\sqrt{2}}$ & $D^{+}\to U_{1/2}^{0}\eta_{s}l^{+}\nu$  & $-b_{1}\left(V_{\text{cs}}\right){}^{*}$ & $D_{s}^{+}\to U_{-1}^{-}\pi^{+}l^{+}\nu$  & $-b_{1}\left(V_{\text{cs}}\right){}^{*}$\tabularnewline
$D^{0}\to U_{-1/2}^{-}\eta_{s}l^{+}\nu$  & $b_{1}\left(V_{\text{cs}}\right){}^{*}$ & $D^{+}\to U_{-1/2}^{-}\pi^{+}l^{+}\nu$  & $b_{1}\left(V_{\text{cs}}\right){}^{*}$ & $D_{s}^{+}\to U_{-1/2}^{-}K^{+}l^{+}\nu$  & $-b_{1}\left(V_{\text{cs}}\right){}^{*}$\tabularnewline
$D^{0}\to U_{0}^{-}K^{0}l^{+}\nu$  & $-b_{1}\left(V_{\text{cs}}\right){}^{*}$ &  &  &  & \tabularnewline
\hline 
\end{tabular}
\end{table}
\begin{table}
\caption{Semileptonic $D$ decays into a light tetraquark $T_{I_{z}}^{Q}$
in the 27 representation and a light meson.}
\label{tab:ampDtoT27Mlnu}%
\begin{tabular}{|c|c|c|c|c|c|}
\hline 
\hline 
channel  & amplitude  & channel  & amplitude  & channel  & amplitude \tabularnewline
\hline 
$D^{0}\to T_{-1}^{-}\pi^{0}l^{+}\nu$  & $\frac{b_{2}\left(V_{\text{cd}}\right){}^{*}}{\sqrt{2}}$ & $D^{+}\to T_{0}^{\prime0}\eta_{q}l^{+}\nu$  & $\frac{b_{2}\left(V_{\text{cd}}\right){}^{*}}{\sqrt{2}}$ & $D_{s}^{+}\to T_{-1}^{\prime-}K^{+}l^{+}\nu$  & $-b_{2}\left(V_{\text{cd}}\right){}^{*}$\tabularnewline
$D^{0}\to T_{-1}^{-}\eta_{q}l^{+}\nu$  & $\frac{b_{2}\left(V_{\text{cd}}\right){}^{*}}{\sqrt{2}}$ & $D^{+}\to T_{0}^{\prime0}\eta_{s}l^{+}\nu$  & $-b_{2}\left(V_{\text{cd}}\right){}^{*}$ & $D_{s}^{+}\to T_{-1/2}^{0}\pi^{0}l^{+}\nu$  & $\frac{b_{2}\left(V_{\text{cd}}\right){}^{*}}{\sqrt{2}}$\tabularnewline
$D^{0}\to T_{-1}^{-}\eta_{s}l^{+}\nu$  & $-b_{2}\left(V_{\text{cd}}\right){}^{*}$ & $D^{+}\to T_{-1}^{\prime-}\pi^{+}l^{+}\nu$  & $b_{2}\left(V_{\text{cd}}\right){}^{*}$ & $D_{s}^{+}\to T_{-1/2}^{0}\eta_{q}l^{+}\nu$  & $\frac{b_{2}\left(V_{\text{cd}}\right){}^{*}}{\sqrt{2}}$\tabularnewline
$D^{0}\to T_{-2}^{--}\pi^{+}l^{+}\nu$  & $b_{2}\left(V_{\text{cd}}\right){}^{*}$ & $D^{+}\to T_{-1/2}^{\prime-}K^{+}l^{+}\nu$  & $b_{2}\left(V_{\text{cd}}\right){}^{*}$ & $D_{s}^{+}\to T_{-1/2}^{0}\eta_{s}l^{+}\nu$  & $-b_{2}\left(V_{\text{cd}}\right){}^{*}$\tabularnewline
$D^{0}\to T_{-3/2}^{--}K^{+}l^{+}\nu$  & $b_{2}\left(V_{\text{cd}}\right){}^{*}$ & $D^{+}\to T_{1}^{\prime+}\pi^{-}l^{+}\nu$  & $b_{2}\left(V_{\text{cd}}\right){}^{*}$ & $D_{s}^{+}\to T_{-3/2}^{-}\pi^{+}l^{+}\nu$  & $b_{2}\left(V_{\text{cd}}\right){}^{*}$\tabularnewline
$D^{0}\to T_{0}^{\prime0}\pi^{-}l^{+}\nu$  & $b_{2}\left(V_{\text{cd}}\right){}^{*}$ & $D^{+}\to T_{0}^{\prime\prime0}\pi^{0}l^{+}\nu$  & $-\frac{b_{2}\left(V_{\text{cd}}\right){}^{*}}{\sqrt{2}}$ & $D_{s}^{+}\to T_{0}^{\prime\prime0}K^{0}l^{+}\nu$  & $-b_{2}\left(V_{\text{cd}}\right){}^{*}$\tabularnewline
$D^{0}\to T_{-1}^{\prime-}\pi^{0}l^{+}\nu$  & $-\frac{b_{2}\left(V_{\text{cd}}\right){}^{*}}{\sqrt{2}}$ & $D^{+}\to T_{0}^{\prime\prime0}\eta_{q}l^{+}\nu$  & $\frac{b_{2}\left(V_{\text{cd}}\right){}^{*}}{\sqrt{2}}$ & $D_{s}^{+}\to T_{1/2}^{\prime+}\pi^{-}l^{+}\nu$  & $b_{2}\left(V_{\text{cd}}\right){}^{*}$\tabularnewline
$D^{0}\to T_{-1}^{\prime-}\eta_{q}l^{+}\nu$  & $\frac{b_{2}\left(V_{\text{cd}}\right){}^{*}}{\sqrt{2}}$ & $D^{+}\to T_{0}^{\prime\prime0}\eta_{s}l^{+}\nu$  & $-b_{2}\left(V_{\text{cd}}\right){}^{*}$ & $D_{s}^{+}\to T_{-1/2}^{\prime0}\pi^{0}l^{+}\nu$  & $-\frac{b_{2}\left(V_{\text{cd}}\right){}^{*}}{\sqrt{2}}$\tabularnewline
$D^{0}\to T_{-1}^{\prime-}\eta_{s}l^{+}\nu$  & $-b_{2}\left(V_{\text{cd}}\right){}^{*}$ & $D^{+}\to T_{1/2}^{\prime0}K^{0}l^{+}\nu$  & $b_{2}\left(V_{\text{cd}}\right){}^{*}$ & $D_{s}^{+}\to T_{-1/2}^{\prime0}\eta_{q}l^{+}\nu$  & $\frac{b_{2}\left(V_{\text{cd}}\right){}^{*}}{\sqrt{2}}$\tabularnewline
$D^{0}\to T_{-1/2}^{\prime-}K^{0}l^{+}\nu$  & $b_{2}\left(V_{\text{cd}}\right){}^{*}$ & $D^{+}\to T_{1/2}^{\prime+}K^{-}l^{+}\nu$  & $b_{2}\left(V_{\text{cd}}\right){}^{*}$ & $D_{s}^{+}\to T_{-1/2}^{\prime0}\eta_{s}l^{+}\nu$  & $-b_{2}\left(V_{\text{cd}}\right){}^{*}$\tabularnewline
$D^{0}\to T_{-1/2}^{0}K^{-}l^{+}\nu$  & $b_{2}\left(V_{\text{cd}}\right){}^{*}$ & $D^{+}\to T_{-1/2}^{\prime0}\overline{K}^{0}l^{+}\nu$  & $b_{2}\left(V_{\text{cd}}\right){}^{*}$ & $D_{s}^{+}\to T_{0}^{+}K^{-}l^{+}\nu$  & $b_{2}\left(V_{\text{cd}}\right){}^{*}$\tabularnewline
$D^{0}\to T_{-3/2}^{-}\overline{K}^{0}l^{+}\nu$  & $b_{2}\left(V_{\text{cd}}\right){}^{*}$ & $D_{s}^{+}\to T_{-1}^{-}K^{+}l^{+}\nu$  & $-b_{2}\left(V_{\text{cd}}\right){}^{*}$ & $D_{s}^{+}\to T_{-1}^{0}\overline{K}^{0}l^{+}\nu$  & $b_{2}\left(V_{\text{cd}}\right){}^{*}$\tabularnewline
$D^{+}\to T_{0}^{\prime0}\pi^{0}l^{+}\nu$  & $\frac{b_{2}\left(V_{\text{cd}}\right){}^{*}}{\sqrt{2}}$ & $D_{s}^{+}\to T_{0}^{\prime0}K^{0}l^{+}\nu$  & $-b_{2}\left(V_{\text{cd}}\right){}^{*}$ &  & \tabularnewline
\hline 
$D^{0}\to T_{0}^{0}K^{-}l^{+}\nu$  & $-b_{2}\left(V_{\text{cs}}\right){}^{*}$ & $D^{0}\to T_{0}^{-}K^{0}l^{+}\nu$  & $b_{2}\left(V_{\text{cs}}\right){}^{*}$ & $D^{+}\to T_{1}^{0}K^{0}l^{+}\nu$  & $b_{2}\left(V_{\text{cs}}\right){}^{*}$\tabularnewline
$D^{0}\to T_{-1}^{-}\overline{K}^{0}l^{+}\nu$  & $-b_{2}\left(V_{\text{cs}}\right){}^{*}$ & $D^{+}\to T_{0}^{\prime0}\overline{K}^{0}l^{+}\nu$  & $-b_{2}\left(V_{\text{cs}}\right){}^{*}$ & $D_{s}^{+}\to T_{0}^{0}\pi^{0}l^{+}\nu$  & $-\frac{b_{2}\left(V_{\text{cs}}\right){}^{*}}{\sqrt{2}}$\tabularnewline
$D^{0}\to T_{-1/2}^{-}\pi^{0}l^{+}\nu$  & $\frac{b_{2}\left(V_{\text{cs}}\right){}^{*}}{\sqrt{2}}$ & $D^{+}\to T_{1/2}^{0}\pi^{0}l^{+}\nu$  & $\frac{b_{2}\left(V_{\text{cs}}\right){}^{*}}{\sqrt{2}}$ & $D_{s}^{+}\to T_{0}^{0}\eta_{q}l^{+}\nu$  & $-\frac{b_{2}\left(V_{\text{cs}}\right){}^{*}}{\sqrt{2}}$\tabularnewline
$D^{0}\to T_{-1/2}^{-}\eta_{q}l^{+}\nu$  & $\frac{b_{2}\left(V_{\text{cs}}\right){}^{*}}{\sqrt{2}}$ & $D^{+}\to T_{1/2}^{0}\eta_{q}l^{+}\nu$  & $\frac{b_{2}\left(V_{\text{cs}}\right){}^{*}}{\sqrt{2}}$ & $D_{s}^{+}\to T_{0}^{0}\eta_{s}l^{+}\nu$  & $b_{2}\left(V_{\text{cs}}\right){}^{*}$\tabularnewline
$D^{0}\to T_{-1/2}^{-}\eta_{s}l^{+}\nu$  & $-b_{2}\left(V_{\text{cs}}\right){}^{*}$ & $D^{+}\to T_{1/2}^{0}\eta_{s}l^{+}\nu$  & $-b_{2}\left(V_{\text{cs}}\right){}^{*}$ & $D_{s}^{+}\to T_{-1}^{-}\pi^{+}l^{+}\nu$  & $-b_{2}\left(V_{\text{cs}}\right){}^{*}$\tabularnewline
$D^{0}\to T_{-3/2}^{--}\pi^{+}l^{+}\nu$  & $b_{2}\left(V_{\text{cs}}\right){}^{*}$ & $D^{+}\to T_{-1/2}^{\prime-}\pi^{+}l^{+}\nu$  & $b_{2}\left(V_{\text{cs}}\right){}^{*}$ & $D_{s}^{+}\to T_{-1/2}^{-}K^{+}l^{+}\nu$  & $-b_{2}\left(V_{\text{cs}}\right){}^{*}$\tabularnewline
$D^{0}\to T_{-1}^{--}K^{+}l^{+}\nu$  & $b_{2}\left(V_{\text{cs}}\right){}^{*}$ & $D^{+}\to T_{0}^{-}K^{+}l^{+}\nu$  & $b_{2}\left(V_{\text{cs}}\right){}^{*}$ & $D_{s}^{+}\to T_{0}^{\prime0}\eta_{q}l^{+}\nu$  & $-\sqrt{2}b_{2}\left(V_{\text{cs}}\right){}^{*}$\tabularnewline
$D^{0}\to T_{0}^{\prime0}K^{-}l^{+}\nu$  & $-b_{2}\left(V_{\text{cs}}\right){}^{*}$ & $D^{+}\to T_{1}^{\prime+}K^{-}l^{+}\nu$  & $-b_{2}\left(V_{\text{cs}}\right){}^{*}$ & $D_{s}^{+}\to T_{0}^{\prime0}\eta_{s}l^{+}\nu$  & $2b_{2}\left(V_{\text{cs}}\right){}^{*}$\tabularnewline
$D^{0}\to T_{1/2}^{0}\pi^{-}l^{+}\nu$  & $b_{2}\left(V_{\text{cs}}\right){}^{*}$ & $D^{+}\to T_{3/2}^{+}\pi^{-}l^{+}\nu$  & $b_{2}\left(V_{\text{cs}}\right){}^{*}$ & $D_{s}^{+}\to T_{1/2}^{0}K^{0}l^{+}\nu$  & $-b_{2}\left(V_{\text{cs}}\right){}^{*}$\tabularnewline
$D^{0}\to T_{-1}^{\prime-}\overline{K}^{0}l^{+}\nu$  & $-b_{2}\left(V_{\text{cs}}\right){}^{*}$ & $D^{+}\to T_{0}^{\prime\prime0}\overline{K}^{0}l^{+}\nu$  & $-b_{2}\left(V_{\text{cs}}\right){}^{*}$ & $D_{s}^{+}\to T_{-1}^{\prime-}\pi^{+}l^{+}\nu$  & $-b_{2}\left(V_{\text{cs}}\right){}^{*}$\tabularnewline
$D^{0}\to T_{-1/2}^{\prime-}\pi^{0}l^{+}\nu$  & $-\frac{b_{2}\left(V_{\text{cs}}\right){}^{*}}{\sqrt{2}}$ & $D^{+}\to T_{1/2}^{\prime0}\pi^{0}l^{+}\nu$  & $-\frac{b_{2}\left(V_{\text{cs}}\right){}^{*}}{\sqrt{2}}$ & $D_{s}^{+}\to T_{-1/2}^{\prime-}K^{+}l^{+}\nu$  & $-b_{2}\left(V_{\text{cs}}\right){}^{*}$\tabularnewline
$D^{0}\to T_{-1/2}^{\prime-}\eta_{q}l^{+}\nu$  & $\frac{b_{2}\left(V_{\text{cs}}\right){}^{*}}{\sqrt{2}}$ & $D^{+}\to T_{1/2}^{\prime0}\eta_{q}l^{+}\nu$  & $\frac{b_{2}\left(V_{\text{cs}}\right){}^{*}}{\sqrt{2}}$ & $D_{s}^{+}\to T_{1/2}^{+}K^{-}l^{+}\nu$  & $-b_{2}\left(V_{\text{cs}}\right){}^{*}$\tabularnewline
$D^{0}\to T_{-1/2}^{\prime-}\eta_{s}l^{+}\nu$  & $-b_{2}\left(V_{\text{cs}}\right){}^{*}$ & $D^{+}\to T_{1/2}^{\prime0}\eta_{s}l^{+}\nu$  & $-b_{2}\left(V_{\text{cs}}\right){}^{*}$ & $D_{s}^{+}\to T_{-1/2}^{0}\overline{K}^{0}l^{+}\nu$  & $-b_{2}\left(V_{\text{cs}}\right){}^{*}$\tabularnewline
$D_{s}^{+}\to T_{1}^{\prime+}\pi^{-}l^{+}\nu$  & $-b_{2}\left(V_{\text{cs}}\right){}^{*}$ & $D_{s}^{+}\to T_{0}^{\prime\prime0}\eta_{s}l^{+}\nu$  & $b_{2}\left(V_{\text{cs}}\right){}^{*}$ & $D_{s}^{+}\to T_{1/2}^{\prime+}K^{-}l^{+}\nu$  & $-b_{2}\left(V_{\text{cs}}\right){}^{*}$\tabularnewline
$D_{s}^{+}\to T_{0}^{\prime\prime0}\pi^{0}l^{+}\nu$  & $\frac{b_{2}\left(V_{\text{cs}}\right){}^{*}}{\sqrt{2}}$ & $D_{s}^{+}\to T_{1/2}^{\prime0}K^{0}l^{+}\nu$  & $-b_{2}\left(V_{\text{cs}}\right){}^{*}$ & $D_{s}^{+}\to T_{-1/2}^{\prime0}\overline{K}^{0}l^{+}\nu$  & $-b_{2}\left(V_{\text{cs}}\right){}^{*}$\tabularnewline
$D_{s}^{+}\to T_{0}^{\prime\prime0}\eta_{q}l^{+}\nu$  & $-\frac{b_{2}\left(V_{\text{cs}}\right){}^{*}}{\sqrt{2}}$ &  &  &  & \tabularnewline
\hline 
\end{tabular}
\end{table}

\subsection{Non-leptonic $B$ decays}
We then consider the two-body non-leptonic decay  $B \to T_{10} P$ and $B \to T_{27} P$. The effective Hamiltonian for $B \to T_{10} P$ is
\begin{align}
{\cal H}_{eff} = & a_3 B^{[ij]}  H_{\bar{3}}^{k} (\overline T_{10})_{ikl}P_{j}^{l} + a_6 B^{[ij]}  (H_{6})_{j}^{[kl]} (\overline T_{10})_{ikm}P_{l}^{m}  \nonumber \\
&+a_{15} B^{[ij]}  (H_{\overline 15})_{j}^{\{kl\}} (\overline T_{10})_{ikm}P_{l}^{m}
+ b_{15} B^{[ij]}  (H_{\overline 15})_{m}^{\{kl\}} (\overline T_{10})_{ikl}P_{j}^{m} \nonumber \\
&+ c_{15} B^{[ij]}  (H_{\overline 15})_{j}^{\{kl\}} (\overline T_{10})_{ikl}P_{m}^{m}+ d_{15} B^{[ij]}  (H_{\overline 15})_{j}^{\{kl\}} (\overline T_{10})_{klm}P_{i}^{m},
\end{align}
The effective Hamiltonian for $B \to T_{27} P$ is
\begin{align}
{\cal H}_{eff} =&\  a_{3}B_{i}H_{\bar{3}}^{k}(\overline{T}_{27})_{kj}^{il}P_{l}^{j}+a_{6}B_{i}(H_{6})_{j}^{[kl]}(\overline{T}_{27})_{ml}^{ij}P_{k}^{m}+b_{6}B_{k}(H_{6})_{j}^{[kl]}(\overline{T}_{27})_{ml}^{ji}P_{i}^{m}\nonumber\nonumber \\
 & +a_{15}B_{i}(H_{\overline{15}})_{j}^{\{kl\}}(\overline{T}_{27})_{kl}^{ij}P_{m}^{m}+b_{15}B_{i}(H_{\overline{15}})_{j}^{\{kl\}}(\overline{T}_{27})_{ml}^{ij}P_{k}^{m}+c_{15}B_{i}(H_{\overline{15}})_{j}^{\{kl\}}(\overline{T}_{27})_{kl}^{im}P_{m}^{j}.\nonumber \\
 & +d_{15}B_{i}(H_{\overline{15}})_{j}^{\{kl\}}(\overline{T}_{27})_{kl}^{jn}P_{n}^{i}+e_{15}B_{k}(H_{\overline{15}})_{j}^{\{kl\}}(\overline{T}_{27})_{lm}^{ji}P_{i}^{m}.
\end{align}
\begin{figure}
\includegraphics[width=0.8\columnwidth]{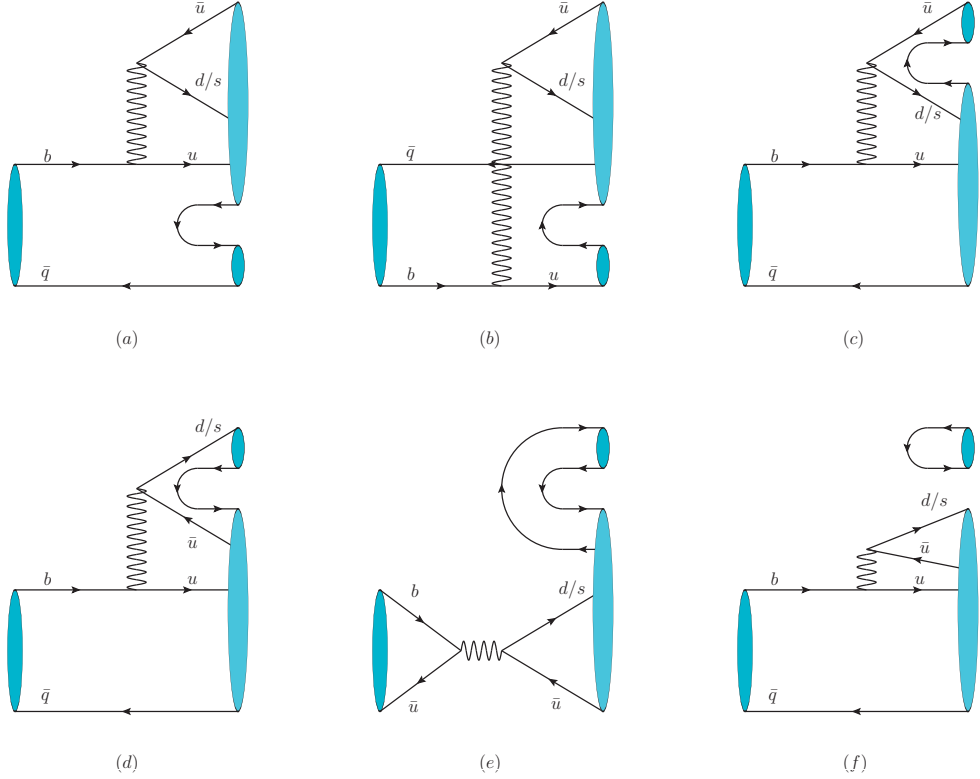} 
\caption{Feynman diagrams for the non-leptonic B decays, where the final states contain a fully-light tetraquark in 10 or 27 representation.}
\label{fig:Bdecay} 
\end{figure}
For simplicity we have used the same set of notations $a_i, b_i, c_i\cdots$ both for the $B \to T_{10} P$ and $B \to T_{27} P$ cases. However, It should be kept in mind that these two sets are in fact independent. Fig.~\ref{fig:Bdecay} shows the Feynman diagrams corresponding to these Hamiltonians. The correspondence between each effective Hamiltonian above and the Feynman diagrams are:
\begin{equation}
a_3\to (e),~~~a_6\to (b,d),~~~a_{15}\to (b,d),~~~c_{15}\to (f),~~~d_{15}\to (a,c)
\end{equation}
for $B \to T_{10} P$ decay and 
\begin{align}
a_3\to (e),~~~a_6\to (b,d),~~~& b_6\to (e),~~~a_{15}\to (f),~~~b_{15}\to (b,d),\nonumber\\
~~~c_{15}\to (c),~~~& d_{15}\to (a),~~~e_{15}\to (e)
\end{align}
for $B \to T_{27} P$ decay.

The decay amplitudes for $b\to d$ and $b\to s$ transitions with a 10 representation tetraquark produced in the final state are listed in Table~\ref{tab:ampBtoT10Mbtod} and Table~\ref{tab:ampBtoT10Mbtos} respectively. 
\begin{table}
\caption{B decays into a light tetraquark $U_{I_{z}}^{Q}$ in the 10 representation
and a light meson ($b\to d$).}
\label{tab:ampBtoT10Mbtod}%
\begin{tabular}{|c|c|c|c|}
\hline 
\hline 
channel  & amplitude  & channel  & amplitude \tabularnewline
\hline 
$B^{-}\to U_{-1/2}^{0}K^{-}$  & $a_{3}+a_{6}-a_{15}+6b_{15}+2d_{15}$ & $\overline{B}^{0}\to U_{0}^{0}\eta_{s}$  & $-a_{3}-a_{6}+a_{15}+2b_{15}+8c_{15}+6d_{15}$\tabularnewline
$B^{-}\to U_{-3/2}^{-}\overline{K}^{0}$  & $a_{3}+a_{6}-a_{15}-2b_{15}+2d_{15}$ & $\overline{B}^{0}\to U_{1/2}^{0}K^{0}$  & $2\left(a_{6}+2a_{15}\right)$\tabularnewline
$B^{-}\to U_{0}^{0}\pi^{-}$  & $-a_{3}-a_{6}+a_{15}-6b_{15}-2d_{15}$ & $\overline{B}^{0}\to U_{-1}^{-}\pi^{+}$  & $a_{3}-a_{6}+3a_{15}-2b_{15}+2d_{15}$\tabularnewline
$B^{-}\to U_{-1}^{-}\pi^{0}$  & $\frac{a_{3}+a_{6}-a_{15}-2b_{15}+2d_{15}}{\sqrt{2}}$ & $\overline{B}^{0}\to U_{-1/2}^{-}K^{+}$  & $a_{3}-a_{6}+3a_{15}-2b_{15}+2d_{15}$\tabularnewline
$B^{-}\to U_{-1}^{-}\eta_{q}$  & $-\frac{a_{3}+a_{6}-a_{15}-2b_{15}+2d_{15}}{\sqrt{2}}$ & $\overline{B}_{s}^{0}\to U_{1/2}^{+}\pi^{-}$  & $a_{3}-a_{6}-5a_{15}+6b_{15}-6d_{15}$\tabularnewline
$B^{-}\to U_{-1}^{-}\eta_{s}$  & $a_{3}+a_{6}-a_{15}-2b_{15}+2d_{15}$ & $\overline{B}_{s}^{0}\to U_{-1/2}^{0}\pi^{0}$  & $\sqrt{2}\left(-a_{3}+a_{6}+a_{15}-2b_{15}+2d_{15}\right)$\tabularnewline
$B^{-}\to U_{-1/2}^{-}K^{0}$  & $-a_{3}-a_{6}+a_{15}+2b_{15}-2d_{15}$ & $\overline{B}_{s}^{0}\to U_{-1/2}^{0}\eta_{q}$  & $-4\sqrt{2}\left(a_{15}+b_{15}+2c_{15}+d_{15}\right)$\tabularnewline
$\overline{B}^{0}\to U_{1/2}^{+}K^{-}$  & $-a_{3}-a_{6}+a_{15}-6b_{15}+6d_{15}$ & $\overline{B}_{s}^{0}\to U_{-1/2}^{0}\eta_{s}$  & $-8c_{15}$\tabularnewline
$\overline{B}^{0}\to U_{1}^{+}\pi^{-}$  & $2\left(a_{6}+2a_{15}\right)$ & $\overline{B}_{s}^{0}\to U_{-3/2}^{-}\pi^{+}$  & $-a_{3}+a_{6}-3a_{15}+2b_{15}-2d_{15}$\tabularnewline
$\overline{B}^{0}\to U_{-1/2}^{0}\overline{K}^{0}$  & $-a_{3}-a_{6}+a_{15}+2b_{15}+6d_{15}$ & $\overline{B}_{s}^{0}\to U_{0}^{0}K^{0}$  & $a_{3}-a_{6}-5a_{15}-2b_{15}-6d_{15}$\tabularnewline
$\overline{B}^{0}\to U_{0}^{0}\pi^{0}$  & $\frac{a_{3}-3a_{6}-a_{15}+6b_{15}+2d_{15}}{\sqrt{2}}$ & $\overline{B}_{s}^{0}\to U_{-1}^{-}K^{+}$  & $-a_{3}+a_{6}-3a_{15}+2b_{15}-2d_{15}$\tabularnewline
$\overline{B}^{0}\to U_{0}^{0}\eta_{q}$  & $\frac{a_{3}+a_{6}+7a_{15}+6b_{15}+16c_{15}+2d_{15}}{\sqrt{2}}$ &  & \tabularnewline
\hline 
\end{tabular}
\end{table}
\begin{table}
\caption{B decays into a light tetraquark $U_{I_{z}}^{Q}$ in the 10 representation
and a light meson ($b\to s$).}
\label{tab:ampBtoT10Mbtos}%
\begin{tabular}{|c|c|c|c|}
\hline 
\hline 
channel  & amplitude  & channel  & amplitude \tabularnewline
\hline 
$B^{-}\to U_{0}^{0}K^{-}$  & $a_{3}+a_{6}-a_{15}+6b_{15}+2d_{15}$ & $\overline{B}^{0}\to U_{-1/2}^{-}\pi^{+}$  & $a_{3}-a_{6}+3a_{15}-2b_{15}+2d_{15}$\tabularnewline
$B^{-}\to U_{1/2}^{0}\pi^{-}$  & $-a_{3}-a_{6}+a_{15}-6b_{15}-2d_{15}$ & $\overline{B}^{0}\to U_{0}^{-}K^{+}$  & $a_{3}-a_{6}+3a_{15}-2b_{15}+2d_{15}$\tabularnewline
$B^{-}\to U_{-1}^{-}\overline{K}^{0}$  & $a_{3}+a_{6}-a_{15}-2b_{15}+2d_{15}$ & $\overline{B}_{s}^{0}\to U_{1/2}^{+}K^{-}$  & $-2\left(a_{6}+2a_{15}\right)$\tabularnewline
$B^{-}\to U_{-1/2}^{-}\pi^{0}$  & $\frac{a_{3}+a_{6}-a_{15}-2b_{15}+2d_{15}}{\sqrt{2}}$ & $\overline{B}_{s}^{0}\to U_{1}^{+}\pi^{-}$  & $a_{3}+a_{6}-a_{15}+6b_{15}-6d_{15}$\tabularnewline
$B^{-}\to U_{-1/2}^{-}\eta_{q}$  & $-\frac{a_{3}+a_{6}-a_{15}-2b_{15}+2d_{15}}{\sqrt{2}}$ & $\overline{B}_{s}^{0}\to U_{-1/2}^{0}\overline{K}^{0}$  & $-2\left(a_{6}+2a_{15}\right)$\tabularnewline
$B^{-}\to U_{-1/2}^{-}\eta_{s}$  & $a_{3}+a_{6}-a_{15}-2b_{15}+2d_{15}$ & $\overline{B}_{s}^{0}\to U_{0}^{0}\pi^{0}$  & $-\sqrt{2}\left(a_{3}+a_{15}+2b_{15}-2d_{15}\right)$\tabularnewline
$B^{-}\to U_{0}^{-}K^{0}$  & $-a_{3}-a_{6}+a_{15}+2b_{15}-2d_{15}$ & $\overline{B}_{s}^{0}\to U_{0}^{0}\eta_{q}$  & $\sqrt{2}\left(a_{6}-2a_{15}-4b_{15}-8c_{15}-4d_{15}\right)$\tabularnewline
$\overline{B}^{0}\to U_{1}^{+}K^{-}$  & $-a_{3}+a_{6}+5a_{15}-6b_{15}+6d_{15}$ & $\overline{B}_{s}^{0}\to U_{0}^{0}\eta_{s}$  & $-2\left(a_{6}+2a_{15}+4c_{15}\right)$\tabularnewline
$\overline{B}^{0}\to U_{0}^{0}\overline{K}^{0}$  & $-a_{3}+a_{6}+5a_{15}+2b_{15}+6d_{15}$ & $\overline{B}_{s}^{0}\to U_{1/2}^{0}K^{0}$  & $a_{3}+a_{6}-a_{15}-2b_{15}-6d_{15}$\tabularnewline
$\overline{B}^{0}\to U_{1/2}^{0}\pi^{0}$  & $\frac{a_{3}-a_{6}+3a_{15}+6b_{15}+2d_{15}}{\sqrt{2}}$ & $\overline{B}_{s}^{0}\to U_{-1}^{-}\pi^{+}$  & $-a_{3}+a_{6}-3a_{15}+2b_{15}-2d_{15}$\tabularnewline
$\overline{B}^{0}\to U_{1/2}^{0}\eta_{q}$  & $\frac{a_{3}-a_{6}+3a_{15}+6b_{15}+16c_{15}+2d_{15}}{\sqrt{2}}$ & $\overline{B}_{s}^{0}\to U_{-1/2}^{-}K^{+}$  & $-a_{3}+a_{6}-3a_{15}+2b_{15}-2d_{15}$\tabularnewline
$\overline{B}^{0}\to U_{1/2}^{0}\eta_{s}$  & $-a_{3}+a_{6}+5a_{15}+2b_{15}+8c_{15}+6d_{15}$ &  & \tabularnewline
\hline 
\end{tabular}
\end{table}
Note that these amplitudes are not all independent, and some of them are proportional to each other. For the $b\to d$ transitions, the decay widths for these channels are related as 
\begin{align}
&\Gamma(B^{-}\to U_{0}^{0}\pi^{-})={}\Gamma(B^{-}\to U_{-1/2}^{0}K^{-}),\nonumber \\
&\Gamma(B^{-}\to U_{-1}^{-}\pi^{0})=\frac{1}{2}\Gamma(B^{-}\to U_{-3/2}^{-}\overline{K}^{0})=\Gamma(B^{-}\to U_{-1}^{-}\eta_{q})\nonumber \\
&=\frac{1}{2}\Gamma(B^{-}\to U_{-1}^{-}\eta_{s})=\frac{1}{2}\Gamma(B^{-}\to U_{-1/2}^{-}K^{0})\nonumber \\
&\Gamma(\overline{B}^{0}\to U_{1}^{+}\pi^{-})={}\Gamma(\overline{B}^{0}\to U_{1/2}^{0}K^{0}),\nonumber \\
&\Gamma(\overline{B}^{0}\to U_{-1}^{-}\pi^{+})={}\Gamma(\overline{B}^{0}\to U_{-1/2}^{-}K^{+})=\Gamma(\overline{B}_{s}^{0}\to U_{-1}^{-}K^{+})=\Gamma(\overline{B}_{s}^{0}\to U_{-3/2}^{-}\pi^{+}).
\end{align}
 For the $b\to s$ transitions, the decay widths for these channels are related as
\begin{align}
&\Gamma(B^{-}\to U_{1/2}^{0}\pi^{-})={}\Gamma(B^{-}\to U_{0}^{0}K^{-}),\nonumber \\
&\Gamma(B^{-}\to U_{-1/2}^{-}\pi^{0})=\frac{1}{2}\Gamma(B^{-}\to U_{-1}^{-}\overline{K}^{0})=\Gamma(B^{-}\to U_{-1/2}^{-}\eta_{q})=\frac{1}{2}\Gamma(B^{-}\to U_{-1/2}^{-}\eta_{s})\nonumber \\
&=\frac{1}{2}\Gamma(B^{-}\to U_{0}^{-}K^{0})=\Gamma(\overline{B}_{s}^{0}\to U_{-1}^{-}\pi^{+}),\nonumber \\
&\Gamma(\overline{B}^{0}\to U_{-1/2}^{-}\pi^{+})={}\Gamma(\overline{B}^{0}\to U_{0}^{-}K^{+})=\Gamma(\overline{B}_{s}^{0}\to U_{-1/2}^{-}K^{+}),\nonumber \\
&\Gamma(\overline{B}_{s}^{0}\to U_{-1/2}^{0}\overline{K}^{0})={}\Gamma(\overline{B}_{s}^{0}\to U_{1/2}^{+}K^{-}).
\end{align}

The decay amplitudes for $b\to d$ and $b\to s$ transitions with a 27 representation tetraquark produced in the final state are listed in Table~\ref{tab:ampBtoT27Mbtod} and Table~\ref{tab:ampBtoT27Mbtos} respectively. 
\begin{table}\tiny
{\caption{B decays into a light tetraquark $T_{I_{z}}^{Q}$ in the 27 representation
and a light meson ($b\to d$).}
\label{tab:ampBtoT27Mbtod}}%
\begin{tabular}{|c|c|c|c|}
\hline
\hline
{channel } & {amplitude } & {channel } & {amplitude }\tabularnewline
\hline 
{$B^{-}\to T_{0}^{0}\pi^{-}$ } & {$4b_{15}-2a_{6}$} & {$\overline{B}^{0}\to T_{1/2}^{+}K^{-}$ } & {$4e_{15}-2b_{6}$}\tabularnewline
{$B^{-}\to T_{-1}^{-}\pi^{0}$ } & {$\frac{a_{3}+3a_{6}+b_{6}-b_{15}+6c_{15}+8d_{15}+3e_{15}}{\sqrt{2}}$} & {$\overline{B}^{0}\to T_{-1/2}^{0}\overline{K}^{0}$ } & {$-2b_{6}+8d_{15}+4e_{15}$}\tabularnewline
{$B^{-}\to T_{-1}^{-}\eta_{q}$ } & {$\frac{a_{3}-a_{6}+16a_{15}+b_{6}+7b_{15}+6c_{15}+8d_{15}+3e_{15}}{\sqrt{2}}$} & {$\overline{B}^{0}\to T_{1}^{\prime+}\pi^{-}$ } & {$a_{3}-a_{6}-b_{6}-b_{15}+6c_{15}-e_{15}$}\tabularnewline
{$B^{-}\to T_{-1}^{-}\eta_{s}$ } & {$-a_{3}+a_{6}+8a_{15}-b_{6}+b_{15}+2c_{15}-3e_{15}$} & {$\overline{B}^{0}\to T_{0}^{\prime\prime0}\pi^{0}$ } & {$\frac{-a_{3}+a_{6}+b_{6}+b_{15}+2c_{15}+e_{15}}{\sqrt{2}}$}\tabularnewline
{$B^{-}\to T_{-1/2}^{-}K^{0}$ } & {$4b_{15}-2a_{6}$} & {$\overline{B}^{0}\to T_{0}^{\prime\prime0}\eta_{q}$ } & {$\frac{a_{3}-a_{6}-b_{6}-b_{15}-2c_{15}-e_{15}}{\sqrt{2}}$}\tabularnewline
{$B^{-}\to T_{-2}^{--}\pi^{+}$ } & {$a_{3}+a_{6}+b_{6}+3b_{15}-2c_{15}+3e_{15}$} & {$\overline{B}^{0}\to T_{0}^{\prime\prime0}\eta_{s}$ } & {$-a_{3}+a_{6}+b_{6}+b_{15}+2c_{15}+e_{15}$}\tabularnewline
{$B^{-}\to T_{-3/2}^{--}K^{+}$ } & {$a_{3}+a_{6}+b_{6}+3b_{15}-2c_{15}+3e_{15}$} & {$\overline{B}^{0}\to T_{1/2}^{\prime0}K^{0}$ } & {$a_{3}-a_{6}-b_{6}-b_{15}-2c_{15}-e_{15}$}\tabularnewline
{$B^{-}\to T_{0}^{\prime0}\pi^{-}$ } & {$a_{3}-a_{6}+b_{6}-b_{15}+6c_{15}+8d_{15}+3e_{15}$} & {$\overline{B}^{0}\to T_{1/2}^{\prime+}K^{-}$ } & {$a_{3}-a_{6}-b_{6}-b_{15}+6c_{15}-e_{15}$}\tabularnewline
{$B^{-}\to T_{-1}^{\prime-}\pi^{0}$ } & {$\frac{-a_{3}+a_{6}-b_{6}+b_{15}+2c_{15}-3e_{15}}{\sqrt{2}}$} & {$\overline{B}^{0}\to T_{-1/2}^{\prime0}\overline{K}^{0}$ } & {$a_{3}-a_{6}-b_{6}-b_{15}-2c_{15}-e_{15}$}\tabularnewline
{$B^{-}\to T_{-1}^{\prime-}\eta_{q}$ } & {$\frac{a_{3}-a_{6}+b_{6}-b_{15}-2c_{15}+3e_{15}}{\sqrt{2}}$} & {$\overline{B}_{s}^{0}\to T_{0}^{0}K^{0}$ } & {$2\left(a_{6}-2b_{15}\right)$}\tabularnewline
{$B^{-}\to T_{-1}^{\prime-}\eta_{s}$ } & {$-a_{3}+a_{6}-b_{6}+b_{15}+2c_{15}-3e_{15}$} & {$\overline{B}_{s}^{0}\to T_{-1}^{-}K^{+}$ } & {$-a_{3}-a_{6}+b_{6}-3b_{15}+2c_{15}+8d_{15}+e_{15}$}\tabularnewline
{$B^{-}\to T_{-1/2}^{\prime-}K^{0}$ } & {$a_{3}-a_{6}+b_{6}-b_{15}-2c_{15}+3e_{15}$} & {$\overline{B}_{s}^{0}\to T_{0}^{\prime0}K^{0}$ } & {$-a_{3}+3a_{6}+b_{6}-3b_{15}+2c_{15}+8d_{15}+e_{15}$}\tabularnewline
{$B^{-}\to T_{-1/2}^{0}K^{-}$ } & {$a_{3}-a_{6}+b_{6}-b_{15}+6c_{15}+8d_{15}+3e_{15}$} & {$\overline{B}_{s}^{0}\to T_{-1}^{\prime-}K^{+}$ } & {$-a_{3}-a_{6}+b_{6}-3b_{15}+2c_{15}+e_{15}$}\tabularnewline
{$B^{-}\to T_{-3/2}^{-}\overline{K}^{0}$ } & {$a_{3}-a_{6}+b_{6}-b_{15}-2c_{15}+3e_{15}$} & {$\overline{B}_{s}^{0}\to T_{1/2}^{+}\pi^{-}$ } & {$4b_{15}-2a_{6}$}\tabularnewline
{$\overline{B}^{0}\to T_{0}^{0}\pi^{0}$ } & {$-\sqrt{2}\left(b_{6}-2e_{15}\right)$} & {$\overline{B}_{s}^{0}\to T_{-1/2}^{0}\pi^{0}$ } & {$\frac{a_{3}+3a_{6}-b_{6}-b_{15}+6c_{15}-e_{15}}{\sqrt{2}}$}\tabularnewline
{$\overline{B}^{0}\to T_{0}^{0}\eta_{q}$ } & {$-\sqrt{2}\left(b_{6}-2e_{15}\right)$} & {$\overline{B}_{s}^{0}\to T_{-1/2}^{0}\eta_{q}$ } & {$\frac{a_{3}-a_{6}+16a_{15}-b_{6}+7b_{15}+6c_{15}-e_{15}}{\sqrt{2}}$}\tabularnewline
{$\overline{B}^{0}\to T_{0}^{0}\eta_{s}$ } & {$2\left(b_{6}-2e_{15}\right)$} & {$\overline{B}_{s}^{0}\to T_{-1/2}^{0}\eta_{s}$ } & {$-a_{3}+a_{6}+8a_{15}+b_{6}+b_{15}+2c_{15}+8d_{15}+e_{15}$}\tabularnewline
{$\overline{B}^{0}\to T_{-1}^{-}\pi^{+}$ } & {$-2b_{6}+8d_{15}+4e_{15}$} & {$\overline{B}_{s}^{0}\to T_{-3/2}^{-}\pi^{+}$ } & {$a_{3}+a_{6}-b_{6}+3b_{15}-2c_{15}-e_{15}$}\tabularnewline
{$\overline{B}^{0}\to T_{-1/2}^{-}K^{+}$ } & {$4e_{15}-2b_{6}$} & {$\overline{B}_{s}^{0}\to T_{0}^{\prime\prime0}K^{0}$ } & {$-a_{3}+a_{6}+b_{6}+b_{15}+2c_{15}+e_{15}$}\tabularnewline
{$\overline{B}^{0}\to T_{1}^{+}\pi^{-}$ } & {$-a_{6}+3\left(b_{15}+e_{15}\right)-b_{6}$} & {$\overline{B}_{s}^{0}\to T_{1/2}^{\prime+}\pi^{-}$ } & {$a_{3}-a_{6}-b_{6}-b_{15}+6c_{15}-e_{15}$}\tabularnewline
{$\overline{B}^{0}\to T_{0}^{\prime0}\pi^{0}$ } & {$\frac{a_{3}+3a_{6}+b_{6}-b_{15}+6c_{15}-8d_{15}-5e_{15}}{\sqrt{2}}$} & {$\overline{B}_{s}^{0}\to T_{-1/2}^{\prime0}\pi^{0}$ } & {$\frac{-a_{3}+a_{6}+b_{6}+b_{15}+2c_{15}+e_{15}}{\sqrt{2}}$}\tabularnewline
{$\overline{B}^{0}\to T_{0}^{\prime0}\eta_{q}$ } & {$\frac{a_{3}-a_{6}+16a_{15}-3b_{6}+7b_{15}+6c_{15}+8d_{15}+3e_{15}}{\sqrt{2}}$} & {$\overline{B}_{s}^{0}\to T_{-1/2}^{\prime0}\eta_{q}$ } & {$\frac{a_{3}-a_{6}-b_{6}-b_{15}-2c_{15}-e_{15}}{\sqrt{2}}$}\tabularnewline
{$\overline{B}^{0}\to T_{0}^{\prime0}\eta_{s}$ } & {$-a_{3}+a_{6}+8a_{15}+3b_{6}+b_{15}+2c_{15}-3e_{15}$} & {$\overline{B}_{s}^{0}\to T_{-1/2}^{\prime0}\eta_{s}$ } & {$-a_{3}+a_{6}+b_{6}+b_{15}+2c_{15}+e_{15}$}\tabularnewline
{$\overline{B}^{0}\to T_{1/2}^{0}K^{0}$ } & {$-2\left(a_{6}-2\left(b_{15}+e_{15}\right)+b_{6}\right)$} & {$\overline{B}_{s}^{0}\to T_{0}^{+}K^{-}$ } & {$a_{3}-a_{6}-b_{6}-b_{15}+6c_{15}-e_{15}$}\tabularnewline
{$\overline{B}^{0}\to T_{-1}^{\prime-}\pi^{+}$ } & {$a_{3}+a_{6}-b_{6}+3b_{15}-2c_{15}-e_{15}$} & {$\overline{B}_{s}^{0}\to T_{-1}^{0}\overline{K}^{0}$ } & {$a_{3}-a_{6}-b_{6}-b_{15}-2c_{15}-e_{15}$}\tabularnewline
{$\overline{B}^{0}\to T_{-1/2}^{\prime-}K^{+}$ } & {$a_{3}+a_{6}-b_{6}+3b_{15}-2c_{15}-e_{15}$} &  & \tabularnewline
\hline 
\end{tabular}{\tiny \par}  
\end{table}
\begin{table}\tiny
{\caption{B decays into a light tetraquark $T_{I_{z}}^{Q}$ in the 27 representation
and a light meson ($b\to s$).}
\label{tab:ampBtoT27Mbtos}}%
\begin{tabular}{|c|c|c|c|}
\hline 
\hline
{channel } & {amplitude } & {channel } & {amplitude }\tabularnewline
\hline 
{$B^{-}\to T_{0}^{0}K^{-}$ } & {$-a_{3}-a_{6}-b_{6}+5b_{15}-6c_{15}-8d_{15}-3e_{15}$} & {$\overline{B}^{0}\to T_{1/2}^{\prime0}\pi^{0}$ } & {$\frac{-a_{3}+a_{6}+b_{6}+b_{15}+2c_{15}+e_{15}}{\sqrt{2}}$}\tabularnewline
{$B^{-}\to T_{-1}^{-}\overline{K}^{0}$ } & {$-a_{3}-a_{6}-b_{6}+5b_{15}+2c_{15}-3e_{15}$} & {$\overline{B}^{0}\to T_{1/2}^{\prime0}\eta_{q}$ } & {$\frac{a_{3}-a_{6}-b_{6}-b_{15}-2c_{15}-e_{15}}{\sqrt{2}}$}\tabularnewline
{$B^{-}\to T_{-1/2}^{-}\pi^{0}$ } & {$\frac{a_{3}+a_{6}+b_{6}+3b_{15}+6c_{15}+8d_{15}+3e_{15}}{\sqrt{2}}$} & {$\overline{B}^{0}\to T_{1/2}^{\prime0}\eta_{s}$ } & {$-a_{3}+a_{6}+b_{6}+b_{15}+2c_{15}+e_{15}$}\tabularnewline
{$B^{-}\to T_{-1/2}^{-}\eta_{q}$ } & {$\frac{a_{3}+a_{6}+16a_{15}+b_{6}+3b_{15}+6c_{15}+8d_{15}+3e_{15}}{\sqrt{2}}$} & {$\overline{B}^{0}\to T_{1}^{0}K^{0}$ } & {$a_{3}-a_{6}-b_{6}-b_{15}-2c_{15}-e_{15}$}\tabularnewline
{$B^{-}\to T_{-1/2}^{-}\eta_{s}$ } & {$-a_{3}-a_{6}+8a_{15}-b_{6}+5b_{15}+2c_{15}-3e_{15}$} & {$\overline{B}_{s}^{0}\to T_{0}^{0}\pi^{0}$ } & {$-\frac{a_{3}+a_{6}+b_{6}+3b_{15}+6c_{15}-5e_{15}}{\sqrt{2}}$}\tabularnewline
{$B^{-}\to T_{-3/2}^{--}\pi^{+}$ } & {$a_{3}+a_{6}+b_{6}+3b_{15}-2c_{15}+3e_{15}$} & {$\overline{B}_{s}^{0}\to T_{0}^{0}\eta_{q}$ } & {$-\frac{a_{3}+a_{6}+16a_{15}+b_{6}+3b_{15}+6c_{15}-5e_{15}}{\sqrt{2}}$}\tabularnewline
{$B^{-}\to T_{-1}^{--}K^{+}$ } & {$a_{3}+a_{6}+b_{6}+3b_{15}-2c_{15}+3e_{15}$} & {$\overline{B}_{s}^{0}\to T_{0}^{0}\eta_{s}$ } & {$a_{3}+a_{6}-8a_{15}+b_{6}-5b_{15}-2c_{15}-8d_{15}-5e_{15}$}\tabularnewline
{$B^{-}\to T_{0}^{\prime0}K^{-}$ } & {$-a_{3}+a_{6}-b_{6}+b_{15}-6c_{15}-8d_{15}-3e_{15}$} & {$\overline{B}_{s}^{0}\to T_{-1}^{-}\pi^{+}$ } & {$-a_{3}-a_{6}-b_{6}-3b_{15}+2c_{15}+5e_{15}$}\tabularnewline
{$B^{-}\to T_{1/2}^{0}\pi^{-}$ } & {$a_{3}-a_{6}+b_{6}-b_{15}+6c_{15}+8d_{15}+3e_{15}$} & {$\overline{B}_{s}^{0}\to T_{-1/2}^{-}K^{+}$ } & {$-a_{3}-a_{6}-b_{6}-3b_{15}+2c_{15}+8d_{15}+5e_{15}$}\tabularnewline
{$B^{-}\to T_{-1}^{\prime-}\overline{K}^{0}$ } & {$-a_{3}+a_{6}-b_{6}+b_{15}+2c_{15}-3e_{15}$} & {$\overline{B}_{s}^{0}\to T_{1}^{+}\pi^{-}$ } & {$3e_{15}-b_{6}$}\tabularnewline
{$B^{-}\to T_{-1/2}^{\prime-}\pi^{0}$ } & {$\frac{-a_{3}+a_{6}-b_{6}+b_{15}+2c_{15}-3e_{15}}{\sqrt{2}}$} & {$\overline{B}_{s}^{0}\to T_{0}^{\prime0}\pi^{0}$ } & {$\sqrt{2}\left(-a_{6}-2\left(b_{15}+2c_{15}+e_{15}\right)+b_{6}\right)$}\tabularnewline
{$B^{-}\to T_{-1/2}^{\prime-}\eta_{q}$ } & {$\frac{a_{3}-a_{6}+b_{6}-b_{15}-2c_{15}+3e_{15}}{\sqrt{2}}$} & {$\overline{B}_{s}^{0}\to T_{0}^{\prime0}\eta_{q}$ } & {$-\sqrt{2}\left(a_{3}+8a_{15}+b_{15}+2c_{15}-3e_{15}\right)$}\tabularnewline
{$B^{-}\to T_{-1/2}^{\prime-}\eta_{s}$ } & {$-a_{3}+a_{6}-b_{6}+b_{15}+2c_{15}-3e_{15}$} & {$\overline{B}_{s}^{0}\to T_{0}^{\prime0}\eta_{s}$ } & {$2\left(a_{3}-4a_{15}-3b_{15}-2c_{15}-4d_{15}-3e_{15}\right)$}\tabularnewline
{$B^{-}\to T_{0}^{-}K^{0}$ } & {$a_{3}-a_{6}+b_{6}-b_{15}-2c_{15}+3e_{15}$} & {$\overline{B}_{s}^{0}\to T_{1/2}^{0}K^{0}$ } & {$-a_{3}+a_{6}-b_{6}+b_{15}+2c_{15}+8d_{15}+5e_{15}$}\tabularnewline
{$\overline{B}^{0}\to T_{0}^{0}\overline{K}^{0}$ } & {$-8d_{15}$} & {$\overline{B}_{s}^{0}\to T_{-1}^{\prime-}\pi^{+}$ } & {$-a_{3}-a_{6}+b_{6}-3b_{15}+2c_{15}+e_{15}$}\tabularnewline
{$\overline{B}^{0}\to T_{-1/2}^{-}\pi^{+}$ } & {$8d_{15}$} & {$\overline{B}_{s}^{0}\to T_{-1/2}^{\prime-}K^{+}$ } & {$-a_{3}-a_{6}+b_{6}-3b_{15}+2c_{15}+e_{15}$}\tabularnewline
{$\overline{B}^{0}\to T_{1}^{+}K^{-}$ } & {$3b_{15}-a_{6}$} & {$\overline{B}_{s}^{0}\to T_{1/2}^{+}K^{-}$ } & {$-a_{3}-a_{6}-b_{6}+5b_{15}-6c_{15}+5e_{15}$}\tabularnewline
{$\overline{B}^{0}\to T_{0}^{\prime0}\overline{K}^{0}$ } & {$-a_{3}-a_{6}+b_{6}+5b_{15}+2c_{15}-8d_{15}+e_{15}$} & {$\overline{B}_{s}^{0}\to T_{-1/2}^{0}\overline{K}^{0}$ } & {$-a_{3}-a_{6}-b_{6}+5b_{15}+2c_{15}+5e_{15}$}\tabularnewline
{$\overline{B}^{0}\to T_{1/2}^{0}\pi^{0}$ } & {$\frac{a_{3}+a_{6}-b_{6}+3b_{15}+6c_{15}-8d_{15}-e_{15}}{\sqrt{2}}$} & {$\overline{B}_{s}^{0}\to T_{1}^{\prime+}\pi^{-}$ } & {$-a_{3}+a_{6}+b_{6}+b_{15}-6c_{15}+e_{15}$}\tabularnewline
{$\overline{B}^{0}\to T_{1/2}^{0}\eta_{q}$ } & {$\frac{a_{3}+a_{6}+16a_{15}-b_{6}+3b_{15}+6c_{15}+8d_{15}-e_{15}}{\sqrt{2}}$} & {$\overline{B}_{s}^{0}\to T_{0}^{\prime\prime0}\pi^{0}$ } & {$\frac{a_{3}-a_{6}-b_{6}-b_{15}-2c_{15}-e_{15}}{\sqrt{2}}$}\tabularnewline
{$\overline{B}^{0}\to T_{1/2}^{0}\eta_{s}$ } & {$-a_{3}-a_{6}+8a_{15}+b_{6}+5b_{15}+2c_{15}+e_{15}$} & {$\overline{B}_{s}^{0}\to T_{0}^{\prime\prime0}\eta_{q}$ } & {$\frac{-a_{3}+a_{6}+b_{6}+b_{15}+2c_{15}+e_{15}}{\sqrt{2}}$}\tabularnewline
{$\overline{B}^{0}\to T_{-1/2}^{\prime-}\pi^{+}$ } & {$a_{3}+a_{6}-b_{6}+3b_{15}-2c_{15}-e_{15}$} & {$\overline{B}_{s}^{0}\to T_{0}^{\prime\prime0}\eta_{s}$ } & {$a_{3}-a_{6}-b_{6}-b_{15}-2c_{15}-e_{15}$}\tabularnewline
{$\overline{B}^{0}\to T_{0}^{-}K^{+}$ } & {$a_{3}+a_{6}-b_{6}+3b_{15}-2c_{15}-e_{15}$} & {$\overline{B}_{s}^{0}\to T_{1/2}^{\prime0}K^{0}$ } & {$-a_{3}+a_{6}+b_{6}+b_{15}+2c_{15}+e_{15}$}\tabularnewline
{$\overline{B}^{0}\to T_{1}^{\prime+}K^{-}$ } & {$-a_{3}+a_{6}+b_{6}+b_{15}-6c_{15}+e_{15}$} & {$\overline{B}_{s}^{0}\to T_{1/2}^{\prime+}K^{-}$ } & {$-a_{3}+a_{6}+b_{6}+b_{15}-6c_{15}+e_{15}$}\tabularnewline
{$\overline{B}^{0}\to T_{3/2}^{+}\pi^{-}$ } & {$a_{3}-a_{6}-b_{6}-b_{15}+6c_{15}-e_{15}$} & {$\overline{B}_{s}^{0}\to T_{-1/2}^{\prime0}\overline{K}^{0}$ } & {$-a_{3}+a_{6}+b_{6}+b_{15}+2c_{15}+e_{15}$}\tabularnewline
{$\overline{B}^{0}\to T_{0}^{\prime\prime0}\overline{K}^{0}$ } & {$-a_{3}+a_{6}+b_{6}+b_{15}+2c_{15}+e_{15}$} &  & \tabularnewline
\hline 
\end{tabular}
\end{table}
These decay channels are also not totally independent. The decay widths of $b\to d$ transitions are related as
\begin{align}
&\Gamma(B^{-}\to T_{0}^{0}\pi^{-})=\Gamma(B^{-}\to T_{-1/2}^{-}K^{0})=\Gamma(\overline{B}_{s}^{0}\to T_{0}^{0}K^{0}),\nonumber\\
&\Gamma(B^{-}\to T_{-2}^{--}\pi^{+})=\Gamma(B^{-}\to T_{-3/2}^{--}K^{+}),\nonumber\\
&\Gamma(B^{-}\to T_{0}^{\prime0}\pi^{-})=\Gamma(B^{-}\to T_{-1/2}^{0}K^{-}),\nonumber\\
&\Gamma(B^{-}\to T_{-1}^{\prime-}\pi^{0})=\Gamma(B^{-}\to T_{-1}^{\prime-}\eta_{q})=\frac{1}{2}\Gamma(B^{-}\to T_{-1}^{\prime-}\eta_{s})=\frac{1}{2}\Gamma(B^{-}\to T_{-1/2}^{\prime-}K^{0})\nonumber\\
&=\frac{1}{2}\Gamma(B^{-}\to T_{-3/2}^{-}\overline{K}^{0}),\nonumber\\
&\Gamma(\overline{B}^{0}\to T_{0}^{0}\pi^{0})=\Gamma(\overline{B}^{0}\to T_{0}^{0}\eta_{q})=\frac{1}{2}\Gamma(\overline{B}^{0}\to T_{0}^{0}\eta_{s})=\frac{1}{2}\Gamma(\overline{B}^{0}\to T_{-1/2}^{-}K^{+})\nonumber\\
&=\frac{1}{2}\Gamma(\overline{B}^{0}\to T_{1/2}^{+}K^{-})=\frac{1}{2}\Gamma(\overline{B}^{0}\to T_{1}^{\prime+}\pi^{-})=\frac{1}{2}\Gamma(\overline{B}_{s}^{0}\to T_{0}^{+}K^{-}),\nonumber\\
&\Gamma(\overline{B}^{0}\to T_{-1}^{-}\pi^{+})=\Gamma(\overline{B}^{0}\to T_{-1/2}^{0}\overline{K}^{0})=\Gamma(\overline{B}^{0}\to T_{-1/2}^{\prime-}K^{+})=\Gamma(\overline{B}_{s}^{0}\to T_{-1}^{\prime-}K^{+}),\nonumber\\
&\Gamma(\overline{B}^{0}\to T_{0}^{\prime\prime0}\pi^{0})=\Gamma(\overline{B}^{0}\to T_{0}^{\prime\prime0}\eta_{q})=\frac{1}{2}\Gamma(\overline{B}^{0}\to T_{0}^{\prime\prime0}\eta_{s})=\frac{1}{2}\Gamma(\overline{B}^{0}\to T_{1/2}^{\prime0}K^{0})\nonumber\\
&=\frac{1}{2}\Gamma(\overline{B}^{0}\to T_{-1/2}^{\prime0}\overline{K}^{0})=\frac{1}{2}\Gamma(\overline{B}_{s}^{0}\to T_{0}^{\prime\prime0}K^{0})=\Gamma(\overline{B}_{s}^{0}\to T_{-1/2}^{\prime0}\eta_{q})=\frac{1}{2}\Gamma(\overline{B}_{s}^{0}\to T_{-1/2}^{\prime0}\eta_{s})\nonumber\\
&=\frac{1}{2}\Gamma(\overline{B}_{s}^{0}\to T_{-1}^{0}\overline{K}^{0})=\Gamma(\overline{B}_{s}^{0}\to T_{-1/2}^{\prime0}\pi^{0}),\nonumber\\
&\Gamma(\overline{B}_{s}^{0}\to T_{1/2}^{+}\pi^{-})=\Gamma(B^{-}\to T_{-1/2}^{-}K^{0})=\Gamma(\overline{B}_{s}^{0}\to T_{0}^{0}K^{0})=\Gamma(\overline{B}^{0}\to T_{1/2}^{\prime+}K^{-})\nonumber\\
&=\Gamma(\overline{B}_{s}^{0}\to T_{0}^{+}K^{-}),\nonumber\\
&\Gamma(\overline{B}_{s}^{0}\to T_{-3/2}^{-}\pi^{+})=\Gamma(\overline{B}^{0}\to T_{-1/2}^{\prime-}K^{+})=\Gamma(\overline{B}_{s}^{0}\to T_{-1}^{\prime-}K^{+}).
\end{align}
The decay widths of $b\to s$ transitions are related as
\begin{align}
&\Gamma(B^{-}\to T_{-3/2}^{--}\pi^{+})=\Gamma(B^{-}\to T_{-1}^{--}K^{+}),\nonumber \\
&\Gamma(B^{-}\to T_{1/2}^{0}\pi^{-})=\Gamma(B^{-}\to T_{0}^{\prime0}K^{-}),\nonumber \\
&\Gamma(B^{-}\to T_{-1}^{\prime-}\overline{K}^{0})=2\Gamma(B^{-}\to T_{-1/2}^{\prime-}\eta_{q})=\Gamma(B^{-}\to T_{-1/2}^{\prime-}\eta_{s}),\nonumber \\
&\Gamma(B^{-}\to T_{-1/2}^{\prime-}\pi^{0})=\frac{1}{2}\Gamma(B^{-}\to T_{-1}^{\prime-}\overline{K}^{0})=\Gamma(B^{-}\to T_{-1/2}^{\prime-}\eta_{q})=\frac{1}{2}\Gamma(B^{-}\to T_{-1/2}^{\prime-}\eta_{s})\nonumber \\
&=\frac{1}{2}\Gamma(B^{-}\to T_{0}^{-}K^{0})=\frac{1}{2}\Gamma(B^{-}\to T_{-1/2}^{\prime-}\eta_{s})
,\nonumber \\
&\Gamma(\overline{B}^{0}\to T_{-1/2}^{-}\pi^{+})=\Gamma(\overline{B}^{0}\to T_{0}^{0}\overline{K}^{0}),\nonumber \\
&\Gamma(\overline{B}^{0}\to T_{-1/2}^{\prime-}\pi^{+})=\Gamma(\overline{B}^{0}\to T_{0}^{-}K^{+})=\Gamma(\overline{B}_{s}^{0}\to T_{-1/2}^{\prime-}K^{+}),\nonumber \\
&\Gamma(\overline{B}^{0}\to T_{3/2}^{+}\pi^{-})=\Gamma(\overline{B}^{0}\to T_{1}^{\prime+}K^{-})=\Gamma(\overline{B}_{s}^{0}\to T_{1/2}^{\prime+}K^{-}),\nonumber \\
&\Gamma(\overline{B}^{0}\to T_{0}^{\prime\prime0}\overline{K}^{0})=2\Gamma(\overline{B}^{0}\to T_{1/2}^{\prime0}\eta_{q})=\Gamma(\overline{B}^{0}\to T_{1/2}^{\prime0}\eta_{s})=2\Gamma(\overline{B}^{0}\to T_{1/2}^{\prime0}\pi^{0})\nonumber \\
&=2\Gamma(\overline{B}_{s}^{0}\to T_{0}^{\prime\prime0}\eta_{q})=\Gamma(\overline{B}_{s}^{0}\to T_{0}^{\prime\prime0}\eta_{s})=\Gamma(\overline{B}^{0}\to T_{1}^{0}K^{0})=\Gamma(\overline{B}_{s}^{0}\to T_{1/2}^{\prime0}K^{0})\nonumber \\
&=\Gamma(\overline{B}_{s}^{0}\to T_{-1/2}^{\prime0}\overline{K}^{0})=\Gamma(\overline{B}^{0}\to T_{1}^{0}K^{0})=2\Gamma(\overline{B}_{s}^{0}\to T_{0}^{\prime\prime0}\pi^{0})=2\Gamma(\overline{B}_{s}^{0}\to T_{0}^{\prime\prime0}\pi^{0}),\nonumber \\
&\Gamma(\overline{B}_{s}^{0}\to T_{-1}^{\prime-}\pi^{+})=\Gamma(\overline{B}^{0}\to T_{0}^{-}K^{+})=\Gamma(\overline{B}_{s}^{0}\to T_{-1/2}^{\prime-}K^{+}),\nonumber \\
&\Gamma(\overline{B}_{s}^{0}\to T_{1}^{\prime+}\pi^{-})=\Gamma(\overline{B}^{0}\to T_{1}^{\prime+}K^{-})=\Gamma(\overline{B}_{s}^{0}\to T_{1/2}^{\prime+}K^{-}).
\end{align}

\subsection{Non-leptonic $D$ decays}

Next we consider the two-body non-leptonic decay  $D \to T_{10} P$ and $D \to T_{27} P$ with the final state containing an fully-light tetraquark in 10 or 27 representation. The effective Hamiltonian for $D \to T_{10} P$ is
\begin{eqnarray}
{\cal H}_{eff} & = &  a_6 D^{[ij]}  (H_{6})_{j}^{[kl]} (\overline T_{10})_{ikm}P_{l}^{m}  +a_{15} D^{[ij]}  (H_{\overline 15})_{j}^{\{kl\}} (\overline T_{10})_{ikm}P_{l}^{m}+ b_{15} D^{[ij]}  (H_{\overline 15})_{m}^{\{kl\}} (\overline T_{10})_{ikl}P_{j}^{m}\nonumber \\
&& + c_{15} D^{[ij]}  (H_{\overline 15})_{j}^{\{kl\}} (\overline T_{10})_{ikl}P_{m}^{m} + d_{15} D^{[ij]}  (H_{\overline 15})_{j}^{\{kl\}} (\overline T_{10})_{klm}P_{i}^{m}, \nonumber
\end{eqnarray}
where $D^{[ij]} = \epsilon^{ijk} D_k$.
The effective Hamiltonian for $D \to T_{27} P$ is
\begin{align}
{\cal H}_{eff} & =a_{6}D_{i}(H_{6})_{j}^{[kl]}(\overline{T}_{27})_{ml}^{ij}P_{k}^{m}+b_{6}D_{k}(H_{6})_{j}^{[kl]}(\overline{T}_{27})_{ml}^{ji}P_{i}^{m} +a_{15}D_{i}(H_{\overline{15}})_{j}^{\{kl\}}(\overline{T}_{27})_{kl}^{ij}P_{m}^{m}\nonumber\\
 &+b_{15}D_{i}(H_{\overline{15}})_{j}^{\{kl\}}(\overline{T}_{27})_{ml}^{ij}P_{k}^{m}+c_{15}D_{i}(H_{\overline{15}})_{j}^{\{kl\}}(\overline{T}_{27})_{kl}^{im}P_{m}^{j}+d_{15}D_{i}(H_{\overline{15}})_{j}^{\{kl\}}(\overline{T}_{27})_{kl}^{jn}P_{n}^{i}\nonumber \\
 & +e_{15}D_{k}(H_{\overline{15}})_{j}^{\{kl\}}(\overline{T}_{27})_{lm}^{ji}P_{i}^{m}.
\end{align}
\begin{figure}
\includegraphics[width=0.8\columnwidth]{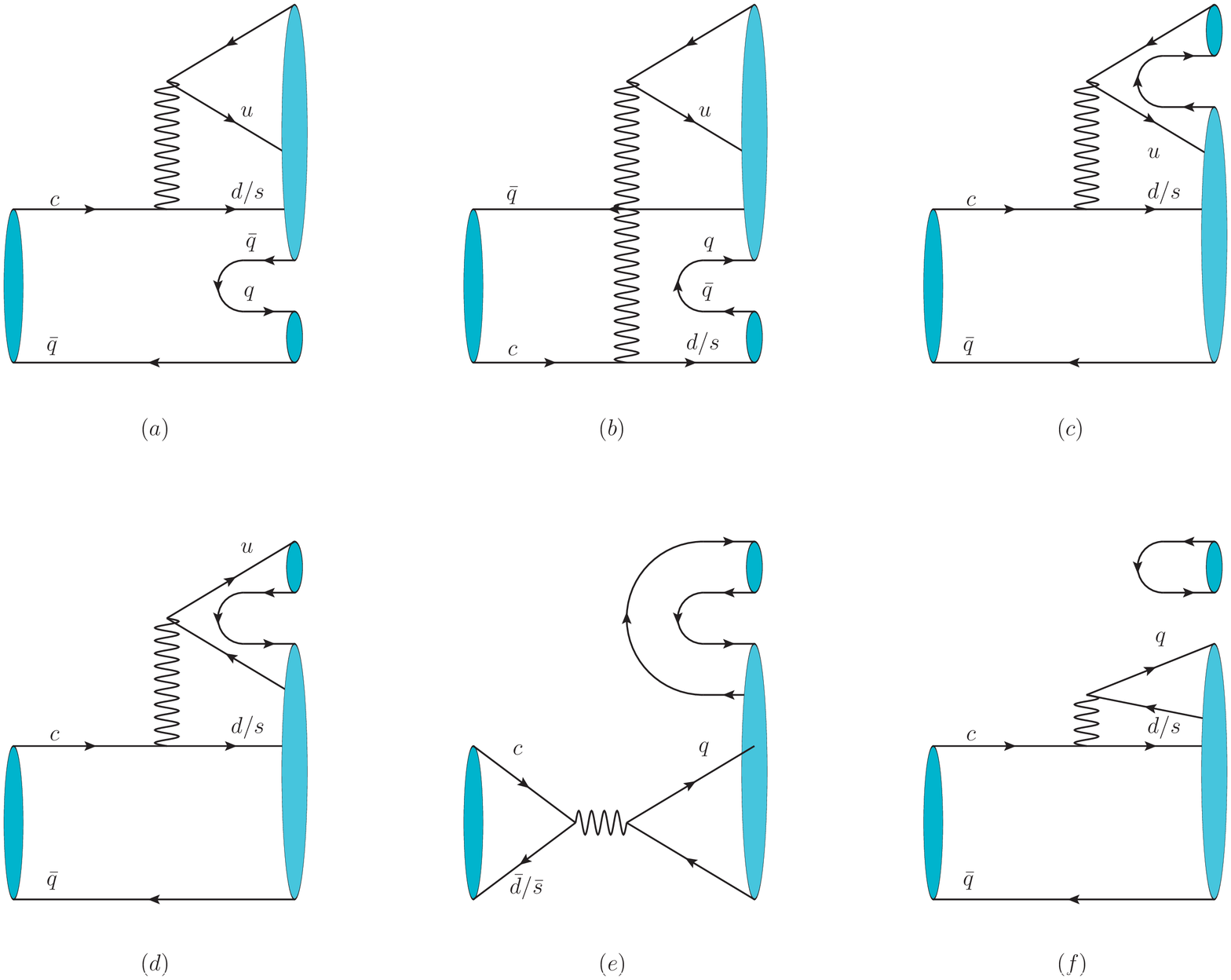} 
\caption{Feynman diagrams for the non-leptonic D decays, where the final states contain a fully-light tetraquark in 10 or 27 representation.}
\label{fig:Ddecay} 
\end{figure}
Fig.~\ref{fig:Ddecay} shows the Feynman diagrams corresponding to these Hamiltonians. The correspondence between each effective Hamiltonian above and the Feynman diagrams are:
\begin{equation}
a_6\to (b,d),~~~a_{15}\to (b,d),~~~c_{15}\to (f),~~~d_{15}\to (a,c)
\end{equation}
for $D \to T_{10} P$ decay and 
\begin{align}
a_6\to (b,d),~~~& b_6\to (e),~~~a_{15}\to (f),~~~b_{15}\to (b,d),\nonumber\\
~~~c_{15}\to (c),~~~& d_{15}\to (a),~~~e_{15}\to (e)
\end{align}
for $D \to T_{27} P$ decay.

The corresponding decay channels are classified by the Cabbibo suppression. For the Cabbibo allowed, singly Cabbibo suppresed and doubly Cabbibo suppressed $D \to T_{10} P$ results, they are listed in Table~\ref{tab:ampDtoT10MNoSup}, Table~\ref{tab:ampDtoT10MSinSup} and Table~\ref{tab:ampDtoT10MDouSup}.
\begin{table}
\caption{D decays into a light tetraquark $U_{I_{z}}^{Q}$ in the 10 representation
and a light meson (No suppression).}
\label{tab:ampDtoT10MNoSup}%
\begin{tabular}{|c|c|c|c|}
\hline 
\hline 
channel  & amplitude  & channel  & amplitude \tabularnewline
\hline 
$D^{0}\to U_{1}^{+}K^{-}$  & $a_{6}-a_{15}-2d_{15}$ & $D^{+}\to U_{1/2}^{0}\pi^{+}$  & $2b_{15}$\tabularnewline
$D^{0}\to U_{0}^{0}\overline{K}^{0}$  & $a_{6}-a_{15}+2b_{15}-2d_{15}$ & $D_{s}^{+}\to U_{3/2}^{++}K^{-}$  & $a_{15}-a_{6}$\tabularnewline
$D^{0}\to U_{1/2}^{0}\pi^{0}$  & $-\frac{a_{6}+a_{15}-2b_{15}}{\sqrt{2}}$ & $D_{s}^{+}\to U_{1/2}^{+}\overline{K}^{0}$  & $a_{15}-a_{6}$\tabularnewline
$D^{0}\to U_{1/2}^{0}\eta_{q}$  & $-\frac{a_{6}+a_{15}+2b_{15}+4c_{15}}{\sqrt{2}}$ & $D_{s}^{+}\to U_{1}^{+}\pi^{0}$  & $\frac{a_{6}+a_{15}-2b_{15}+2d_{15}}{\sqrt{2}}$\tabularnewline
$D^{0}\to U_{1/2}^{0}\eta_{s}$  & $a_{6}-a_{15}-2\left(c_{15}+d_{15}\right)$ & $D_{s}^{+}\to U_{1}^{+}\eta_{q}$  & $\frac{a_{6}+a_{15}+2\left(b_{15}+2c_{15}+d_{15}\right)}{\sqrt{2}}$\tabularnewline
$D^{0}\to U_{-1/2}^{-}\pi^{+}$  & $-a_{6}-a_{15}$ & $D_{s}^{+}\to U_{1}^{+}\eta_{s}$  & $-a_{6}+a_{15}+2c_{15}$\tabularnewline
$D^{0}\to U_{0}^{-}K^{+}$  & $-a_{6}-a_{15}$ & $D_{s}^{+}\to U_{0}^{0}\pi^{+}$  & $a_{6}+a_{15}-2b_{15}+2d_{15}$\tabularnewline
$D^{+}\to U_{1}^{+}\overline{K}^{0}$  & $-2b_{15}$ & $D_{s}^{+}\to U_{1/2}^{0}K^{+}$  & $a_{6}+a_{15}+2d_{15}$\tabularnewline
\hline 
\end{tabular}
\end{table}
\begin{table}
\caption{D decays into a light tetraquark $U_{I_{z}}^{Q}$ in the 10 representation
and a light meson (Single suppression).}
\label{tab:ampDtoT10MSinSup}%
\begin{tabular}{|c|c|c|c|}
\hline 
\hline 
channel  & amplitude  & channel  & amplitude \tabularnewline
\hline 
$D^{0}\to U_{1/2}^{+}K^{-}$  & $\left(-a_{6}+a_{15}+2d_{15}\right)$ & $D^{+}\to U_{1}^{+}\eta_{q}$  & $-\frac{\left(a_{6}+a_{15}+4c_{15}+2d_{15}\right)}{\sqrt{2}}$\tabularnewline
$D^{0}\to U_{1}^{+}\pi^{-}$  & $\left(-a_{6}+a_{15}+2d_{15}\right)$ & $D^{+}\to U_{1}^{+}\eta_{s}$  & $\left(a_{6}-a_{15}-2\left(b_{15}+c_{15}\right)\right)$\tabularnewline
$D^{0}\to U_{-1/2}^{0}\overline{K}^{0}$  & $-\left(a_{6}-a_{15}+2b_{15}-2d_{15}\right)$ & $D^{+}\to U_{0}^{0}\pi^{+}$  & $-\left(a_{6}+a_{15}+2\left(b_{15}+d_{15}\right)\right)$\tabularnewline
$D^{0}\to U_{0}^{0}\pi^{0}$  & $\frac{\left(3a_{6}+a_{15}-2\left(b_{15}+d_{15}\right)\right)}{\sqrt{2}}$ & $D^{+}\to U_{1/2}^{0}K^{+}$  & $-\left(a_{6}+a_{15}-2b_{15}+2d_{15}\right)$\tabularnewline
$D^{0}\to U_{0}^{0}\eta_{q}$  & $\frac{\left(a_{6}+3a_{15}+2\left(b_{15}+4c_{15}+d_{15}\right)\right)}{\sqrt{2}}$ & $D_{s}^{+}\to U_{3/2}^{++}\pi^{-}$  & $\left(a_{6}-a_{15}\right)$\tabularnewline
$D^{0}\to U_{0}^{0}\eta_{s}$  & $\left(-a_{6}+a_{15}+2\left(b_{15}+2c_{15}+d_{15}\right)\right)$ & $D_{s}^{+}\to U_{1/2}^{+}\pi^{0}$  & $-\sqrt{2}\left(a_{6}-b_{15}+d_{15}\right)$\tabularnewline
$D^{0}\to U_{1/2}^{0}K^{0}$  & $-\left(a_{6}-a_{15}+2b_{15}-2d_{15}\right)$ & $D_{s}^{+}\to U_{1/2}^{+}\eta_{q}$  & $-\sqrt{2}\left(a_{15}+b_{15}+2c_{15}+d_{15}\right)$\tabularnewline
$D^{0}\to U_{-1}^{-}\pi^{+}$  & $2\left(a_{6}+a_{15}\right)$ & $D_{s}^{+}\to U_{1/2}^{+}\eta_{s}$  & $-2c_{15}$\tabularnewline
$D^{0}\to U_{-1/2}^{-}K^{+}$  & $2\left(a_{6}+a_{15}\right)$ & $D_{s}^{+}\to U_{1}^{+}K^{0}$  & $\left(a_{6}-a_{15}+2b_{15}\right)$\tabularnewline
$D^{+}\to U_{3/2}^{++}K^{-}$  & $\left(a_{6}-a_{15}\right)$ & $D_{s}^{+}\to U_{-1/2}^{0}\pi^{+}$  & $-\left(a_{6}+a_{15}-2b_{15}+2d_{15}\right)$\tabularnewline
$D^{+}\to U_{1/2}^{+}\overline{K}^{0}$  & $\left(a_{6}-a_{15}+2b_{15}\right)$ & $D_{s}^{+}\to U_{0}^{0}K^{+}$  & $-\left(a_{6}+a_{15}+2\left(b_{15}+d_{15}\right)\right)$\tabularnewline
$D^{+}\to U_{1}^{+}\pi^{0}$  & $-\frac{\left(a_{6}+a_{15}+2d_{15}\right)}{\sqrt{2}}$ &  & \tabularnewline
\hline 
\end{tabular}
\end{table}
\begin{table}
\caption{D decays into a light tetraquark $U_{I_{z}}^{Q}$ in the 10 representation
and a light meson. (Double suppression)}
\label{tab:ampDtoT10MDouSup}%
\begin{tabular}{|c|c|c|c|}
\hline 
\hline 
channel  & amplitude  & channel  & amplitude \tabularnewline
\hline 
$D^{0}\to U_{1/2}^{+}\pi^{-}$  & $\left(a_{6}-a_{15}-2d_{15}\right)$ & $D^{+}\to U_{1/2}^{+}\pi^{0}$  & $\sqrt{2}\left(a_{6}+d_{15}\right)$\tabularnewline
$D^{0}\to U_{-1/2}^{0}\pi^{0}$  & $\sqrt{2}\left(d_{15}-a_{6}\right)$ & $D^{+}\to U_{1/2}^{+}\eta_{q}$  & $\sqrt{2}\left(a_{15}+2c_{15}+d_{15}\right)$\tabularnewline
$D^{0}\to U_{-1/2}^{0}\eta_{q}$  & $-\sqrt{2}\left(a_{15}+2c_{15}+d_{15}\right)$ & $D^{+}\to U_{1/2}^{+}\eta_{s}$  & $2\left(b_{15}+c_{15}\right)$\tabularnewline
$D^{0}\to U_{-1/2}^{0}\eta_{s}$  & $-2\left(b_{15}+c_{15}\right)$ & $D^{+}\to U_{1}^{+}K^{0}$  & $\left(a_{15}-a_{6}\right)$\tabularnewline
$D^{0}\to U_{-3/2}^{-}\pi^{+}$  & $-\left(a_{6}+a_{15}\right)$ & $D^{+}\to U_{-1/2}^{0}\pi^{+}$  & $\left(a_{6}+a_{15}+2d_{15}\right)$\tabularnewline
$D^{0}\to U_{0}^{0}K^{0}$  & $\left(a_{6}-a_{15}+2b_{15}-2d_{15}\right)$ & $D^{+}\to U_{0}^{0}K^{+}$  & $\left(a_{6}+a_{15}-2b_{15}+2d_{15}\right)$\tabularnewline
$D^{0}\to U_{-1}^{-}K^{+}$  & $-\left(a_{6}+a_{15}\right)$ & $D_{s}^{+}\to U_{1/2}^{+}K^{0}$  & $-2b_{15}$\tabularnewline
$D^{+}\to U_{3/2}^{++}\pi^{-}$  & $\left(a_{15}-a_{6}\right)$ & $D_{s}^{+}\to U_{-1/2}^{0}K^{+}$  & $2b_{15}$\tabularnewline
\hline 
\end{tabular}
\end{table}
The relations between these channels are:
\begin{align}
&\Gamma(D^{0}\to U_{-3/2}^{-}\pi^{+})={}\Gamma(D^{0}\to U_{-1}^{-}K^{+}),~~~~
\Gamma(D^{+}\to U_{3/2}^{++}\pi^{-})={}\Gamma(D^{+}\to U_{1}^{+}K^{0}),\nonumber \\
&\Gamma(D_{s}^{+}\to U_{-1/2}^{0}K^{+})={}\Gamma(D_{s}^{+}\to U_{1/2}^{+}K^{0}),~~~~
\Gamma(D^{0}\to U_{1}^{+}\pi^{-})=\Gamma(D^{0}\to U_{1/2}^{+}K^{-}),\nonumber \\
&\Gamma(D^{0}\to U_{1/2}^{0}K^{0})=\Gamma(D^{0}\to U_{-1/2}^{0}\overline{K}^{0}),~~~~
\Gamma(D^{0}\to U_{-1}^{-}\pi^{+})=\Gamma(D^{0}\to U_{-1/2}^{-}K^{+}),\nonumber \\
&\Gamma(D^{+}\to U_{0}^{0}\pi^{+})=\Gamma(D_{s}^{+}\to U_{0}^{0}K^{+}),~~~~
\Gamma(D_{s}^{+}\to U_{3/2}^{++}\pi^{-})=\Gamma(D^{+}\to U_{3/2}^{++}K^{-}),\nonumber \\
&\Gamma(D_{s}^{+}\to U_{1}^{+}K^{0})=\Gamma(D^{+}\to U_{1/2}^{+}\overline{K}^{0}),~~~~
\Gamma(D_{s}^{+}\to U_{-1/2}^{0}\pi^{+})=\Gamma(D^{+}\to U_{1/2}^{0}K^{+}),\nonumber \\
&\Gamma(D^{0}\to U_{-3/2}^{-}\pi^{+})={}\Gamma(D^{0}\to U_{-1}^{-}K^{+}),~~~~
\Gamma(D^{+}\to U_{3/2}^{++}\pi^{-})={}\Gamma(D^{+}\to U_{1}^{+}K^{0}),\nonumber \\
&\Gamma(D_{s}^{+}\to U_{-1/2}^{0}K^{+})={}\Gamma(D_{s}^{+}\to U_{1/2}^{+}K^{0}).
\end{align}
For the Cabbibo allowed, singly Cabbibo suppresed and doubly Cabbibo suppressed $D \to T_{27} P$ results, they are listed in Table~\ref{tab:ampDtoT27MNoSup}, Table~\ref{tab:ampDtoT27MSinSup} and Table~\ref{tab:ampDtoT27MDouSup}.
\begin{table}
\caption{D decays into a light tetraquark $T_{I_{z}}^{Q}$ in the 27 representation
and a light meson (No suppression).}
\label{tab:ampDtoT27MNoSup}%
\begin{tabular}{|c|c|c|c|}
\hline 
\hline 
channel  & amplitude  & channel  & amplitude \tabularnewline
\hline 
$D^{0}\to T_{0}^{0}\overline{K}^{0}$  & $-2c_{15}$ & $D^{+}\to T_{3/2}^{+}\eta_{s}$  & $a_{6}+2a_{15}+b_{15}$\tabularnewline
$D^{0}\to T_{-1/2}^{-}\pi^{+}$  & $2c_{15}$ & $D^{+}\to T_{1/2}^{\prime0}\pi^{+}$  & $b_{15}-a_{6}$\tabularnewline
$D^{0}\to T_{1}^{+}K^{-}$  & $a_{6}+b_{15}$ & $D^{+}\to T_{1}^{0}K^{+}$  & $b_{15}-a_{6}$\tabularnewline
$D^{0}\to T_{0}^{\prime0}\overline{K}^{0}$  & $a_{6}+b_{6}+b_{15}-2c_{15}-e_{15}$ & $D^{+}\to T_{1/2}^{\prime0}\pi^{+}$  & $b_{15}-a_{6}$\tabularnewline
$D^{0}\to T_{1/2}^{0}\pi^{0}$  & $\frac{-a_{6}-b_{6}+b_{15}-2c_{15}+2d_{15}+e_{15}}{\sqrt{2}}$ & $D^{+}\to T_{1}^{0}K^{+}$  & $b_{15}-a_{6}$\tabularnewline
$D^{0}\to T_{1/2}^{0}\eta_{q}$  & $\frac{-a_{6}+4a_{15}-b_{6}+b_{15}+2c_{15}+2d_{15}+e_{15}}{\sqrt{2}}$ & $D^{+}\to T_{3/2}^{+}\eta_{s}$  & $a_{6}+2a_{15}+b_{15}$\tabularnewline
$D^{0}\to T_{1/2}^{0}\eta_{s}$  & $a_{6}+2a_{15}+b_{6}+b_{15}-e_{15}$ & $D_{s}^{+}\to T_{0}^{0}\pi^{+}$  & $-2c_{15}$\tabularnewline
$D^{0}\to T_{-1/2}^{\prime-}\pi^{+}$  & $-a_{6}-b_{6}+b_{15}+e_{15}$ & $D_{s}^{+}\to T_{1}^{+}\pi^{0}$  & $\frac{b_{6}+e_{15}}{\sqrt{2}}$\tabularnewline
$D^{0}\to T_{0}^{-}K^{+}$  & $-a_{6}-b_{6}+b_{15}+e_{15}$ & $D_{s}^{+}\to T_{1}^{+}\eta_{q}$  & $\frac{b_{6}+e_{15}}{\sqrt{2}}$\tabularnewline
$D^{0}\to T_{1}^{\prime+}K^{-}$  & $b_{6}-2d_{15}-e_{15}$ & $D_{s}^{+}\to T_{0}^{\prime0}\pi^{+}$  & $a_{6}+b_{6}-b_{15}-2c_{15}+e_{15}$\tabularnewline
$D^{0}\to T_{3/2}^{+}\pi^{-}$  & $-b_{6}+2d_{15}+e_{15}$ & $D_{s}^{+}\to T_{1/2}^{0}K^{+}$  & $a_{6}+b_{6}-b_{15}+2d_{15}+e_{15}$\tabularnewline
$D^{0}\to T_{0}^{\prime\prime0}\overline{K}^{0}$  & $b_{6}-e_{15}$ & $D_{s}^{+}\to T_{1/2}^{+}\overline{K}^{0}$  & $-2c_{15}$\tabularnewline
$D^{0}\to T_{1/2}^{\prime0}\pi^{0}$  & $\frac{b_{6}-e_{15}}{\sqrt{2}}$ & $D_{s}^{+}\to T_{2}^{++}\pi^{-}$  & $b_{6}+e_{15}$\tabularnewline
$D^{0}\to T_{1/2}^{\prime0}\eta_{q}$  & $\frac{e_{15}-b_{6}}{\sqrt{2}}$ & $D_{s}^{+}\to T_{1}^{\prime+}\pi^{0}$  & $\frac{a_{6}-b_{6}-b_{15}+2c_{15}-e_{15}}{\sqrt{2}}$\tabularnewline
$D^{0}\to T_{1/2}^{\prime0}\eta_{s}$  & $b_{6}-e_{15}$ & $D_{s}^{+}\to T_{1}^{\prime+}\eta_{q}$  & $\frac{a_{6}-4a_{15}+b_{6}-b_{15}-2c_{15}+e_{15}}{\sqrt{2}}$\tabularnewline
$D^{0}\to T_{1}^{0}K^{0}$  & $e_{15}-b_{6}$ & $D_{s}^{+}\to T_{1}^{\prime+}\eta_{s}$  & $-a_{6}-2a_{15}-b_{6}-b_{15}-2d_{15}-e_{15}$\tabularnewline
$D^{+}\to T_{1/2}^{0}\pi^{+}$  & $2\left(c_{15}+d_{15}\right)$ & $D_{s}^{+}\to T_{3/2}^{+}K^{0}$  & $b_{6}+2d_{15}+e_{15}$\tabularnewline
$D^{+}\to T_{2}^{++}K^{-}$  & $a_{6}+b_{15}$ & $D_{s}^{+}\to T_{0}^{\prime\prime0}\pi^{+}$  & $a_{6}-b_{15}$\tabularnewline
$D^{+}\to T_{1}^{\prime+}\overline{K}^{0}$  & $a_{6}+b_{15}-2\left(c_{15}+d_{15}\right)$ & $D_{s}^{+}\to T_{1/2}^{\prime0}K^{+}$  & $a_{6}-b_{15}$\tabularnewline
$D^{+}\to T_{3/2}^{+}\pi^{0}$  & $\frac{-a_{6}+b_{15}-2\left(c_{15}+d_{15}\right)}{\sqrt{2}}$ & $D_{s}^{+}\to T_{3/2}^{++}K^{-}$  & $a_{6}+b_{6}+b_{15}+e_{15}$\tabularnewline
$D^{+}\to T_{3/2}^{+}\eta_{q}$  & $\frac{-a_{6}+4a_{15}+b_{15}+2c_{15}+2d_{15}}{\sqrt{2}}$ & $D_{s}^{+}\to T_{1/2}^{\prime+}\overline{K}^{0}$  & $a_{6}+b_{6}+b_{15}-2c_{15}+e_{15}$\tabularnewline
\hline 
\end{tabular}
\end{table}
\begin{table}\tiny
{\caption{D decays into a light tetraquark $T_{I_{z}}^{Q}$ in the 27 representation
and a light meson (Single suppression).}
\label{tab:ampDtoT27MSinSup}}%
\begin{tabular}{|c|c|c|c|}
\hline 
\hline 
{channel } & {amplitude } & {channel } & {amplitude }\tabularnewline
\hline 
{$D^{0}\to T_{0}^{0}\pi^{0}$ } & {$\frac{\left(a_{6}+b_{6}-b_{15}-2d_{15}-e_{15}\right)}{\sqrt{2}}$} & {$D^{+}\to T_{1/2}^{+}\overline{K}^{0}$ } & {$-2d_{15}$}\tabularnewline
{$D^{0}\to T_{0}^{0}\eta_{q}$ } & {$\frac{\left(a_{6}-4a_{15}+b_{6}-b_{15}-2d_{15}-e_{15}\right)}{\sqrt{2}}$} & {$D^{+}\to T_{2}^{++}\pi^{-}$ } & {$-\left(a_{6}+b_{6}+b_{15}+e_{15}\right)$}\tabularnewline
{$D^{0}\to T_{0}^{0}\eta_{s}$ } & {$-\left(a_{6}+2a_{15}+b_{6}+b_{15}+2c_{15}-e_{15}\right)$} & {$D^{+}\to T_{1}^{\prime+}\pi^{0}$ } & {$\frac{\left(3a_{6}+b_{6}-b_{15}+2c_{15}+4d_{15}+e_{15}\right)}{\sqrt{2}}$}\tabularnewline
{$D^{0}\to T_{-1}^{-}\pi^{+}$ } & {$\left(a_{6}+b_{6}-b_{15}-2c_{15}-e_{15}\right)$} & {$D^{+}\to T_{1}^{\prime+}\eta_{q}$ } & {$\frac{\left(a_{6}-8a_{15}-b_{6}-3b_{15}-2c_{15}-4d_{15}-e_{15}\right)}{\sqrt{2}}$}\tabularnewline
{$D^{0}\to T_{-1/2}^{-}K^{+}$ } & {$\left(a_{6}+b_{6}-b_{15}+2c_{15}-e_{15}\right)$} & {$D^{+}\to T_{1}^{\prime+}\eta_{s}$ } & {$-\left(a_{6}+4a_{15}-b_{6}+b_{15}+2c_{15}-e_{15}\right)$}\tabularnewline
{$D^{0}\to T_{1}^{+}\pi^{-}$ } & {$-\left(a_{6}+b_{15}\right)$} & {$D^{+}\to T_{3/2}^{+}K^{0}$ } & {$-\left(a_{6}+b_{6}+b_{15}-2c_{15}+e_{15}\right)$}\tabularnewline
{$D^{0}\to T_{0}^{\prime0}\pi^{0}$ } & {$\frac{\left(3a_{6}+b_{6}-b_{15}+2c_{15}-4d_{15}-e_{15}\right)}{\sqrt{2}}$} & {$D^{+}\to T_{0}^{\prime\prime0}\pi^{+}$ } & {$2\left(a_{6}-b_{15}\right)$}\tabularnewline
{$D^{0}\to T_{0}^{\prime0}\eta_{q}$ } & {$\frac{\left(a_{6}-8a_{15}+3b_{6}-3b_{15}-2c_{15}-4d_{15}-3e_{15}\right)}{\sqrt{2}}$} & {$D^{+}\to T_{1/2}^{\prime0}K^{+}$ } & {$2\left(a_{6}-b_{15}\right)$}\tabularnewline
{$D^{0}\to T_{0}^{\prime0}\eta_{s}$ } & {$-\left(a_{6}+4a_{15}+3b_{6}+b_{15}+2c_{15}-3e_{15}\right)$} & {$D^{+}\to T_{3/2}^{++}K^{-}$ } & {$\left(a_{6}-b_{6}+b_{15}-e_{15}\right)$}\tabularnewline
{$D^{0}\to T_{1/2}^{0}K^{0}$ } & {$-\left(a_{6}-b_{6}+b_{15}-2c_{15}+e_{15}\right)$} & {$D^{+}\to T_{1/2}^{\prime+}\overline{K}^{0}$ } & {$\left(a_{6}-b_{6}+b_{15}-2c_{15}-4d_{15}-e_{15}\right)$}\tabularnewline
{$D^{0}\to T_{-1}^{\prime-}\pi^{+}$ } & {$2\left(a_{6}+b_{6}-b_{15}-e_{15}\right)$} & {$D_{s}^{+}\to T_{0}^{0}K^{+}$ } & {$-\left(a_{6}+b_{6}-b_{15}+2c_{15}+2d_{15}+e_{15}\right)$}\tabularnewline
{$D^{0}\to T_{-1/2}^{\prime-}K^{+}$ } & {$2\left(a_{6}+b_{6}-b_{15}-e_{15}\right)$} & {$D_{s}^{+}\to T_{0}^{\prime0}K^{+}$ } & {$-\left(3a_{6}+b_{6}-3b_{15}+2c_{15}+4d_{15}+e_{15}\right)$}\tabularnewline
{$D^{0}\to T_{1/2}^{+}K^{-}$ } & {$\left(a_{6}+b_{6}+b_{15}-2d_{15}-e_{15}\right)$} & {$D_{s}^{+}\to T_{1/2}^{+}\pi^{0}$ } & {$\frac{\left(a_{6}+b_{6}-b_{15}+e_{15}\right)}{\sqrt{2}}$}\tabularnewline
{$D^{0}\to T_{-1/2}^{0}\overline{K}^{0}$ } & {$\left(a_{6}+b_{6}+b_{15}-2c_{15}-e_{15}\right)$} & {$D_{s}^{+}\to T_{1/2}^{+}\eta_{q}$ } & {$\frac{\left(a_{6}-4a_{15}+b_{6}-b_{15}+e_{15}\right)}{\sqrt{2}}$}\tabularnewline
{$D^{0}\to T_{1}^{\prime+}\pi^{-}$ } & {$2\left(b_{6}-2d_{15}-e_{15}\right)$} & {$D_{s}^{+}\to T_{1/2}^{+}\eta_{s}$ } & {$-\left(a_{6}+2a_{15}+b_{6}+b_{15}+2c_{15}+2d_{15}+e_{15}\right)$}\tabularnewline
{$D^{0}\to T_{0}^{\prime\prime0}\pi^{0}$ } & {$\sqrt{2}\left(e_{15}-b_{6}\right)$} & {$D_{s}^{+}\to T_{-1/2}^{0}\pi^{+}$ } & {$\left(a_{6}+b_{6}-b_{15}-2c_{15}+e_{15}\right)$}\tabularnewline
{$D^{0}\to T_{0}^{\prime\prime0}\eta_{q}$ } & {$\sqrt{2}\left(b_{6}-e_{15}\right)$} & {$D_{s}^{+}\to T_{1}^{\prime+}K^{0}$ } & {$\left(a_{6}-b_{6}+b_{15}-2c_{15}-4d_{15}-e_{15}\right)$}\tabularnewline
{$D^{0}\to T_{0}^{\prime\prime0}\eta_{s}$ } & {$-2\left(b_{6}-e_{15}\right)$} & {$D_{s}^{+}\to T_{0}^{\prime\prime0}K^{+}$ } & {$-2\left(a_{6}-b_{15}\right)$}\tabularnewline
{$D^{0}\to T_{1/2}^{\prime0}K^{0}$ } & {$2\left(b_{6}-e_{15}\right)$} & {$D_{s}^{+}\to T_{3/2}^{++}\pi^{-}$ } & {$-\left(a_{6}-b_{6}+b_{15}-e_{15}\right)$}\tabularnewline
{$D^{0}\to T_{1/2}^{\prime+}K^{-}$ } & {$2\left(b_{6}-2d_{15}-e_{15}\right)$} & {$D_{s}^{+}\to T_{1/2}^{\prime+}\pi^{0}$ } & {$\frac{\left(3a_{6}-b_{6}-b_{15}+2c_{15}-e_{15}\right)}{\sqrt{2}}$}\tabularnewline
{$D^{0}\to T_{-1/2}^{\prime0}\overline{K}^{0}$ } & {$2\left(b_{6}-e_{15}\right)$} & {$D_{s}^{+}\to T_{1/2}^{\prime+}\eta_{q}$ } & {$\frac{\left(a_{6}-8a_{15}+b_{6}-3b_{15}-2c_{15}+e_{15}\right)}{\sqrt{2}}$}\tabularnewline
{$D^{+}\to T_{0}^{0}\pi^{+}$ } & {$-2d_{15}$} & {$D_{s}^{+}\to T_{1/2}^{\prime+}\eta_{s}$ } & {$-\left(a_{6}+4a_{15}+b_{6}+b_{15}+2c_{15}+4d_{15}+e_{15}\right)$}\tabularnewline
{$D^{+}\to T_{1}^{+}\pi^{0}$ } & {$-\frac{\left(b_{6}+e_{15}\right)}{\sqrt{2}}$} & {$D_{s}^{+}\to T_{-1/2}^{\prime0}\pi^{+}$ } & {$2\left(a_{6}-b_{15}\right)$}\tabularnewline
{$D^{+}\to T_{1}^{+}\eta_{q}$ } & {$-\frac{\left(b_{6}+e_{15}\right)}{\sqrt{2}}$} & {$D_{s}^{+}\to T_{1}^{++}K^{-}$ } & {$\left(a_{6}+b_{6}+b_{15}+e_{15}\right)$}\tabularnewline
{$D^{+}\to T_{0}^{\prime0}\pi^{+}$ } & {$\left(a_{6}-b_{6}-b_{15}-2c_{15}-4d_{15}-e_{15}\right)$} & {$D_{s}^{+}\to T_{0}^{+}\overline{K}^{0}$ } & {$\left(a_{6}+b_{6}+b_{15}-2c_{15}+e_{15}\right)$}\tabularnewline
{$D^{+}\to T_{1/2}^{0}K^{+}$ } & {$\left(a_{6}-b_{6}-b_{15}+2c_{15}-e_{15}\right)$} &  & \tabularnewline
\hline 
\end{tabular}
\end{table}
\begin{table}
\caption{D decays into a light tetraquark $T_{I_{z}}^{Q}$ in the 27 representation
and a light meson (Double suppression).}
\label{tab:ampDtoT27MDouSup}%
\begin{tabular}{|c|c|c|c|}
\hline 
\hline 
channel  & amplitude  & channel  & amplitude \tabularnewline
\hline 
$D^{0}\to T_{0}^{0}K^{0}$  & $\left(a_{6}+b_{15}\right)$ & $D^{+}\to T_{-1/2}^{0}\pi^{+}$  & $-\left(b_{6}+2d_{15}+e_{15}\right)$\tabularnewline
$D^{0}\to T_{-1}^{-}K^{+}$  & $-\left(a_{6}+b_{6}-b_{15}+2c_{15}-e_{15}\right)$ & $D^{+}\to T_{1}^{\prime+}K^{0}$  & $\left(a_{6}+b_{6}+b_{15}-2c_{15}+e_{15}\right)$\tabularnewline
$D^{0}\to T_{0}^{\prime0}K^{0}$  & $\left(a_{6}-b_{6}+b_{15}-2c_{15}+e_{15}\right)$ & $D^{+}\to T_{0}^{\prime\prime0}K^{+}$  & $-\left(a_{6}-b_{15}\right)$\tabularnewline
$D^{0}\to T_{-1}^{\prime-}K^{+}$  & $-\left(a_{6}+b_{6}-b_{15}-e_{15}\right)$ & $D^{+}\to T_{3/2}^{++}\pi^{-}$  & $-\left(a_{6}+b_{6}+b_{15}+e_{15}\right)$\tabularnewline
$D^{0}\to T_{1/2}^{+}\pi^{-}$  & $-\left(a_{6}+b_{15}\right)$ & $D^{+}\to T_{1/2}^{\prime+}\pi^{0}$  & $\frac{\left(2a_{6}+b_{6}+2d_{15}+e_{15}\right)}{\sqrt{2}}$\tabularnewline
$D^{0}\to T_{-1/2}^{0}\pi^{0}$  & $\frac{\left(2a_{6}+b_{6}-2d_{15}-e_{15}\right)}{\sqrt{2}}$ & $D^{+}\to T_{1/2}^{\prime+}\eta_{q}$  & $-\frac{\left(4a_{15}+b_{6}+2b_{15}+2d_{15}+e_{15}\right)}{\sqrt{2}}$\tabularnewline
$D^{0}\to T_{-1/2}^{0}\eta_{q}$  & $-\frac{\left(4a_{15}-b_{6}+2b_{15}+2d_{15}+e_{15}\right)}{\sqrt{2}}$ & $D^{+}\to T_{1/2}^{\prime+}\eta_{s}$  & $-\left(2a_{15}-b_{6}+2c_{15}-e_{15}\right)$\tabularnewline
$D^{0}\to T_{-1/2}^{0}\eta_{s}$  & $-\left(2a_{15}+b_{6}+2c_{15}-e_{15}\right)$ & $D^{+}\to T_{-1/2}^{\prime0}\pi^{+}$  & $\left(a_{6}-b_{15}\right)$\tabularnewline
$D^{0}\to T_{-3/2}^{-}\pi^{+}$  & $\left(a_{6}+b_{6}-b_{15}-e_{15}\right)$ & $D^{+}\to T_{1}^{++}K^{-}$  & $-\left(b_{6}+e_{15}\right)$\tabularnewline
$D^{0}\to T_{0}^{\prime\prime0}K^{0}$  & $-\left(b_{6}-e_{15}\right)$ & $D^{+}\to T_{0}^{+}\overline{K}^{0}$  & $-\left(b_{6}+2d_{15}+e_{15}\right)$\tabularnewline
$D^{0}\to T_{1/2}^{\prime+}\pi^{-}$  & $\left(b_{6}-2d_{15}-e_{15}\right)$ & $D^{+}\to T_{1/2}^{+}\eta_{s}$  & $\left(b_{6}+e_{15}\right)$\tabularnewline
$D^{0}\to T_{-1/2}^{\prime0}\pi^{0}$  & $-\frac{\left(b_{6}-e_{15}\right)}{\sqrt{2}}$ & $D_{s}^{+}\to T_{1/2}^{+}K^{0}$  & $\left(a_{6}+b_{15}\right)$\tabularnewline
$D^{0}\to T_{-1/2}^{\prime0}\eta_{q}$  & $\frac{\left(b_{6}-e_{15}\right)}{\sqrt{2}}$ & $D_{s}^{+}\to T_{-1/2}^{0}K^{+}$  & $-\left(a_{6}-b_{15}+2\left(c_{15}+d_{15}\right)\right)$\tabularnewline
$D^{0}\to T_{-1/2}^{\prime0}\eta_{s}$  & $-\left(b_{6}-e_{15}\right)$ & $D_{s}^{+}\to T_{1/2}^{\prime+}K^{0}$  & $\left(a_{6}+b_{15}-2\left(c_{15}+d_{15}\right)\right)$\tabularnewline
$D^{0}\to T_{0}^{+}K^{-}$  & $\left(b_{6}-2d_{15}-e_{15}\right)$ & $D_{s}^{+}\to T_{-1/2}^{\prime0}K^{+}$  & $-\left(a_{6}-b_{15}\right)$\tabularnewline
$D^{0}\to T_{-1}^{0}\overline{K}^{0}$  & $\left(b_{6}-e_{15}\right)$ & $D_{s}^{+}\to T_{1}^{++}\pi^{-}$  & $-\left(a_{6}+b_{15}\right)$\tabularnewline
$D^{+}\to T_{0}^{0}K^{+}$  & $\left(b_{6}+e_{15}\right)$ & $D_{s}^{+}\to T_{0}^{+}\pi^{0}$  & $\sqrt{2}a_{6}$\tabularnewline
$D^{+}\to T_{0}^{\prime0}K^{+}$  & $-\left(a_{6}-b_{6}-b_{15}+2c_{15}-e_{15}\right)$ & $D_{s}^{+}\to T_{0}^{+}\eta_{q}$  & $-\sqrt{2}\left(2a_{15}+b_{15}\right)$\tabularnewline
$D^{+}\to T_{1/2}^{+}\pi^{0}$  & $-\frac{\left(b_{6}+e_{15}\right)}{\sqrt{2}}$ & $D_{s}^{+}\to T_{0}^{+}\eta_{s}$  & $-2\left(a_{15}+c_{15}+d_{15}\right)$\tabularnewline
$D^{+}\to T_{1/2}^{+}\eta_{q}$  & $-\frac{\left(b_{6}+e_{15}\right)}{\sqrt{2}}$ & $D_{s}^{+}\to T_{-1}^{0}\pi^{+}$  & $\left(a_{6}-b_{15}\right)$\tabularnewline
\hline 
\end{tabular}
\end{table}
The relations between these channels are:
\begin{align}
&\Gamma(D^{0}\to T_{-1/2}^{-}\pi^{+})=\Gamma(D^{0}\to T_{0}^{0}\overline{K}^{0})=\Gamma(D_{s}^{+}\to T_{1/2}^{+}\overline{K}^{0}),\nonumber \\
&\Gamma(D^{0}\to T_{-1/2}^{\prime-}\pi^{+})=\Gamma(D^{0}\to T_{0}^{-}K^{+}),\nonumber \\
&\Gamma(D^{0}\to T_{3/2}^{+}\pi^{-})=\Gamma(D^{0}\to T_{1}^{\prime+}K^{-}),\nonumber \\
&\Gamma(D^{0}\to T_{0}^{\prime\prime0}\overline{K}^{0})=2\Gamma(D^{0}\to T_{1/2}^{\prime0}\eta_{q})=\Gamma(D^{0}\to T_{1/2}^{\prime0}\eta_{s})=2\Gamma(D^{0}\to T_{1/2}^{\prime0}\pi^{0})\nonumber \\
&=\Gamma(D^{0}\to T_{1}^{0}K^{0}),\nonumber \\
&\Gamma(D^{+}\to T_{1/2}^{\prime0}\pi^{+})=\Gamma(D^{+}\to T_{1}^{0}K^{+})=\Gamma(D_{s}^{+}\to T_{1/2}^{\prime0}K^{+})=\Gamma(D_{s}^{+}\to T_{0}^{\prime\prime0}\pi^{+}),\nonumber \\
&\Gamma(D_{s}^{+}\to T_{0}^{0}\pi^{+})=\Gamma(D^{0}\to T_{0}^{0}\overline{K}^{0})=\Gamma(D_{s}^{+}\to T_{1/2}^{+}\overline{K}^{0}),\nonumber \\
&\Gamma(D_{s}^{+}\to T_{1}^{+}\pi^{0})=\Gamma(D_{s}^{+}\to T_{1}^{+}\eta_{q})=\frac{1}{2}\Gamma(D_{s}^{+}\to T_{2}^{++}\pi^{-}),\nonumber\\
&\Gamma(D^{0}\to T_{-1}^{\prime-}\pi^{+})=\Gamma(D^{0}\to T_{-1/2}^{\prime-}K^{+}),~~~~
\Gamma(D^{0}\to T_{1}^{\prime+}\pi^{-})=\Gamma(D^{0}\to T_{1/2}^{\prime+}K^{-}),\nonumber \\
&\Gamma(D^{0}\to T_{0}^{\prime\prime0}\pi^{0})=\Gamma(D^{0}\to T_{0}^{\prime\prime0}\eta_{q})=\frac{1}{2}\Gamma(D^{0}\to T_{0}^{\prime\prime0}\eta_{s})=\frac{1}{2}\Gamma(D^{0}\to T_{1/2}^{\prime0}K^{0})\nonumber \\
&=\frac{1}{2}\Gamma(D^{0}\to T_{-1/2}^{\prime0}\overline{K}^{0}),\nonumber \\
&\Gamma(D^{+}\to T_{0}^{0}\pi^{+})=\Gamma(D^{+}\to T_{1/2}^{+}\overline{K}^{0}),~~~~
\Gamma(D^{+}\to T_{1}^{+}\pi^{0})=\Gamma(D^{+}\to T_{1}^{+}\eta_{q}),\nonumber \\
&\Gamma(D^{+}\to T_{2}^{++}\pi^{-})=\Gamma(D_{s}^{+}\to T_{1}^{++}K^{-}),~~~~
\Gamma(D^{+}\to T_{3/2}^{+}K^{0})=\Gamma(D_{s}^{+}\to T_{0}^{+}\overline{K}^{0}),\nonumber \\
&\Gamma(D^{+}\to T_{0}^{\prime\prime0}\pi^{+})=\Gamma(D^{+}\to T_{1/2}^{\prime0}K^{+})=\Gamma(D_{s}^{+}\to T_{0}^{\prime\prime0}K^{+})=\Gamma(D_{s}^{+}\to T_{-1/2}^{\prime0}\pi^{+}),\nonumber \\
&\Gamma(D_{s}^{+}\to T_{1}^{\prime+}K^{0})=\Gamma(D^{+}\to T_{1/2}^{\prime+}\overline{K}^{0}),~~~~
\Gamma(D_{s}^{+}\to T_{3/2}^{++}\pi^{-})=\Gamma(D^{+}\to T_{3/2}^{++}K^{-}),\nonumber \\
&\Gamma(D^{0}\to T_{1/2}^{+}\pi^{-})=\Gamma(D^{0}\to T_{0}^{0}K^{0})=\Gamma(D_{s}^{+}\to T_{1/2}^{+}K^{0})=\Gamma(D^{0}\to T_{0}^{+}K^{-}),\nonumber \\
&\Gamma(D^{0}\to T_{-3/2}^{-}\pi^{+})=\Gamma(D^{0}\to T_{-1}^{\prime-}K^{+}),\nonumber \\
&\Gamma(D^{0}\to T_{0}^{\prime\prime0}K^{0})=2\Gamma(D^{0}\to T_{-1/2}^{\prime0}\eta_{q})=\Gamma(D^{0}\to T_{-1/2}^{\prime0}\eta_{s})=\Gamma(D^{0}\to T_{-1}^{0}\overline{K}^{0})\nonumber \\
&=2\Gamma(D^{0}\to T_{-1/2}^{\prime0}\pi^{0}),\nonumber \\
&\Gamma(D^{+}\to T_{0}^{0}K^{+})=2\Gamma(D^{+}\to T_{1/2}^{+}\eta_{q})=\Gamma(D^{+}\to T_{1/2}^{+}\eta_{s})=\Gamma(D^{+}\to T_{1}^{++}K^{-}),\nonumber \\
&\Gamma(D^{+}\to T_{1/2}^{+}\pi^{0})=\frac{1}{2}\Gamma(D^{+}\to T_{0}^{0}K^{+})=\Gamma(D^{+}\to T_{1/2}^{+}\eta_{q})=\frac{1}{2}\Gamma(D^{+}\to T_{1/2}^{+}\eta_{s})\nonumber \\
&=\frac{1}{2}\Gamma(D^{+}\to T_{1}^{++}K^{-}),\nonumber \\
&\Gamma(D^{+}\to T_{-1/2}^{0}\pi^{+})=\Gamma(D^{+}\to T_{0}^{+}\overline{K}^{0}),\nonumber \\
&\Gamma(D^{+}\to T_{-1/2}^{\prime0}\pi^{+})=\Gamma(D^{+}\to T_{0}^{\prime\prime0}K^{+})=\Gamma(D_{s}^{+}\to T_{-1/2}^{\prime0}K^{+}),\nonumber \\
&\Gamma(D_{s}^{+}\to T_{1}^{++}\pi^{-})=\Gamma(D^{0}\to T_{0}^{0}K^{0})=\Gamma(D_{s}^{+}\to T_{1/2}^{+}K^{0}),\nonumber \\
&\Gamma(D_{s}^{+}\to T_{-1}^{0}\pi^{+})=\Gamma(D^{+}\to T_{0}^{\prime\prime0}K^{+})=\Gamma(D_{s}^{+}\to T_{-1/2}^{\prime0}K^{+}).
\end{align}

\subsection{Non-leptonic $B_c$ decays}
In this subsection we consider the two-body non-leptonic decay  of $B_c$. Now there are two possible transitions, one of them is $B_c \to T_{10}/T_{27} D$ where $b\to d/s\bar u u$, while another one is $B_c \to T_{10}/T_{27} P$ where $b \bar c \to d/s \bar u$ which is induced by the operator $H_8$, with $(H_8)^2_1=V_{ud}^*$, $(H_8)^3_1=V_{us}^*$.  The effective Hamiltonian of $B_c \to T_{10}/T_{27} D$ is
\begin{eqnarray}
\mathcal{H}_{eff}=c_1 B_c (H_{15})_k^{\{ij\}} (T_{10})_{\{ijm\}} \epsilon^{klm} D_l+c_2 B_c (H_{15})_k^{\{ij\}} (T_{27})_{\{ij\}}^{\{kl\}} D_l.
\end{eqnarray}
\begin{figure}
\includegraphics[width=0.5\columnwidth]{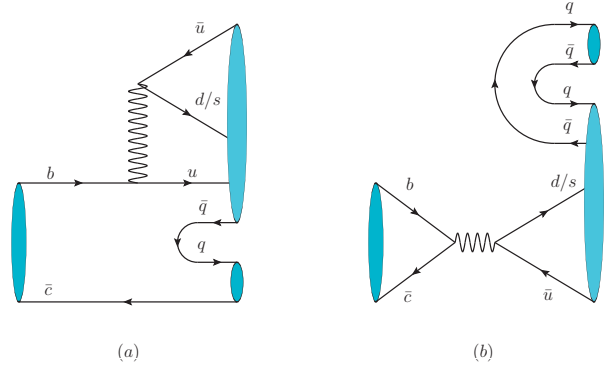} 
\caption{Feynman diagrams for the non-leptonic decays $B_c \to T_{10}/T_{27} D$ (a) and $B_c \to T_{10}/T_{27} P$ (b).}
\label{fig:Bcdecay} 
\end{figure}
Fig.~\ref{fig:Bcdecay} (a) shows the Feynman diagram corresponding to these two Hamiltonians. 
For $B_c \to T_{27} P$ decay.
The amplitudes are listed in Table~\ref{tab:ampBctoT10D} and Table~\ref{tab:ampBctoT27D} respectively, where the left two columns correspond to $b\to d$, while the right two correspond to $b\to s$. 
\begin{table}
\caption{Bc decays into a light tetraquark $T_{I_{z}}^{Q}$ in the 10 representation
and a charmed meson. The left two columns correspond to $b\to d$, while the right two correspond to $b\to s$. }
\label{tab:ampBctoT10D}%
\begin{tabular}{|c|c|c|c|}
\hline 
\hline
channel  & amplitude  & channel  & amplitude \tabularnewline
\hline 
$B_{c}\to U_{-1/2}^{0}D_{s}^{-}$  & $-8c_{1}$ & $B_{c}\to U_{0}^{0}D_{s}^{-}$  & $-8c_{1}$\tabularnewline
$B_{c}\to U_{0}^{0}D^{-}$  & $8c_{1}$ & $B_{c}\to U_{1/2}^{0}D^{-}$  & $8c_{1}$\tabularnewline
\hline 
\end{tabular}
\end{table}
\begin{table}
\caption{Bc decays into a light tetraquark $T_{I_{z}}^{Q}$ in the 27 representation
and a charmed meson. The left two columns correspond to $b\to d$, while the right two correspond to $b\to s$. }
\label{tab:ampBctoT27D}%
\begin{tabular}{|c|c|c|c|}
\hline
\hline 
channel  & amplitude  & channel  & amplitude \tabularnewline
\hline 
$B_{c}\to T_{-1}^{-}\overline{D}^{0}$  & $8c_{2}$ & $B_{c}\to T_{0}^{0}D_{s}^{-}$  & $-8c_{2}$\tabularnewline
$B_{c}\to T_{0}^{\prime0}D^{-}$  & $8c_{2}$ & $B_{c}\to T_{-1/2}^{-}\overline{D}^{0}$  & $8c_{2}$\tabularnewline
$B_{c}\to T_{-1/2}^{0}D_{s}^{-}$  & $8c_{2}$ & $B_{c}\to T_{0}^{\prime0}D_{s}^{-}$  & $-8c_{2}$\tabularnewline
 &  & $B_{c}\to T_{1/2}^{0}D^{-}$  & $8c_{2}$\tabularnewline
\hline 
\end{tabular}
\end{table}

The effective Hamiltonian of $B_c \to T_{10}/T_{27} P$ is
\begin{eqnarray}
\mathcal{H}_{eff}=d_1 B_c (H_{8})_i^j (T_{10})_{\{jkm\}} \epsilon^{ilm} P_l^k+d_2 B_c (H_{8})_i^j (T_{27})_{\{jk\}}^{\{il\}} P_l^k.
\end{eqnarray}
These two terms are described by the Feynman diagram of Fig.~\ref{fig:Bcdecay} (b).
The amplitudes are listed in Table~\ref{tab:ampBctoT10P} and Table~\ref{tab:ampBctoT27P} respectively. Since these amplitudes have very simple forms, so the relationships among them are obvious and we will not explicitly list them here.
\begin{table}
\caption{Bc decays into a light tetraquark $T_{I_{z}}^{Q}$ in the 10 representation
and a light meson. }
\label{tab:ampBctoT10P}%
\begin{tabular}{|c|c|c|c|}
\hline 
\hline 
channel  & amplitude  & channel  & amplitude \tabularnewline
\hline 
$B_{c}\to U_{-1/2}^{0}K^{-}$  & $-d_{1}\left(V_{\text{ud}}\right){}^{*}$ & $B_{c}\to U_{-1}^{-}\eta_{q}$  & $\frac{d_{1}\left(V_{\text{ud}}\right){}^{*}}{\sqrt{2}}$\tabularnewline
$B_{c}\to U_{-3/2}^{-}\overline{K}^{0}$  & $-d_{1}\left(V_{\text{ud}}\right){}^{*}$ & $B_{c}\to U_{-1}^{-}\eta_{s}$  & $-d_{1}\left(V_{\text{ud}}\right){}^{*}$\tabularnewline
$B_{c}\to U_{0}^{0}\pi^{-}$  & $d_{1}\left(V_{\text{ud}}\right){}^{*}$ & $B_{c}\to U_{-1/2}^{-}\pi^{0}$  & $-\frac{d_{1}\left(V_{\text{us}}\right){}^{*}}{\sqrt{2}}$\tabularnewline
$B_{c}\to U_{0}^{0}K^{-}$  & $-d_{1}\left(V_{\text{us}}\right){}^{*}$ & $B_{c}\to U_{-1/2}^{-}K^{0}$  & $d_{1}\left(V_{\text{ud}}\right){}^{*}$\tabularnewline
$B_{c}\to U_{1/2}^{0}\pi^{-}$  & $d_{1}\left(V_{\text{us}}\right){}^{*}$ & $B_{c}\to U_{-1/2}^{-}\eta_{q}$  & $\frac{d_{1}\left(V_{\text{us}}\right){}^{*}}{\sqrt{2}}$\tabularnewline
$B_{c}\to U_{-1}^{-}\pi^{0}$  & $-\frac{d_{1}\left(V_{\text{ud}}\right){}^{*}}{\sqrt{2}}$ & $B_{c}\to U_{-1/2}^{-}\eta_{s}$  & $-d_{1}\left(V_{\text{us}}\right){}^{*}$\tabularnewline
$B_{c}\to U_{-1}^{-}\overline{K}^{0}$  & $-d_{1}\left(V_{\text{us}}\right){}^{*}$ & $B_{c}\to U_{0}^{-}K^{0}$  & $d_{1}\left(V_{\text{us}}\right){}^{*}$\tabularnewline
\hline 
\end{tabular}
\end{table}
\begin{table}
\caption{Bc decays into a light tetraquark $T_{I_{z}}^{Q}$ in the 27 representation
and a light meson. }
\label{tab:ampBctoT27P}%
\begin{tabular}{|c|c|c|c|}
\hline 
\hline 
channel  & amplitude  & channel  & amplitude \tabularnewline
\hline 
$B_{c}\to T_{0}^{0}K^{-}$  & $-d_{2}\left(V_{\text{us}}\right){}^{*}$ & $B_{c}\to T_{-1}^{\prime-}\eta_{q}$  & $\frac{d_{2}\left(V_{\text{ud}}\right){}^{*}}{\sqrt{2}}$\tabularnewline
$B_{c}\to T_{-1}^{-}\pi^{0}$  & $\frac{d_{2}\left(V_{\text{ud}}\right){}^{*}}{\sqrt{2}}$ & $B_{c}\to T_{-1}^{\prime-}\eta_{s}$  & $-d_{2}\left(V_{\text{ud}}\right){}^{*}$\tabularnewline
$B_{c}\to T_{-1}^{-}\overline{K}^{0}$  & $-d_{2}\left(V_{\text{us}}\right){}^{*}$ & $B_{c}\to T_{-1/2}^{\prime-}\pi^{0}$  & $-\frac{d_{2}\left(V_{\text{us}}\right){}^{*}}{\sqrt{2}}$\tabularnewline
$B_{c}\to T_{-1}^{-}\eta_{q}$  & $\frac{d_{2}\left(V_{\text{ud}}\right){}^{*}}{\sqrt{2}}$ & $B_{c}\to T_{-1/2}^{\prime-}K^{0}$  & $d_{2}\left(V_{\text{ud}}\right){}^{*}$\tabularnewline
$B_{c}\to T_{-1}^{-}\eta_{s}$  & $-d_{2}\left(V_{\text{ud}}\right){}^{*}$ & $B_{c}\to T_{-1/2}^{\prime-}\eta_{q}$  & $\frac{d_{2}\left(V_{\text{us}}\right){}^{*}}{\sqrt{2}}$\tabularnewline
$B_{c}\to T_{-1/2}^{-}\pi^{0}$  & $\frac{d_{2}\left(V_{\text{us}}\right){}^{*}}{\sqrt{2}}$ & $B_{c}\to T_{-1/2}^{\prime-}\eta_{s}$  & $-d_{2}\left(V_{\text{us}}\right){}^{*}$\tabularnewline
$B_{c}\to T_{-1/2}^{-}\eta_{q}$  & $\frac{d_{2}\left(V_{\text{us}}\right){}^{*}}{\sqrt{2}}$ & $B_{c}\to T_{0}^{-}K^{0}$  & $d_{2}\left(V_{\text{us}}\right){}^{*}$\tabularnewline
$B_{c}\to T_{-1/2}^{-}\eta_{s}$  & $-d_{2}\left(V_{\text{us}}\right){}^{*}$ & $B_{c}\to T_{-1/2}^{0}K^{-}$  & $d_{2}\left(V_{\text{ud}}\right){}^{*}$\tabularnewline
$B_{c}\to T_{-2}^{--}\pi^{+}$  & $d_{2}\left(V_{\text{ud}}\right){}^{*}$ & $B_{c}\to T_{-3/2}^{-}\overline{K}^{0}$  & $d_{2}\left(V_{\text{ud}}\right){}^{*}$\tabularnewline
$B_{c}\to T_{-3/2}^{--}\pi^{+}$  & $d_{2}\left(V_{\text{us}}\right){}^{*}$ & $B_{c}\to T_{0}^{\prime0}K^{-}$  & $-d_{2}\left(V_{\text{us}}\right){}^{*}$\tabularnewline
$B_{c}\to T_{-3/2}^{--}K^{+}$  & $d_{2}\left(V_{\text{ud}}\right){}^{*}$ & $B_{c}\to T_{1/2}^{0}\pi^{-}$  & $d_{2}\left(V_{\text{us}}\right){}^{*}$\tabularnewline
$B_{c}\to T_{-1}^{--}K^{+}$  & $d_{2}\left(V_{\text{us}}\right){}^{*}$ & $B_{c}\to T_{-1}^{\prime-}\pi^{0}$  & $-\frac{d_{2}\left(V_{\text{ud}}\right){}^{*}}{\sqrt{2}}$\tabularnewline
$B_{c}\to T_{0}^{\prime0}\pi^{-}$  & $d_{2}\left(V_{\text{ud}}\right){}^{*}$ & $B_{c}\to T_{-1}^{\prime-}\overline{K}^{0}$  & $-d_{2}\left(V_{\text{us}}\right){}^{*}$\tabularnewline
\hline 
\end{tabular}
\end{table}

\section{Golden Decay Channels}
In this section, with all the decay channels listed above, we can choose the golden decay channels from them. Compared with other channels,  the golden decay channels  should have greater chance to be observed in the experiments. Generally, the criterion for choosing the golden channels is based on three requirements. The first one is to see whether the decay channel offers a large enough phase space to produce the fully-light tetraquark. For $B$ or $B_c$ decays this is not the problem, while for $D$ decays only the channels with the pion in the final state will be considered. Secondly, the channels with $\pi^0, \eta_q, \eta_s$ in the final state will not be considered. Although these neutral particles are able to be detected by the current experimental techniques, their high background prevent us to choose them as final states appearing in the golden channels. On the other hand, the channels with $K^0, {\overline K}^0$ should be kept since their corresponding excited states $K^{0*}, {\overline K}^{0*}$ can decay to charged pions.  Finally, for $D$ decays we only consider the Cabbibo allowed channels. The golden channels for semi-leptonic $B, D$ decays are listed in Table~\ref{tab:GoldenBtoTUlnu} and Table~\ref{tab:GoldenDtoTUlnu}. The golden channels for non-leptonic $B\to T/U\ P$ decays are listed in Table~\ref{tab:GoldenBtoU} and Table~\ref{tab:GoldenBtoT}. The golden channels for $D$ decays are listed in Table~\ref{tab:GoldenDdecay}, which only contain the Cabbibo allowed channels. For the $B_c \to U/T\ D$ decays, there are no preferred channels among them.

Furthermore, one should also check the possible final states of the fully-light tetraquark strong decays. The most suitable case is to consider the two-body decays $U/T\to PP$. For $T_{27}$ states, the two-body strong decay can be described by a simple effective Hamiltonian:
\begin{equation}
{\cal H}_{str}^{T\to PP}\sim (T_{27})_{kl}^{ij}P_i^k P_j^l,
\end{equation}
while for $T_{10}$ states, there are no such two-body processes. Although the effective Hamiltonian for decays into three pseudoscalar mesons do exist, due to the limited phase space we will not consider such three-body decays for the $T_{10}$ states. We only choose the $T\to PP$ channels with the two final states being both charged, and  being either double pions or one pion and one kaon for $B$ decays but only double pions for $D$ decays. The channels satisfying these requirements are
\begin{align}
& T_{-3/2}^{--} \to\pi^{-}K^{-},\ \ T_{-2}^{--}\to\pi^{-}\pi^{-},\ \ T_{1/2}^{0}\to\pi^{+}K^{-},\ \ T_{-1/2}^{0}\to\pi^{-}K^{+},\nonumber \\
 & T_{0}^{{'0}}\to\pi^{+}\pi^{-},\ \ T_{3/2}^{++}\to\pi^{+}K^{+},\ \  T_{2}^{++}\to\pi^{+}\pi^{+}.
\end{align}
In Table~\ref{tab:BestGoldenBtoT} we list the best channels for reconstructing the $T_{27}$ from $B/D$ decays. There is only one suitable $D$ decay channel $D_s^+\to\pi^+\pi^-\pi^+$ and the resonant contribution can be from $T_0^{\prime 0}\to\pi^+\pi^-$ or $T_2^{++}\to\pi^+\pi^+$. The BarBar collaboration presented a Dalitz plot analysis for $D_s^+\to\pi^+\pi^-\pi^+$ \cite{Aubert:2008ao}, where the S-wave contribution in the $\pi^+\pi^-$ channel is measured. The possible candidates for such scalar particles are $a_0(980)$ or $f_0(980)$, which can have 4-quark structure, and for example, be a tetraquark $T_0^{\prime 0}$ in 27 state. On the other hand, the tetraquark $T_2^{++}$ has the flavor structure $uu\bar d \bar d$, from which we can enumerate its mass to be around $2m_{\pi}$. However, as shown in the Fig. 3(d) of Ref.~\cite{Aubert:2008ao}, at this region there is a large peak produced by the interference between the S-wave and $f_2(1270)$. This implies that a more sensitive measurement in the $2m_{\pi}$ mass region for the $\pi^+\pi^+$ channel is required  to detect the potentially existed $T_2^{++}$ .
\begin{table}
\caption{Golden channels of $B\to$$U_{I_{z}}^{Q}/T_{I_{z}}^{Q}Pl\bar{\nu}$.}
\label{tab:GoldenBtoTUlnu}%
\begin{tabular}{|cccc|}
\hline 
\hline 
$B\to$$U_{I_{z}}^{Q}Pl\bar{\nu}$ &  &  & \tabularnewline
\hline 
$B^{-}\to U_{1/2}^{+}K^{-}l^{-}\bar{\nu}$  & $\overline{B}^{0}\to U_{3/2}^{++}K^{-}l^{-}\bar{\nu}$  & $\overline{B}_{s}^{0}\to U_{3/2}^{++}\pi^{-}l^{-}\bar{\nu}$  & $\overline{B}_{s}^{0}\to U_{0}^{0}K^{+}l^{-}\bar{\nu}$ \tabularnewline
$B^{-}\to U_{1}^{+}\pi^{-}l^{-}\bar{\nu}$  & $\overline{B}^{0}\to U_{1/2}^{+}\overline{K}^{0}l^{-}\bar{\nu}$  & $\overline{B}_{s}^{0}\to U_{1}^{+}K^{0}l^{-}\bar{\nu}$  & $\overline{B}_{s}^{0}\to U_{-1/2}^{0}\pi^{+}l^{-}\bar{\nu}$ \tabularnewline
$\overline{B}^{0}\to U_{1/2}^{0}K^{+}l^{-}\bar{\nu}$  & $\overline{B}^{0}\to U_{0}^{0}\pi^{+}l^{-}\bar{\nu}$  & $\overline{B}^{0}\to U_{1/2}^{+}\overline{K}^{0}l^{-}\bar{\nu}$  & $B^{-}\to U_{1/2}^{0}\overline{K}^{0}l^{-}\bar{\nu}$ \tabularnewline
$B^{-}\to U_{1/2}^{0}K^{0}l^{-}\bar{\nu}$  &  &  & \tabularnewline
\hline 
\hline 
$B\to$$T_{I_{z}}^{Q}Pl\bar{\nu}$ &  &  & \tabularnewline
\hline 
$B^{-}\to T_{-1}^{-}\pi^{+}l^{-}\bar{\nu}$  & $\overline{B}^{0}\to T_{2}^{++}\pi^{-}l^{-}\bar{\nu}$  & $\overline{B}_{s}^{0}\to T_{-1/2}^{0}\pi^{+}l^{-}\bar{\nu}$  & $\overline{B}_{s}^{0}\to T_{0}^{+}\overline{K}^{0}l^{-}\bar{\nu}$ \tabularnewline
$B^{-}\to T_{-1/2}^{-}K^{+}l^{-}\bar{\nu}$  & $\overline{B}^{0}\to T_{3/2}^{+}K^{0}l^{-}\bar{\nu}$  & $\overline{B}_{s}^{0}\to T_{1}^{\prime+}K^{0}l^{-}\bar{\nu}$  & $\overline{B}_{s}^{0}\to T_{0}^{0}K^{+}l^{-}\bar{\nu}$ \tabularnewline
$B^{-}\to T_{1}^{+}\pi^{-}l^{-}\bar{\nu}$  & $\overline{B}^{0}\to T_{3/2}^{++}K^{-}l^{-}\bar{\nu}$  & $\overline{B}_{s}^{0}\to T_{3/2}^{++}\pi^{-}l^{-}\bar{\nu}$  & $\overline{B}_{s}^{0}\to T_{0}^{\prime0}K^{+}l^{-}\bar{\nu}$ \tabularnewline
$B^{-}\to T_{1/2}^{+}K^{-}l^{-}\bar{\nu}$  & $\overline{B}^{0}\to T_{1/2}^{\prime+}\overline{K}^{0}l^{-}\bar{\nu}$  & $\overline{B}_{s}^{0}\to T_{1}^{++}K^{-}l^{-}\bar{\nu}$  & $\overline{B}^{0}\to T_{1/2}^{0}K^{+}l^{-}\bar{\nu}$ \tabularnewline
$\overline{B}^{0}\to T_{0}^{\prime0}\pi^{+}l^{-}\bar{\nu}$  &  &  & \tabularnewline
\hline 
\end{tabular}
\end{table}
\begin{table}
\caption{Golden channels of $D\to$$U_{I_{z}}^{Q}/T_{I_{z}}^{Q}Pl\bar{\nu}$.}
\label{tab:GoldenDtoTUlnu}%
\begin{tabular}{|cccc|}
\hline 
\hline 
$D\to$$U_{I_{z}}^{Q}Pl\bar{\nu}$ &  &  & \tabularnewline
\hline 
$D^{0}\to U_{1/2}^{0}\pi^{-}l^{+}\nu$  & $D^{+}\to U_{-1/2}^{-}\pi^{+}l^{+}\nu$  & $D_{s}^{+}\to U_{1}^{+}\pi^{-}l^{+}\nu$  & $D_{s}^{+}\to U_{-1}^{-}\pi^{+}l^{+}\nu$ \tabularnewline
\hline 
\hline 
$D\to$$T_{I_{z}}^{Q}Pl\bar{\nu}$ &  &  & \tabularnewline
\hline 
$D^{0}\to T_{-3/2}^{--}\pi^{+}l^{+}\nu$  & $D_{s}^{+}\to T_{1}^{\prime+}\pi^{-}l^{+}\nu$  & $D^{+}\to T_{3/2}^{+}\pi^{-}l^{+}\nu$  & $D_{s}^{+}\to T_{-1}^{\prime-}\pi^{+}l^{+}\nu$ \tabularnewline
$D^{0}\to T_{1/2}^{0}\pi^{-}l^{+}\nu$  & $D^{+}\to T_{-1/2}^{\prime-}\pi^{+}l^{+}\nu$  & $D_{s}^{+}\to T_{-1}^{-}\pi^{+}l^{+}\nu$  &  \tabularnewline
\hline 
\end{tabular}
\end{table}
\begin{table}
\caption{Golden channels of $B\to$$U_{I_{z}}^{Q}P$.}
\label{tab:GoldenBtoU}%
\begin{tabular}{|cccc|}
\hline 
\hline 
$B\to$$U_{I_{z}}^{Q}P$$\ (b\to d)$ &  &  & \tabularnewline
\hline 
$B^{-}\to U_{-1/2}^{0}K^{-}$  & $\overline{B}^{0}\to U_{-1}^{-}\pi^{+}$  & $\overline{B}^{0}\to U_{1}^{+}\pi^{-}$  & $B^{-}\to U_{-1/2}^{-}K^{0}$ \tabularnewline
$B^{-}\to U_{-3/2}^{-}\overline{K}^{0}$  & $\overline{B}^{0}\to U_{-1/2}^{-}K^{+}$  & $\overline{B}_{s}^{0}\to U_{-3/2}^{-}\pi^{+}$  & $\overline{B}^{0}\to U_{1/2}^{+}K^{-}$ \tabularnewline
$B^{-}\to U_{0}^{0}\pi^{-}$  & $\overline{B}_{s}^{0}\to U_{1/2}^{+}\pi^{-}$  & $\overline{B}_{s}^{0}\to U_{-1}^{-}K^{+}$  & \tabularnewline
\hline 
\hline 
$B\to$$U_{I_{z}}^{Q}P$$\ (b\to s)$ &  &  & \tabularnewline
\hline 
$B^{-}\to U_{0}^{0}K^{-}$  & $\overline{B}^{0}\to U_{1}^{+}K^{-}$  & $\overline{B}_{s}^{0}\to U_{1/2}^{+}K^{-}$  & $\overline{B}_{s}^{0}\to U_{-1/2}^{-}K^{+}$ \tabularnewline
$B^{-}\to U_{1/2}^{0}\pi^{-}$  & $\overline{B}^{0}\to U_{-1/2}^{-}\pi^{+}$  & $\overline{B}_{s}^{0}\to U_{1}^{+}\pi^{-}$  & $\overline{B}_{s}^{0}\to U_{-1}^{-}\pi^{+}$ \tabularnewline
$B^{-}\to U_{0}^{-}K^{0}$  & $\overline{B}^{0}\to U_{0}^{-}K^{+}$  &  & \tabularnewline
\hline 
\end{tabular}
\end{table}
\begin{table}
\caption{Golden channels of $B\to$$T_{I_{z}}^{Q}P$.}
\label{tab:GoldenBtoT}%
\begin{tabular}{|cccc|}
\hline 
\hline 
$B\to$$T_{I_{z}}^{Q}P$$\ (b\to d)$ &  &  & \tabularnewline
\hline 
{$B^{-}\to T_{0}^{0}\pi^{-}$} & {{$\overline{B}^{0}\to T_{1/2}^{+}K^{-}$ } } & {{$\overline{B}^{0}\to T_{-1}^{-}\pi^{+}$ } } & {{$\overline{B}^{0}\to T_{1/2}^{\prime+}K^{-}$ } }\tabularnewline
{{$B^{-}\to T_{-1/2}^{-}K^{0}$ } } & {{$B^{-}\to T_{-3/2}^{-}\overline{K}^{0}$ } } & {{$\overline{B}^{0}\to T_{-1/2}^{-}K^{+}$ } } & {{$\overline{B}_{s}^{0}\to T_{-1}^{-}K^{+}$ } }\tabularnewline
{{$B^{-}\to T_{-2}^{--}\pi^{+}$ } } & {{$\overline{B}^{0}\to T_{1}^{\prime+}\pi^{-}$ } } & {{$\overline{B}^{0}\to T_{1}^{+}\pi^{-}$ } } & {{$\overline{B}_{s}^{0}\to T_{-1}^{\prime-}K^{+}$ } }\tabularnewline
{{$B^{-}\to T_{-3/2}^{--}K^{+}$ } } & {{$B^{-}\to T_{-1/2}^{\prime-}K^{0}$ } } & {{$\overline{B}^{0}\to T_{-1}^{\prime-}\pi^{+}$ } } & {{$\overline{B}_{s}^{0}\to T_{1/2}^{+}\pi^{-}$ } }\tabularnewline
{{$B^{-}\to T_{0}^{\prime0}\pi^{-}$ } } & {{$B^{-}\to T_{-1/2}^{0}K^{-}$ } } & {{$\overline{B}^{0}\to T_{-1/2}^{\prime-}K^{+}$ } } & {{$\overline{B}_{s}^{0}\to T_{-3/2}^{-}\pi^{+}$ } }\tabularnewline
{{$\overline{B}_{s}^{0}\to T_{1/2}^{\prime+}\pi^{-}$ } } & {{$\overline{B}_{s}^{0}\to T_{0}^{+}K^{-}$ } } &  & \tabularnewline
\hline 
\hline 
$B\to$$T_{I_{z}}^{Q}P$$\ (b\to s)$ &  &  & \tabularnewline
\hline 
{{$B^{-}\to T_{0}^{0}K^{-}$ } } & {{$B^{-}\to T_{1/2}^{0}\pi^{-}$ } } & {{$\overline{B}^{0}\to T_{0}^{-}K^{+}$ } } & {{$\overline{B}_{s}^{0}\to T_{1}^{+}\pi^{-}$ } }\tabularnewline
{{$B^{-}\to T_{-1}^{-}\overline{K}^{0}$ } } & {{$B^{-}\to T_{0}^{-}K^{0}$ } } & {{$\overline{B}^{0}\to T_{1}^{\prime+}K^{-}$ } } & {{$\overline{B}_{s}^{0}\to T_{-1}^{\prime-}\pi^{+}$ } }\tabularnewline
{{$B^{-}\to T_{-3/2}^{--}\pi^{+}$ } } & {{$\overline{B}^{0}\to T_{-1/2}^{-}\pi^{+}$ } } & {{$\overline{B}^{0}\to T_{3/2}^{+}\pi^{-}$ } } & {{$\overline{B}_{s}^{0}\to T_{-1/2}^{\prime-}K^{+}$ } }\tabularnewline
{{$B^{-}\to T_{-1}^{--}K^{+}$ } } & {{$\overline{B}^{0}\to T_{1}^{+}K^{-}$ } } & {{$\overline{B}_{s}^{0}\to T_{-1}^{-}\pi^{+}$ } } & {{$\overline{B}_{s}^{0}\to T_{1/2}^{+}K^{-}$ } }\tabularnewline
{{$B^{-}\to T_{0}^{\prime0}K^{-}$ } } & {{$\overline{B}^{0}\to T_{-1/2}^{\prime-}\pi^{+}$ } } & {{$\overline{B}_{s}^{0}\to T_{-1/2}^{-}K^{+}$ } } & {{$\overline{B}_{s}^{0}\to T_{1}^{\prime+}\pi^{-}$ } }\tabularnewline
{{$\overline{B}_{s}^{0}\to T_{1/2}^{\prime+}K^{-}$ } } & $B^{-}\to T_{-1}^{\prime-}\overline{K}^{0}$  &  & \tabularnewline
\hline 
\end{tabular}
\end{table}
\begin{table}
\caption{Golden channels for $D$ decays into a light tetraquark in 10 or 27 representation and a light meson.}
\label{tab:GoldenDdecay}%
\begin{tabular}{|c|cc|}
\hline 
\hline 
$D\to U_{I_{z}}^{Q}P$ & $D\to T_{I_{z}}^{Q}P$ & \tabularnewline
\hline 
$D^{0}\to U_{-1/2}^{-}\pi^{+}$  & $D^{0}\to T_{-1/2}^{\prime-}\pi^{+}$  & $D^{+}\to T_{1/2}^{\prime0}\pi^{+}$ \tabularnewline
$D^{+}\to U_{1/2}^{0}\pi^{+}$  & $D^{0}\to T_{-1/2}^{-}\pi^{+}$  & $D_{s}^{+}\to T_{0}^{0}\pi^{+}$ \tabularnewline
$D_{s}^{+}\to U_{0}^{0}\pi^{+}$  & $D^{0}\to T_{3/2}^{+}\pi^{-}$  & $D_{s}^{+}\to T_{0}^{\prime0}\pi^{+}$ \tabularnewline
 & $D^{+}\to T_{1/2}^{0}\pi^{+}$  & $D_{s}^{+}\to T_{2}^{++}\pi^{-}$ \tabularnewline
 & $D^{+}\to T_{1/2}^{\prime0}\pi^{+}$  & $D_{s}^{+}\to T_{0}^{\prime\prime0}\pi^{+}$ \tabularnewline
\hline 
\end{tabular}
\end{table}
\begin{table}
\caption{The best three-body decay channels for reconstructing $T_{27}$ from $B/D$ decays. Note that there is no $D\to (T_{I_{z}}^{Q}\to PP)Pl\bar{\nu}$ channels with all the three final mesons being pions.}
\label{tab:BestGoldenBtoT}%
\begin{tabular}{|ccc|}
\hline 
\hline 
$B\to T_{I_{z}}^{Q}Pl\bar{\nu}$  &  & \tabularnewline
\hline 
$\overline{B}^{0}\to(T_{0}^{\prime0}\to\pi^{+}\pi^{-})\pi^{+}l^{-}\bar{\nu}$  & $\overline{B}^{0}\to(T_{2}^{++}\to\pi^{+}\pi^{-})\pi^{-}l^{-}\bar{\nu}$  & $\overline{B}_{s}^{0}\to(T_{-1/2}^{0}\to\pi^{-}K^{+})\pi^{+}l^{-}\bar{\nu}$ \tabularnewline
$\overline{B}^{0}\to(T_{3/2}^{++}\to\pi^{+}K^{+})K^{-}l^{-}\bar{\nu}$  & $\overline{B}_{s}^{0}\to(T_{3/2}^{++}\to\pi^{+}K^{+})\pi^{-}l^{-}\bar{\nu}$  & $\overline{B}_{s}^{0}\to(T_{0}^{\prime0}\to\pi^{+}\pi^{-})K^{+}l^{-}\bar{\nu}$ \tabularnewline
$\overline{B}^{0}\to(T_{1/2}^{0}\to\pi^{+}K^{-})K^{+}l^{-}\bar{\nu}$  &  & \tabularnewline
\hline 
\hline 
$B/D\to T_{I_{z}}^{Q}P$  &  & \tabularnewline
\hline 
$B^{-}\to(T_{-2}^{--}\to\pi^{-}\pi^{-})\pi^{+}$  & $B^{-}\to(T_{-3/2}^{--}\to\pi^{-}K^{-})K^{+}$ & $B^{-}\to(T_{0}^{\prime0}\to\pi^{+}\pi^{-})\pi^{-}$\tabularnewline
$B^{-}\to(T_{-1/2}^{0}\to\pi^{-}K^{+})K^{-}$  & $B^{-}\to(T_{1/2}^{0}\to\pi^{+}K^{-})\pi^{-}$ & $B^{-}\to(T_{-3/2}^{--}\to\pi^{-}K^{-})\pi^{+}$\tabularnewline
$B^{-}\to(T_{0}^{\prime0}\to\pi^{+}\pi^{-})K^{-}$ &  & \tabularnewline
\hline 
$D_{s}^{+}\to(T_{0}^{\prime0}\to\pi^{+}\pi^{-})\pi^{+}$  & $D_{s}^{+}\to(T_{2}^{++}\to\pi^{+}\pi^{+})\pi^{-}$  & \tabularnewline
\hline 
\end{tabular}
\end{table}

\section{Conclusions}
\label{sec:conclusions}
In this work we performed a SU(3) analysis for both semi-leptonic and non-leptonic heavy meson weak decays into a pseudoscalar meson and a fully-light tetraquark. We firstly gave a tensor reduction for the SU(3) representation of fully-light tetraquark and point out that the $ss\bar u\bar d$ tetraquark that we mostly interested in belongs to the 10 or 27 representation. Accordingly, we used SU(3) symmetry to analyze the decays $B/D \to U/T~P~l\nu$, $B/D \to U/T~P $ and $B_c \to U/T~P/D$, with $U/T$ represents a fully-light tetraquark in 10 or 27 representation. For each decay modes we listed the decay amplitudes of all the channels and the relations among them. Finally, from the decay amplitudes of all the channels we listed the golden decay channels which are expected to have more possibilities to be observed in experiments. This study will provide a guidance for searching for the light four-quark states in the future experiments.

\section*{Acknowledgements}
The authors are very grateful to Prof. Wei Wang and Prof. Xiao-Rui Lyu for useful discussions. This work is
supported in part by the DFG and the NSFC through funds provided to the Sino-German CRC 110 ``Symmetries and the Emergence of
Structure in QCD'', the National Science Foundation of China under contract No.12005294 and  the National Science Foundation of China under contract No.12065020. 

\begin{appendix}
\section{Flavor wave functions of the fully-light tetraquarks}\label{flavorWFs}

In this appendix, we list all the flavor wave functions of the fully-light tetraquarks in the nine irreducible SU(3) representations.

27 Representation:
\begin{align}
(T_{27})_{11}^{11}=&\  \frac{1}{20} (ds\bar{d}\bar{s}+ds\bar{s}\bar{d}+sd\bar{d}\bar{s}+sd\bar{s}\bar{d}-3 du\bar{d}\bar{u}-3 du\bar{u}\bar{d}-3 ud\bar{d}\bar{u}-3 ud\bar{u}\bar{d}\nonumber\\ &\ +2 dd\bar{d}\bar{d}-3 su\bar{s}\bar{u}-3 su\bar{u}\bar{s}-3 us\bar{s}\bar{u}-3 us\bar{u}\bar{s}+2 ss\bar{s}\bar{s}+6 uu\bar{u}\bar{u}),\nonumber\\ (T_{27})_{11}^{12}=&\  \frac{1}{10} (-ds\bar{s}\bar{u}-ds\bar{u}\bar{s}-sd\bar{s}\bar{u}-sd\bar{u}\bar{s}-2 dd\bar{d}\bar{u}-2 dd\bar{u}\bar{d}+3 du\bar{u}\bar{u}+3 ud\bar{u}\bar{u}),\nonumber\\ (T_{27})_{11}^{13}=&\  \frac{1}{10} (-ds\bar{d}\bar{u}-ds\bar{u}\bar{d}-sd\bar{d}\bar{u}-sd\bar{u}\bar{d}-2 ss\bar{s}\bar{u}-2 ss\bar{u}\bar{s}+3 su\bar{u}\bar{u}+3 us\bar{u}\bar{u}),\nonumber\\ (T_{27})_{11}^{22}=&\  dd\bar{u}\bar{u},\nonumber\\ (T_{27})_{11}^{23}=&\  \frac{1}{2} (ds\bar{u}\bar{u}+sd\bar{u}\bar{u}),\nonumber\\ (T_{27})_{11}^{33}=&\  ss\bar{u}\bar{u},\nonumber\\ (T_{27})_{12}^{11}=&\  \frac{1}{10} (-su\bar{d}\bar{s}-su\bar{s}\bar{d}-us\bar{d}\bar{s}-us\bar{s}\bar{d}-2 du\bar{d}\bar{d}-2 ud\bar{d}\bar{d}+3 uu\bar{d}\bar{u}+3 uu\bar{u}\bar{d}),\nonumber\\ (T_{27})_{12}^{12}=&\  \frac{1}{40} (-ds\bar{d}\bar{s}-ds\bar{s}\bar{d}-sd\bar{d}\bar{s}-sd\bar{s}\bar{d}+7 du\bar{d}\bar{u}+7 du\bar{u}\bar{d}+7 ud\bar{d}\bar{u}+7 ud\bar{u}\bar{d}\nonumber\\ &\ -6 dd\bar{d}\bar{d}-su\bar{s}\bar{u}-su\bar{u}\bar{s}-us\bar{s}\bar{u}-us\bar{u}\bar{s}+2 ss\bar{s}\bar{s}-6 uu\bar{u}\bar{u}),\nonumber\\ (T_{27})_{12}^{13}=&\  \frac{1}{10} (2 su\bar{d}\bar{u}+2 su\bar{u}\bar{d}+2 us\bar{d}\bar{u}+2 us\bar{u}\bar{d}-ds\bar{d}\bar{d}-sd\bar{d}\bar{d}-ss\bar{d}\bar{s}-ss\bar{s}\bar{d}),\nonumber\\ (T_{27})_{12}^{22}=&\  \frac{1}{10} (-ds\bar{s}\bar{u}-ds\bar{u}\bar{s}-sd\bar{s}\bar{u}-sd\bar{u}\bar{s}+3 dd\bar{d}\bar{u}+3 dd\bar{u}\bar{d}-2 du\bar{u}\bar{u}-2 ud\bar{u}\bar{u}),\nonumber\\ (T_{27})_{12}^{23}=&\  \frac{1}{10} (2 ds\bar{d}\bar{u}+2 ds\bar{u}\bar{d}+2 sd\bar{d}\bar{u}+2 sd\bar{u}\bar{d}-ss\bar{s}\bar{u}-ss\bar{u}\bar{s}-su\bar{u}\bar{u}-us\bar{u}\bar{u}),\nonumber\\ (T_{27})_{12}^{33}=&\  \frac{1}{2} (ss\bar{d}\bar{u}+ss\bar{u}\bar{d}),\nonumber\\ (T_{27})_{13}^{11}=&\  \frac{1}{10} (-du\bar{d}\bar{s}-du\bar{s}\bar{d}-ud\bar{d}\bar{s}-ud\bar{s}\bar{d}-2 su\bar{s}\bar{s}-2 us\bar{s}\bar{s}+3 uu\bar{s}\bar{u}+3 uu\bar{u}\bar{s}),\nonumber\\ (T_{27})_{13}^{12}=&\  \frac{1}{10} (2 du\bar{s}\bar{u}+2 du\bar{u}\bar{s}+2 ud\bar{s}\bar{u}+2 ud\bar{u}\bar{s}-dd\bar{d}\bar{s}-dd\bar{s}\bar{d}-ds\bar{s}\bar{s}-sd\bar{s}\bar{s}),\nonumber\\ (T_{27})_{13}^{13}=&\  \frac{1}{40} (-ds\bar{d}\bar{s}-ds\bar{s}\bar{d}-sd\bar{d}\bar{s}-sd\bar{s}\bar{d}-du\bar{d}\bar{u}-du\bar{u}\bar{d}-ud\bar{d}\bar{u}-ud\bar{u}\bar{d}\nonumber\\ &\ +2 dd\bar{d}\bar{d}+7 su\bar{s}\bar{u}+7 su\bar{u}\bar{s}+7 us\bar{s}\bar{u}+7 us\bar{u}\bar{s}-6 ss\bar{s}\bar{s}-6 uu\bar{u}\bar{u}),\nonumber\\ (T_{27})_{13}^{22}=&\  \frac{1}{2} (dd\bar{s}\bar{u}+dd\bar{u}\bar{s}),\nonumber\\ (T_{27})_{13}^{23}=&\  \frac{1}{10} (2 ds\bar{s}\bar{u}+2 ds\bar{u}\bar{s}+2 sd\bar{s}\bar{u}+2 sd\bar{u}\bar{s}-dd\bar{d}\bar{u}-dd\bar{u}\bar{d}-du\bar{u}\bar{u}-ud\bar{u}\bar{u}),\nonumber\\ (T_{27})_{13}^{33}=&\  \frac{1}{10} (-ds\bar{d}\bar{u}-ds\bar{u}\bar{d}-sd\bar{d}\bar{u}-sd\bar{u}\bar{d}+3 ss\bar{s}\bar{u}+3 ss\bar{u}\bar{s}-2 su\bar{u}\bar{u}-2 us\bar{u}\bar{u}),\nonumber\\ (T_{27})_{22}^{11}=&\  uu\bar{d}\bar{d},\nonumber\\ (T_{27})_{22}^{12}=&\  \frac{1}{10} (-su\bar{d}\bar{s}-su\bar{s}\bar{d}-us\bar{d}\bar{s}-us\bar{s}\bar{d}+3 du\bar{d}\bar{d}+3 ud\bar{d}\bar{d}-2 uu\bar{d}\bar{u}-2 uu\bar{u}\bar{d}),\nonumber\\ (T_{27})_{22}^{13}=&\  \frac{1}{2} (su\bar{d}\bar{d}+us\bar{d}\bar{d}),\nonumber\\ (T_{27})_{22}^{22}=&\  \frac{1}{20} (-3 ds\bar{d}\bar{s}-3 ds\bar{s}\bar{d}-3 sd\bar{d}\bar{s}-3 sd\bar{s}\bar{d}-3 du\bar{d}\bar{u}-3 du\bar{u}\bar{d}-3 ud\bar{d}\bar{u}\nonumber\\ &\ -3 ud\bar{u}\bar{d}+6 dd\bar{d}\bar{d}+su\bar{s}\bar{u}+su\bar{u}\bar{s}+us\bar{s}\bar{u}+us\bar{u}\bar{s}+2 ss\bar{s}\bar{s}+2 uu\bar{u}\bar{u}),\nonumber\\ (T_{27})_{22}^{23}=&\  \frac{1}{10} (-su\bar{d}\bar{u}-su\bar{u}\bar{d}-us\bar{d}\bar{u}-us\bar{u}\bar{d}+3 ds\bar{d}\bar{d}+3 sd\bar{d}\bar{d}-2 ss\bar{d}\bar{s}-2 ss\bar{s}\bar{d}),\nonumber\\ (T_{27})_{22}^{33}=&\  ss\bar{d}\bar{d},\nonumber\\ (T_{27})_{23}^{11}=&\  \frac{1}{2} (uu\bar{d}\bar{s}+uu\bar{s}\bar{d}),\nonumber\\ (T_{27})_{23}^{12}=&\  \frac{1}{10} (2 du\bar{d}\bar{s}+2 du\bar{s}\bar{d}+2 ud\bar{d}\bar{s}+2 ud\bar{s}\bar{d}-su\bar{s}\bar{s}-us\bar{s}\bar{s}-uu\bar{s}\bar{u}-uu\bar{u}\bar{s}),\nonumber\\ (T_{27})_{23}^{13}=&\  \frac{1}{10} (2 su\bar{d}\bar{s}+2 su\bar{s}\bar{d}+2 us\bar{d}\bar{s}+2 us\bar{s}\bar{d}-du\bar{d}\bar{d}-ud\bar{d}\bar{d}-uu\bar{d}\bar{u}-uu\bar{u}\bar{d}),\nonumber\\ (T_{27})_{23}^{22}=&\  \frac{1}{10} (-du\bar{s}\bar{u}-du\bar{u}\bar{s}-ud\bar{s}\bar{u}-ud\bar{u}\bar{s}+3 dd\bar{d}\bar{s}+3 dd\bar{s}\bar{d}-2 ds\bar{s}\bar{s}-2 sd\bar{s}\bar{s}),\nonumber\\ (T_{27})_{23}^{23}=&\  \frac{1}{40} (7 ds\bar{d}\bar{s}+7 ds\bar{s}\bar{d}+7 sd\bar{d}\bar{s}+7 sd\bar{s}\bar{d}-du\bar{d}\bar{u}-du\bar{u}\bar{d}-ud\bar{d}\bar{u}-ud\bar{u}\bar{d}\nonumber\\ &\ -6 dd\bar{d}\bar{d}-su\bar{s}\bar{u}-su\bar{u}\bar{s}-us\bar{s}\bar{u}-us\bar{u}\bar{s}-6 ss\bar{s}\bar{s}+2 uu\bar{u}\bar{u}),\nonumber\\ (T_{27})_{23}^{33}=&\  \frac{1}{10} (-su\bar{d}\bar{u}-su\bar{u}\bar{d}-us\bar{d}\bar{u}-us\bar{u}\bar{d}-2 ds\bar{d}\bar{d}-2 sd\bar{d}\bar{d}+3 ss\bar{d}\bar{s}+3 ss\bar{s}\bar{d}),\nonumber\\ (T_{27})_{33}^{11}=&\  uu\bar{s}\bar{s},\nonumber\\ (T_{27})_{33}^{12}=&\  \frac{1}{2} (du\bar{s}\bar{s}+ud\bar{s}\bar{s}),\nonumber\\ (T_{27})_{33}^{13}=&\  \frac{1}{10} (-du\bar{d}\bar{s}-du\bar{s}\bar{d}-ud\bar{d}\bar{s}-ud\bar{s}\bar{d}+3 su\bar{s}\bar{s}+3 us\bar{s}\bar{s}-2 uu\bar{s}\bar{u}-2 uu\bar{u}\bar{s}),\nonumber\\ (T_{27})_{33}^{22}=&\  dd\bar{s}\bar{s},\nonumber\\ (T_{27})_{33}^{23}=&\  \frac{1}{10} (-du\bar{s}\bar{u}-du\bar{u}\bar{s}-ud\bar{s}\bar{u}-ud\bar{u}\bar{s}-2 dd\bar{d}\bar{s}-2 dd\bar{s}\bar{d}+3 ds\bar{s}\bar{s}+3 sd\bar{s}\bar{s}),\nonumber\\ (T_{27})_{33}^{33}=&\  \frac{1}{20} (-3 ds\bar{d}\bar{s}-3 ds\bar{s}\bar{d}-3 sd\bar{d}\bar{s}-3 sd\bar{s}\bar{d}+du\bar{d}\bar{u}+du\bar{u}\bar{d}+ud\bar{d}\bar{u}+ud\bar{u}\bar{d}\nonumber\\ &\ +2 dd\bar{d}\bar{d}-3 su\bar{s}\bar{u}-3 su\bar{u}\bar{s}-3 us\bar{s}\bar{u}-3 us\bar{u}\bar{s}+6 ss\bar{s}\bar{s}+2 uu\bar{u}\bar{u}).
\end{align}
$10$ and $\overline{10}$ Representations:
\begin{align}
(T_{10})^{111}=&\  uu\bar{d}\bar{s}-uu\bar{s}\bar{d},\nonumber\\ (T_{10})^{112}=&\  \frac{1}{3} (du\bar{d}\bar{s}-du\bar{s}\bar{d}+ud\bar{d}\bar{s}-ud\bar{s}\bar{d}+uu\bar{s}\bar{u}-uu\bar{u}\bar{s}),\nonumber\\ (T_{10})^{113}=&\  \frac{1}{3} (su\bar{d}\bar{s}-su\bar{s}\bar{d}+us\bar{d}\bar{s}-us\bar{s}\bar{d}-uu\bar{d}\bar{u}+uu\bar{u}\bar{d}),\nonumber\\ (T_{10})^{122}=&\  \frac{1}{3} (du\bar{s}\bar{u}-du\bar{u}\bar{s}+ud\bar{s}\bar{u}-ud\bar{u}\bar{s}+dd\bar{d}\bar{s}-dd\bar{s}\bar{d}),\nonumber\\ (T_{10})^{222}=&\  dd\bar{s}\bar{u}-dd\bar{u}\bar{s},\nonumber\\ (T_{10})^{123}=&\  \frac{1}{6} (ds\bar{d}\bar{s}-ds\bar{s}\bar{d}+sd\bar{d}\bar{s}-sd\bar{s}\bar{d}-du\bar{d}\bar{u}+du\bar{u}\bar{d}-ud\bar{d}\bar{u}+ud\bar{u}\bar{d}+su\bar{s}\bar{u}\nonumber\\ &\ -su\bar{u}\bar{s}+us\bar{s}\bar{u}-us\bar{u}\bar{s}),\nonumber\\ (T_{10})^{133}=&\  \frac{1}{3} (-su\bar{d}\bar{u}+su\bar{u}\bar{d}-us\bar{d}\bar{u}+us\bar{u}\bar{d}+ss\bar{d}\bar{s}-ss\bar{s}\bar{d}),\nonumber\\ (T_{10})^{223}=&\  \frac{1}{3} (ds\bar{s}\bar{u}-ds\bar{u}\bar{s}+sd\bar{s}\bar{u}-sd\bar{u}\bar{s}-dd\bar{d}\bar{u}+dd\bar{u}\bar{d}),\nonumber\\ (T_{10})^{233}=&\  \frac{1}{3} (-ds\bar{d}\bar{u}+ds\bar{u}\bar{d}-sd\bar{d}\bar{u}+sd\bar{u}\bar{d}+ss\bar{s}\bar{u}-ss\bar{u}\bar{s}),\nonumber\\ (T_{10})^{333}=&\  ss\bar{u}\bar{d}-ss\bar{d}\bar{u},\\
(T_{\overline{10}})_{111}=&\  ds\bar{u}\bar{u}-sd\bar{u}\bar{u},\nonumber\\ (T_{\overline{10}})_{112}=&\  \frac{1}{3} (ds\bar{d}\bar{u}+ds\bar{u}\bar{d}-sd\bar{d}\bar{u}-sd\bar{u}\bar{d}+su\bar{u}\bar{u}-us\bar{u}\bar{u}),\nonumber\\ (T_{\overline{10}})_{113}=&\  \frac{1}{3} (ds\bar{s}\bar{u}+ds\bar{u}\bar{s}-sd\bar{s}\bar{u}-sd\bar{u}\bar{s}-du\bar{u}\bar{u}+ud\bar{u}\bar{u}),\nonumber\\ (T_{\overline{10}})_{122}=&\  \frac{1}{3} (su\bar{d}\bar{u}+su\bar{u}\bar{d}-us\bar{d}\bar{u}-us\bar{u}\bar{d}+ds\bar{d}\bar{d}-sd\bar{d}\bar{d}),\nonumber\\ (T_{\overline{10}})_{222}=&\  su\bar{d}\bar{d}-us\bar{d}\bar{d},\nonumber\\ (T_{\overline{10}})_{123}=&\  \frac{1}{6} (ds\bar{d}\bar{s}+ds\bar{s}\bar{d}-sd\bar{d}\bar{s}-sd\bar{s}\bar{d}-du\bar{d}\bar{u}-du\bar{u}\bar{d}+ud\bar{d}\bar{u}+ud\bar{u}\bar{d}+su\bar{s}\bar{u}\nonumber\\ &\ +su\bar{u}\bar{s}-us\bar{s}\bar{u}-us\bar{u}\bar{s}),\nonumber\\ (T_{\overline{10}})_{133}=&\  \frac{1}{3} (-du\bar{s}\bar{u}-du\bar{u}\bar{s}+ud\bar{s}\bar{u}+ud\bar{u}\bar{s}+ds\bar{s}\bar{s}-sd\bar{s}\bar{s}),\nonumber\\ (T_{\overline{10}})_{223}=&\  \frac{1}{3} (su\bar{d}\bar{s}+su\bar{s}\bar{d}-us\bar{d}\bar{s}-us\bar{s}\bar{d}-du\bar{d}\bar{d}+ud\bar{d}\bar{d}),\nonumber\\ (T_{\overline{10}})_{233}=&\  \frac{1}{3} (-du\bar{d}\bar{s}-du\bar{s}\bar{d}+ud\bar{d}\bar{s}+ud\bar{s}\bar{d}+su\bar{s}\bar{s}-us\bar{s}\bar{s}),\nonumber\\ (T_{\overline{10}})_{333}=&\  ud\bar{s}\bar{s}-du\bar{s}\bar{s}.
\end{align}
Four $8$ Representations:
\begin{align}
(T_{8}^{(1)})_{1}^{1}=&\  \frac{1}{12} (-2 ds\bar{d}\bar{s}-2 ds\bar{s}\bar{d}-2 sd\bar{d}\bar{s}-2 sd\bar{s}\bar{d}+du\bar{d}\bar{u}+du\bar{u}\bar{d}+ud\bar{d}\bar{u}+ud\bar{u}\bar{d}\nonumber\\ &\ -4 dd\bar{d}\bar{d}+su\bar{s}\bar{u}+su\bar{u}\bar{s}+us\bar{s}\bar{u}+us\bar{u}\bar{s}-4 ss\bar{s}\bar{s}+8 uu\bar{u}\bar{u}),\nonumber\\ (T_{8}^{(1)})_{1}^{2}=&\  \frac{1}{4} (su\bar{d}\bar{s}+su\bar{s}\bar{d}+us\bar{d}\bar{s}+us\bar{s}\bar{d}+2 du\bar{d}\bar{d}+2 ud\bar{d}\bar{d}+2 uu\bar{d}\bar{u}+2 uu\bar{u}\bar{d}),\nonumber\\ (T_{8}^{(1)})_{1}^{3}=&\  \frac{1}{4} (du\bar{d}\bar{s}+du\bar{s}\bar{d}+ud\bar{d}\bar{s}+ud\bar{s}\bar{d}+2 su\bar{s}\bar{s}+2 us\bar{s}\bar{s}+2 uu\bar{s}\bar{u}+2 uu\bar{u}\bar{s}),\nonumber\\ (T_{8}^{(1)})_{2}^{2}=&\  \frac{1}{12} (ds\bar{d}\bar{s}+ds\bar{s}\bar{d}+sd\bar{d}\bar{s}+sd\bar{s}\bar{d}+du\bar{d}\bar{u}+du\bar{u}\bar{d}+ud\bar{d}\bar{u}+ud\bar{u}\bar{d}+8 dd\bar{d}\bar{d}\nonumber\\ &\ -2 su\bar{s}\bar{u}-2 su\bar{u}\bar{s}-2 us\bar{s}\bar{u}-2 us\bar{u}\bar{s}-4 ss\bar{s}\bar{s}-4 uu\bar{u}\bar{u}),\nonumber\\ (T_{8}^{(1)})_{2}^{3}=&\  \frac{1}{4} (du\bar{s}\bar{u}+du\bar{u}\bar{s}+ud\bar{s}\bar{u}+ud\bar{u}\bar{s}+2 dd\bar{d}\bar{s}+2 dd\bar{s}\bar{d}+2 ds\bar{s}\bar{s}+2 sd\bar{s}\bar{s}),\nonumber\\ (T_{8}^{(1)})_{3}^{3}=&\  \frac{1}{12} (ds\bar{d}\bar{s}+ds\bar{s}\bar{d}+sd\bar{d}\bar{s}+sd\bar{s}\bar{d}-2 du\bar{d}\bar{u}-2 du\bar{u}\bar{d}-2 ud\bar{d}\bar{u}-2 ud\bar{u}\bar{d}\nonumber\\ &\ -4 dd\bar{d}\bar{d}+su\bar{s}\bar{u}+su\bar{u}\bar{s}+us\bar{s}\bar{u}+us\bar{u}\bar{s}+8 ss\bar{s}\bar{s}-4 uu\bar{u}\bar{u}),\nonumber\\ (T_{8}^{(1)})_{3}^{2}=&\  \frac{1}{4} (su\bar{d}\bar{u}+su\bar{u}\bar{d}+us\bar{d}\bar{u}+us\bar{u}\bar{d}+2 ds\bar{d}\bar{d}+2 sd\bar{d}\bar{d}+2 ss\bar{d}\bar{s}+2 ss\bar{s}\bar{d}),\nonumber\\ (T_{8}^{(1)})_{3}^{1}=&\  \frac{1}{4} (ds\bar{d}\bar{u}+ds\bar{u}\bar{d}+sd\bar{d}\bar{u}+sd\bar{u}\bar{d}+2 ss\bar{s}\bar{u}+2 ss\bar{u}\bar{s}+2 su\bar{u}\bar{u}+2 us\bar{u}\bar{u}),\nonumber\\ (T_{8}^{(1)})_{2}^{1}=&\  \frac{1}{4} (ds\bar{s}\bar{u}+ds\bar{u}\bar{s}+sd\bar{s}\bar{u}+sd\bar{u}\bar{s}+2 dd\bar{d}\bar{u}+2 dd\bar{u}\bar{d}+2 du\bar{u}\bar{u}+2 ud\bar{u}\bar{u}),\\
(T_{8}^{(2)})_{1}^{1}=&\  -\frac{3}{2} (du\bar{d}\bar{u}-du\bar{u}\bar{d}+ud\bar{d}\bar{u}-ud\bar{u}\bar{d}+su\bar{s}\bar{u}-su\bar{u}\bar{s}+us\bar{s}\bar{u}-us\bar{u}\bar{s}),\nonumber\\ (T_{8}^{(2)})_{1}^{2}=&\  \frac{3}{2} (su\bar{d}\bar{s}-su\bar{s}\bar{d}+us\bar{d}\bar{s}-us\bar{s}\bar{d}+2 uu\bar{d}\bar{u}-2 uu\bar{u}\bar{d}),\nonumber\\ (T_{8}^{(2)})_{1}^{3}=&\  -\frac{3}{2} (du\bar{d}\bar{s}-du\bar{s}\bar{d}+ud\bar{d}\bar{s}-ud\bar{s}\bar{d}-2 uu\bar{s}\bar{u}+2 uu\bar{u}\bar{s}),\nonumber\\ (T_{8}^{(2)})_{2}^{2}=&\  \frac{3}{2} (ds\bar{d}\bar{s}-ds\bar{s}\bar{d}+sd\bar{d}\bar{s}-sd\bar{s}\bar{d}+du\bar{d}\bar{u}-du\bar{u}\bar{d}+ud\bar{d}\bar{u}-ud\bar{u}\bar{d}),\nonumber\\ (T_{8}^{(2)})_{2}^{3}=&\  -\frac{3}{2} (-du\bar{s}\bar{u}+du\bar{u}\bar{s}-ud\bar{s}\bar{u}+ud\bar{u}\bar{s}+2 dd\bar{d}\bar{s}-2 dd\bar{s}\bar{d}),\nonumber\\ (T_{8}^{(2)})_{3}^{3}=&\  -\frac{3}{2} (ds\bar{d}\bar{s}-ds\bar{s}\bar{d}+sd\bar{d}\bar{s}-sd\bar{s}\bar{d}-su\bar{s}\bar{u}+su\bar{u}\bar{s}-us\bar{s}\bar{u}+us\bar{u}\bar{s}),\nonumber\\ (T_{8}^{(2)})_{3}^{2}=&\  \frac{3}{2} (su\bar{d}\bar{u}-su\bar{u}\bar{d}+us\bar{d}\bar{u}-us\bar{u}\bar{d}+2 ss\bar{d}\bar{s}-2 ss\bar{s}\bar{d}),\nonumber\\ (T_{8}^{(2)})_{3}^{1}=&\  -\frac{3}{2} (ds\bar{d}\bar{u}-ds\bar{u}\bar{d}+sd\bar{d}\bar{u}-sd\bar{u}\bar{d}+2 ss\bar{s}\bar{u}-2 ss\bar{u}\bar{s}),\nonumber\\ (T_{8}^{(2)})_{2}^{1}=&\  -\frac{3}{2} (ds\bar{s}\bar{u}-ds\bar{u}\bar{s}+sd\bar{s}\bar{u}-sd\bar{u}\bar{s}+2 dd\bar{d}\bar{u}-2 dd\bar{u}\bar{d}),\\
(T_{8}^{(3)})_{1}^{1}=&\  -\frac{3}{2} (du\bar{d}\bar{u}+du\bar{u}\bar{d}-ud\bar{d}\bar{u}-ud\bar{u}\bar{d}+su\bar{s}\bar{u}+su\bar{u}\bar{s}-us\bar{s}\bar{u}-us\bar{u}\bar{s}),\nonumber\\ (T_{8}^{(3)})_{1}^{2}=&\  -\frac{3}{2} (su\bar{d}\bar{s}+su\bar{s}\bar{d}-us\bar{d}\bar{s}-us\bar{s}\bar{d}+2 du\bar{d}\bar{d}-2 ud\bar{d}\bar{d}),\nonumber\\ (T_{8}^{(3)})_{1}^{3}=&\  -\frac{3}{2} (du\bar{d}\bar{s}+du\bar{s}\bar{d}-ud\bar{d}\bar{s}-ud\bar{s}\bar{d}+2 su\bar{s}\bar{s}-2 us\bar{s}\bar{s}),\nonumber\\ (T_{8}^{(3)})_{2}^{2}=&\  \frac{3}{2} (ds\bar{d}\bar{s}+ds\bar{s}\bar{d}-sd\bar{d}\bar{s}-sd\bar{s}\bar{d}+du\bar{d}\bar{u}+du\bar{u}\bar{d}-ud\bar{d}\bar{u}-ud\bar{u}\bar{d}),\nonumber\\ (T_{8}^{(3)})_{2}^{3}=&\  \frac{3}{2} (du\bar{s}\bar{u}+du\bar{u}\bar{s}-ud\bar{s}\bar{u}-ud\bar{u}\bar{s}+2 ds\bar{s}\bar{s}-2 sd\bar{s}\bar{s}),\nonumber\\ (T_{8}^{(3)})_{3}^{3}=&\  -\frac{3}{2} (ds\bar{d}\bar{s}+ds\bar{s}\bar{d}-sd\bar{d}\bar{s}-sd\bar{s}\bar{d}-su\bar{s}\bar{u}-su\bar{u}\bar{s}+us\bar{s}\bar{u}+us\bar{u}\bar{s}),\nonumber\\ (T_{8}^{(3)})_{3}^{2}=&\  -\frac{3}{2} (-su\bar{d}\bar{u}-su\bar{u}\bar{d}+us\bar{d}\bar{u}+us\bar{u}\bar{d}+2 ds\bar{d}\bar{d}-2 sd\bar{d}\bar{d}),\nonumber\\ (T_{8}^{(3)})_{3}^{1}=&\  -\frac{3}{2} (ds\bar{d}\bar{u}+ds\bar{u}\bar{d}-sd\bar{d}\bar{u}-sd\bar{u}\bar{d}-2 su\bar{u}\bar{u}+2 us\bar{u}\bar{u}),\nonumber\\ (T_{8}^{(3)})_{2}^{1}=&\  \frac{3}{2} (ds\bar{s}\bar{u}+ds\bar{u}\bar{s}-sd\bar{s}\bar{u}-sd\bar{u}\bar{s}+2 du\bar{u}\bar{u}-2 ud\bar{u}\bar{u}),\\
(T_{8}^{(4)})_{1}^{1}=&\  \frac{1}{6} (2 ds\bar{d}\bar{s}-2 ds\bar{s}\bar{d}-2 sd\bar{d}\bar{s}+2 sd\bar{s}\bar{d}-du\bar{d}\bar{u}+du\bar{u}\bar{d}+ud\bar{d}\bar{u}-ud\bar{u}\bar{d}\nonumber\\ &\ -su\bar{s}\bar{u}+su\bar{u}\bar{s}+us\bar{s}\bar{u}-us\bar{u}\bar{s}),\nonumber\\ (T_{8}^{(4)})_{1}^{2}=&\  \frac{1}{2} (su\bar{d}\bar{s}-su\bar{s}\bar{d}-us\bar{d}\bar{s}+us\bar{s}\bar{d}),\nonumber\\ (T_{8}^{(4)})_{1}^{3}=&\  \frac{1}{2} (-du\bar{d}\bar{s}+du\bar{s}\bar{d}+ud\bar{d}\bar{s}-ud\bar{s}\bar{d}),\nonumber\\ (T_{8}^{(4)})_{2}^{2}=&\  \frac{1}{6} (-ds\bar{d}\bar{s}+ds\bar{s}\bar{d}+sd\bar{d}\bar{s}-sd\bar{s}\bar{d}-du\bar{d}\bar{u}+du\bar{u}\bar{d}+ud\bar{d}\bar{u}-ud\bar{u}\bar{d}\nonumber\\ &\ +2 su\bar{s}\bar{u}-2 su\bar{u}\bar{s}-2 us\bar{s}\bar{u}+2 us\bar{u}\bar{s}),\nonumber\\ (T_{8}^{(4)})_{2}^{3}=&\  \frac{1}{2} (-du\bar{s}\bar{u}+du\bar{u}\bar{s}+ud\bar{s}\bar{u}-ud\bar{u}\bar{s}),\nonumber\\ (T_{8}^{(4)})_{3}^{3}=&\  \frac{1}{6} (-ds\bar{d}\bar{s}+ds\bar{s}\bar{d}+sd\bar{d}\bar{s}-sd\bar{s}\bar{d}+2 du\bar{d}\bar{u}-2 du\bar{u}\bar{d}-2 ud\bar{d}\bar{u}+2 ud\bar{u}\bar{d}\nonumber\\ &\ -su\bar{s}\bar{u}+su\bar{u}\bar{s}+us\bar{s}\bar{u}-us\bar{u}\bar{s}),\nonumber\\ (T_{8}^{(4)})_{3}^{2}=&\  \frac{1}{2} (-su\bar{d}\bar{u}+su\bar{u}\bar{d}+us\bar{d}\bar{u}-us\bar{u}\bar{d}),\nonumber\\ (T_{8}^{(4)})_{3}^{1}=&\  \frac{1}{2} (-ds\bar{d}\bar{u}+ds\bar{u}\bar{d}+sd\bar{d}\bar{u}-sd\bar{u}\bar{d}),\nonumber\\ (T_{8}^{(4)})_{2}^{1}=&\  \frac{1}{2} (ds\bar{s}\bar{u}-ds\bar{u}\bar{s}-sd\bar{s}\bar{u}+sd\bar{u}\bar{s}).
\end{align}
Two singlets:
\begin{align}
(T_{1}^{(1)})=&\  \frac{1}{2} (ds\bar{d}\bar{s}+ds\bar{s}\bar{d}+sd\bar{d}\bar{s}+sd\bar{s}\bar{d}+du\bar{d}\bar{u}+du\bar{u}\bar{d}+ud\bar{d}\bar{u}+ud\bar{u}\bar{d}\nonumber\\ &\ +2 dd\bar{d}\bar{d}+su\bar{s}\bar{u}+su\bar{u}\bar{s}+us\bar{s}\bar{u}+us\bar{u}\bar{s}+2 ss\bar{s}\bar{s}+2 uu\bar{u}\bar{u}),\nonumber\\ (T_{1}^{(2)})=&\  \frac{1}{2} (ds\bar{d}\bar{s}-ds\bar{s}\bar{d}-sd\bar{d}\bar{s}+sd\bar{s}\bar{d}+du\bar{d}\bar{u}-du\bar{u}\bar{d}-ud\bar{d}\bar{u}+ud\bar{u}\bar{d}\nonumber\\ &\ +su\bar{s}\bar{u}-su\bar{u}\bar{s}-us\bar{s}\bar{u}+us\bar{u}\bar{s}).
\end{align}
\end{appendix}

\end{document}